%% file: SUS-21-006_temp.tex
\pdfoutput=1
\documentclass[11pt,twoside,a4paper,cmspaper,final,collab]{cms-tdr}
\def\svnVersion{7445b35}\def\svnDate{2024/04/26}\def\cmsCernNoTag{CERN-EP-2023-209}\def\cmsCernDate{\today}\def\cmsMessage{Published in Physical Review D as \href{http://dx.doi.org/10.1103/PhysRevD.109.072007}{\doi{10.1103/PhysRevD.109.072007}.}}
\begin{document}\cmsNoteHeader{SUS-21-006}

\newlength\cmsFigWidth
\ifthenelse{\boolean{cms@external}}{\setlength\cmsFigWidth{0.49\textwidth}}{\setlength\cmsFigWidth{0.9\textwidth}}
\ifthenelse{\boolean{cms@external}}{\providecommand{\cmsAppendix}{}}{\providecommand{\cmsAppendix}{Appendix~}}
\ifthenelse{\boolean{cms@external}}{\providecommand{\NA}{\ensuremath{\cdots}\xspace}}{\providecommand{\NA}{\ensuremath{\text{---}}\xspace}}
\providecommand{\cmsTable}[1]{\resizebox{\textwidth}{!}{#1}}
\newlength\cmsTabSkip\setlength{\cmsTabSkip}{1.5ex}
\newcommand{\njets}{\ensuremath{N_{\text{jet}}}\xspace}
\newcommand{\nbjets}{\ensuremath{N_{\PQb\text{-jet}}}\xspace}
\newcommand{\nelec}{\ensuremath{N_{\Pe}}\xspace}
\newcommand{\nmuon}{\ensuremath{N_{\Pgm}}\xspace}
\newcommand{\nDTk}{\ensuremath{N_{\mathrm{DTk}}}\xspace}
\newcommand{\nshort}{\ensuremath{N_{\text{short}}}\xspace}
\newcommand{\nlong}{\ensuremath{N_{\mathrm{long}}}\xspace}
\newcommand{\lsp}{\PSGczDo}
\newcommand{\chargino}{\PSGcpmDo}
\newcommand{\mchargino}{\ensuremath{m_\chargino}\xspace}
\newcommand{\mlsp}{\ensuremath{m_{\lsp}}\xspace}
\newcommand{\mgluino}{\ensuremath{m_{\PSg}}\xspace}
\newcommand{\msquark}{\ensuremath{m_{\PSQ}}\xspace}
\newcommand{\mDTkell}{\ensuremath{m_{\text{DTk},\ell}}\xspace}
\newcommand{\wjets}{{{\PW}\!+jets}\xspace}
\newcommand{\zjets}{{{\PZ}\!+jets}\xspace}
\newcommand{\dR}{\ensuremath{\Delta R}\xspace}
\newcommand{\imini}{\ensuremath{I}\xspace}
\newcommand{\rscale}{\ensuremath{\mu_{\text{R}}}\xspace}
\newcommand{\fscale}{\ensuremath{\mu_{\text{F}}}\xspace}
\newcommand{\sieie}{\ensuremath{\sigma_{i \eta i \eta}}\xspace}
\newcommand{\dedx}{\ensuremath{\rd E/\rd x}\xspace}
\newcommand{\fdep}{\ensuremath{f_{\text{dep}}}\xspace}
\newcommand{\fdeplow}{\ensuremath{\fdep^{\text{low}}}\xspace}
\newcommand{\fdephi}{\ensuremath{\fdep^{\text{high}}}\xspace}
\newcommand{\edep}{\ensuremath{E_{\text{dep}}}\xspace}
\newcommand{\thilow}{\ensuremath{\theta^{\text{high}}_{\text{low}}}\xspace}
\newcommand{\harmonic}{\ensuremath{I_{\mathrm{h}}}\xspace}
\newcommand{\isorel}{\ensuremath{I_{\text{rel}}}\xspace}
\newcommand{\klowhi}{\ensuremath{\kappa^{\text{low}}_{\text{high}}}\xspace}
\newcommand{\crreal}{\ensuremath{\mathrm{CR}^{\text{genuine}}}\xspace}
\newcommand{\nCRreal}{\ensuremath{N_{\mathrm{CR}^{\text{genuine}}_{i}}}\xspace}
\newcommand{\kmuon}{\ensuremath{\kappa^{\mu\,\text{veto}}_{\mu\,\text{match}}}\xspace}
\newcommand{\crrealmu}{\ensuremath{\mathrm{CR}^{\text{genuine-}\mu}}\xspace}
\newcommand{\nCRrealmu}{\ensuremath{N_{\mathrm{CR}^{\text{genuine-}\mu}_{i}}}\xspace}
\newcommand{\crfake}{\ensuremath{\mathrm{CR}^{\text{spurious}}}\xspace}
\newcommand{\nbdthi}{\ensuremath{N^{\text{QCD}}_{\text{BDT}^{\text{high}}}}\xspace}
\newcommand{\nbdtlo}{\ensuremath{N^{\text{QCD}}_{\text{BDT}^{\text{low}}}}\xspace}
\newcommand{\nfake}{\ensuremath{N^{\text{spurious}}_{\text{SR}_i}}\xspace}
\newcommand{\nCRfake}{\ensuremath{N_{\text{CR}_{i}^{\text{spurious}}}}\xspace}
\newcommand{\nfakehi}{\ensuremath{(N^{\text{spurious}}_{\text{SR}_i})_{\dedx\text{-high}}}\xspace}
\newcommand{\nfakelo}{\ensuremath{(N^{\text{spurious}}_{\text{SR}_i})_{\dedx\text{-low}}}\xspace}
\newcommand{\njisr}{\ensuremath{N_{\text{jet}}^{\text{ISR}}}\xspace}
\newcommand{\mdedx}{\ensuremath{m_{\text{DTk;\,\dedx}}}\xspace}
\newcommand{\dmplus}{\ensuremath{\Delta m^{\pm}}\xspace}
\newcommand{\dmstrong}{\ensuremath{\Delta m_{(\PSg,\PSQ)-\mathrm{LSP}}}\xspace}
\newcommand{\ptmisshard}{\ensuremath{p_{\mathrm{T,hard}}^{\mathrm{miss}}}\xspace}
\newcommand{\dedxlohi}{\ensuremath{\phi^{\text{high}}_{\text{low}}}\xspace}
\newcommand{\charginogen}{\ensuremath{\widetilde{\chi}^{\pm}_{1,\text{gen}}}\xspace}

\newcolumntype{R}{>{$}r<{$}}
\newcolumntype{L}{>{$}l<{$}}
\newcolumntype{M}{R@{$\;$}L}
\newcolumntype{S}{r@{$\,\pm\,$}r}

\cmsNoteHeader{SUS-21-006}
\title{Search for supersymmetry in final states with disappearing tracks in
  proton-proton collisions at \texorpdfstring{$\sqrt{s} = 13\TeV$}{sqrt(s) = 13 TeV} }

\date{\today}

\abstract{A search is presented for charged, long-lived supersymmetric particles in final states with one or more disappearing tracks. The search is based on data from proton-proton collisions at a center-of-mass energy of 13\TeV collected with the CMS detector at the CERN LHC between 2016 and 2018, corresponding to an integrated luminosity of 137\fbinv. The search is performed over final states  characterized by varying numbers of jets, \mbox{\PQb-tagged} jets, electrons, and muons. The length of signal-candidate tracks in the plane perpendicular to the beam axis is used to characterize the lifetimes of wino- and higgsino-like charginos produced in the context of the minimal supersymmetric standard model. The \dedx energy loss of signal-candidate tracks is used to increase the sensitivity to charginos with a large mass and thus a small Lorentz boost. The observed results are found to be statistically consistent with the background-only hypothesis.  Limits on the pair production cross section of gluinos and squarks are presented in the framework of simplified models of supersymmetric particle production and decay, and for electroweakino production based on models of wino and higgsino dark matter. The limits presented are the most stringent to date for scenarios with light third-generation squarks and a wino- or higgsino-like dark matter candidate capable of explaining the observed dark matter relic density.}

\hypersetup{
pdfauthor={CMS Collaboration},
pdftitle={Search for supersymmetry in final states with disappearing tracks in proton-proton collisions at sqrt(s) = 13 TeV},
pdfsubject={CMS},
pdfkeywords={CMS, SUSY, disappearing, tracks}}

\maketitle 

\section{Introduction}
\label{sec:introduction}

Supersymmetry
(SUSY)~\cite{Ramond:1971gb,Golfand:1971iw,Neveu:1971rx,Volkov:1972jx,Wess:1973kz,Wess:1974tw,Fayet:1974pd,Fayet:1976cr,Nilles:1983ge},
a well-motivated extension of the standard model (SM) of particle physics,
addresses open questions of the SM such as dark matter (DM)~\cite{Zwicky:1937zza,Rubin:1970zza}
and the fine tuning~\cite{Susskind:1978ms,tHooft:1979rat,Veltman:1980mj} of the electroweak sector.
Studies from the ATLAS~\cite{Aad:2008zzm} and CMS~\cite{Chatrchyan:2008aa}
Collaborations at the CERN LHC have placed significant constraints on
the minimal supersymmetric SM (MSSM),
with no direct evidence as yet for the existence of SUSY particles.
Nonetheless,
important regions of the MSSM parameter space remain unexplored,
such as regions with a nearly pure wino- or higgsino-like
lightest supersymmetric particle (LSP)
that are consistent with the observed DM relic density~\cite{Planck:2018vyg}.

A DM candidate can arise in SUSY if the LSP is stable,
as occurs in models with \mbox{$R$-parity~\cite{Fayet:1974pd,Farrar:1978xj}} conservation.
In viable realizations of the $R$-parity conserving MSSM,
the LSP is a stable neutralino~\lsp,
namely an electroweakino mass eigenstate composed of a mixture
of bino, wino, and higgsino states.
To be consistent with the observed DM relic density,
a nearly pure wino (higgsino) LSP must have a mass of
approximately 2 (1)\TeV in such realizations~\cite{Delgado:2020url}.
The constraint arises from the value of the co-annihilation cross section
required to sufficiently reduce the DM content of the early universe
before thermal freeze-out~\cite{Griest:1990kh,Jungman:1995df}.
The co-annihilation cross section depends on the masses of the LSP and its co-annihilation partner.
For a pure wino- or higgsino-like LSP,
this partner is a chargino \chargino that is nearly degenerate in mass with the LSP,
having a mass only a few hundred MeV larger.
Because of the limited kinematic phase space in $\chargino\to\lsp\PGppm$ decays,
the \chargino then  has a macroscopic lifetime,
with a decay length at the LHC on the order of several centimeters or more.

Long-lived charginos associated with pure wino or higgsino DM 
give rise to the distinctive experimental signature known as a disappearing track (DTk).
In $\chargino\to\lsp\PGppm$ decays,
the resulting pion has a momentum of only a few hundred~MeV, 
too low, in general, for it to be reconstructed.
The \lsp in these decays escapes undetected with most of the momentum.
The chargino is reconstructed as a track with hits up to its point of decay,
beyond which no hits or associated signals are recorded.
This leads to the DTk signature in which a reconstructed track
emanating from the beam collision region
ends abruptly within the sensitive tracking volume,
with a continuation that has ``disappeared.''
The DTk signature has been used in recent searches to set
the most stringent chargino mass limits to date for certain scenarios, e.g.,
wino or higgsino LSP scenarios with minimal neutralino
mixing~\cite{CMS:2019ybf,CMS:2020atg,ATLAS:2022rme}.

This paper presents a search for long-lived charginos,
probing a comprehensive set of final states relevant to the $R$-parity conserving MSSM.
The analysis is based on a sample of proton-proton ($\Pp\Pp$) collision events collected
at center-of-mass energy $\sqrt{s}=13\TeV$ with the CMS detector in 2016--2018,
corresponding to an integrated luminosity of 137\fbinv.
It expands on previous searches~\cite{CMS:2019ybf,CMS:2020atg}
by introducing electron+DTk and muon+DTk channels,
by re-analyzing the fully hadronic channel,
and by making use of a machine-learning-based track classification method
to improve the DTk selection efficiency and background rejection.
The addition of the electron+DTk and muon+DTk channels increases sensitivity to
new-physics scenarios with leptonic final states.
Additional sensitivity to high-mass LSP states
is achieved by employing a measure of the \dedx ionization energy loss
of signal-candidate tracks,
which takes large values for candidates with a small Lorentz boost.
The analysis targets both the electroweak and strong production mechanisms,
making use of jet and \PQb-tagged jet multiplicities to distinguish between 
different production modes and decay chains.
The incorporation of new final states and observables improves the sensitivity to
relatively unexplored regions of the MSSM,
including scenarios with pure wino or higgsino DM with masses of 1\TeV or larger,
or with light third-generation squarks that yield leptons in the final state.

This paper is structured as follows.
The signal models considered for the analysis are presented in Section~\ref{sec:sms}.
The CMS detector is described in Section~\ref{sec:detector}
and the event reconstruction in Section~\ref{sec:reconstruction}.
Section~\ref{sec:mc} discusses the simulated event samples used in the study.
The DTk selection procedure is presented in Section~\ref{sec:tracks}
and the definition of the search regions (SRs) in Section~\ref{sec:events}.
The methods used to estimate SM backgrounds are presented in Section~\ref{sec:background}.
Systematic uncertainties are discussed in Section~\ref{sec:systematic}.
The results and summary are presented in Sections~\ref{sec:results}
and~\ref{sec:summary}, respectively. 
Tabulated results for the study are provided
in the HEPData record for this analysis~\cite{hepdata}.

\section{Simplified signal models}
\label{sec:sms}

\begin{figure*}[t]
\centering
\includegraphics[width=0.32\linewidth]{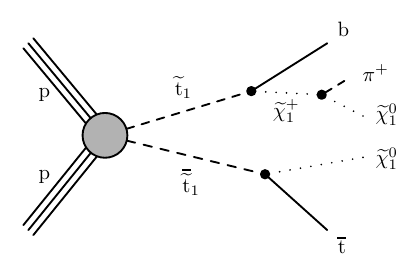}
\includegraphics[width=0.32\linewidth]{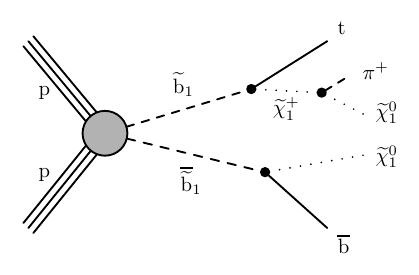} 
\includegraphics[width=0.32\linewidth]{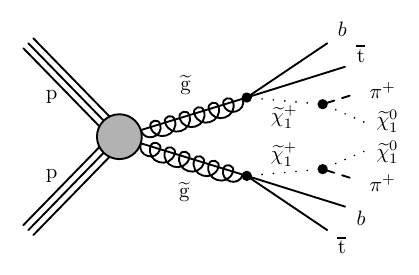}
\includegraphics[width=0.32\linewidth]{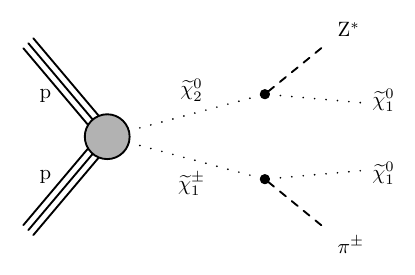} \,
\includegraphics[width=0.32\linewidth]{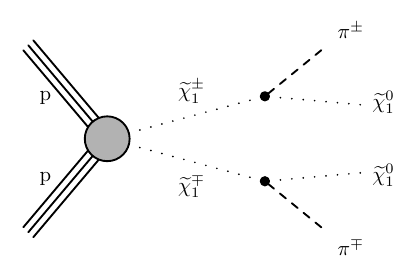} \, 
\includegraphics[width=0.32\linewidth]{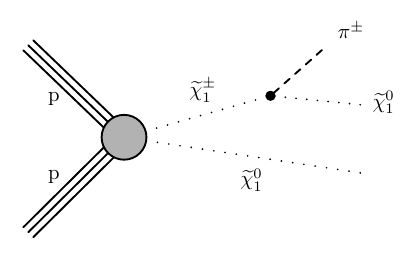} \\
\caption{
Representative diagrams for the simplified models considered in this analysis.
From left to right: T6btLL, T6tbLL, and T5btbtLL (upper);
and TChiWZ, TChiWW, and TChiW (lower).
The shaded circles at the production vertices represent a sum over perturbative terms.
}
\label{fig:SMSFD}
\end{figure*}

A number of simplified models of SUSY~\cite{bib-sms-2,bib-sms-4,Chatrchyan:2013sza}
are used to survey and characterize the sensitivity of the search.
These models,
depicted by representative production diagrams in Fig.~\ref{fig:SMSFD}
and summarized in Table~\ref{tab:SMSNAMES},
are effective-Lagrangian descriptions defined by the SUSY particle masses and
the production and decay processes.
A simplified model focuses on a specific small set of particles and interactions,
with all other new-physics particles considered to be irrelevant to that particular process.
One model, denoted T6btLL, features top squark pair production in which the
top squark decays as $\sTop\to\PQt\lsp$ and $\sTop\to\PQb\chargino$,
each with 50\% probability.
An analogous model with bottom squark production is labeled~T6tbLL.
A third model, denoted T5btbtLL,
features gluino pair production in which the gluino decays as
$\PSg\to\bbbar\lsp$,
$\ttbar\lsp$,
$\PQt\PAQb\PSGcmDo$,
or $\PAQt\PQb\PSGcpDo$,
each with 25\% probability.
Various proper decay lengths $c\tau$ of the chargino are considered.
For the models with gluino and squark pair production,
we focus on a benchmark model with $c\tau=10\cm$,
corresponding roughly to pure wino and higgsino-like states.
We also consider a model with $c\tau=200\cm$,
which can be realized in a scenario with a bino-like LSP and a wino coannihilation partner.
The chargino and LSP are required to be mass degenerate within a few hundred~MeV.
We set this mass difference to 180\MeV.
This choice has no direct impact on the acceptance because
the decay products of the chargino are not reconstructed.

Sensitivity to the direct production of either a nearly pure wino DM candidate 
or a nearly pure higgsino DM candidate is considered
through the topologies denoted TChiWZ, TChiWW, and TChiW.
The \PSGczDt particle depicted for the TChiWZ diagram
[Fig.~\ref{fig:SMSFD} (lower left)]
is the second lightest neutralino, only present in the higgsino scenario.
Relationships among the electroweakino masses,
the \chargino lifetime,
and the \chargino decay width
are constrained by radiative corrections that account for a large difference
between the LSP mass and the SUSY-breaking scale.
They are characterized in Refs.~\cite{Ibe:2012sx,Ibe:2022lkl} 
for the wino scenario and in 
Refs.~\cite{Nagata:2014wma,Fukuda:2017jmk} for the higgsino scenario.
The mass difference 
$\Delta m^{0} = \Delta m(\PSGczDt, \lsp )$ between the two neutral states
is taken to be twice the mass difference 
$\dmplus = \Delta m(\chargino, \lsp )$ between the charged and lightest states.
This choice corresponds to large $\tan\beta$,
where $\tan\beta$ is the ratio of the vacuum expectation values of the
neutral components of the two Higgs doublets in the MSSM~\cite{Martin:1997ns}.
The lifetime of the \PSGczDt,
while not directly relevant for this study,
is taken from a tree-level calculation based on the program SUSYHIT\,1.5a~\cite{Allanach:2001kg}.
The free parameters of the model are
\dmplus and the chargino mass~\mchargino.   

\section{The CMS detector and analysis triggers}
\label{sec:detector}

\begin{table*}[t]
\topcaption{
Overview of the simplified models of supersymmetry considered in this analysis.
}
\centering
\renewcommand{\arraystretch}{1.15}
\cmsTable{
\begin{scotch}{lll}
Model name & Description	                                                         & Processes \\ \hline
T6btLL     & Top squark associated \chargino production                                  & $\sTop\sTop$; $\sTop\to\PQt\lsp$, $\sTop\to\PQb\chargino$ \\
T6tbLL     & Bottom squark associated \chargino production                               & $\sBot\sBot$; $\sBot\to\PQb\lsp$, $\sBot\to\PQt\chargino$ \\
T5btbtLL   & Gluino associated \chargino production                                      & $\PSg\PSg$; $\PSg\to\bbbar\lsp$, $\ttbar\lsp$, $\PQt\PAQb\PSGcmDo$, $\PAQt\PQb\PSGcpDo$ \\ [\cmsTabSkip]
TChiWZ     & \multicolumn{1}{c}{\multirow{3}{*}{Electroweakino production for wino or higgsino LSP}} & $\PSGczDt\chargino$; $\PSGczDt\to {\PZst} \lsp$, $\chargino\to\lsp\pi^{\pm}$      \\
TChiWW     & \multicolumn{1}{c}{}                                                        & $\tilde{\chi}_{1}^{+}\tilde{\chi}_{1}^{-}$; $\chargino\to\lsp\pi^{\pm}$      \\
TChiW      & \multicolumn{1}{c}{}                                                        & $\chargino \lsp$; $\chargino\to\lsp\pi^{\pm}$, $\lsp$ \\
\end{scotch}
}
\label{tab:SMSNAMES}
\end{table*}

The CMS detector is structured around 
a cylindrical superconducting solenoid with an inner diameter of 6\unit{m}.
The solenoid provides a nearly uniform 3.8\unit{T} magnetic field within its volume.
This volume contains a number of nested particle detectors:
from the innermost region outward,
a silicon pixel and strip tracking detector,
a lead tungstate crystal electromagnetic calorimeter (ECAL),
and a brass and scintillator hadron calorimeter (HCAL).
The ECAL and HCAL,
each composed of a barrel and two endcap sections,
cover the pseudorapidity range $\abs{\eta}<3.0$.
Forward calorimeters extend the coverage to $3.0<\abs{\eta}<5.2$.
Muons are measured with gas-ionization detectors
embedded in a steel flux-return yoke outside the solenoid, allowing
muon reconstruction within $\abs{\eta}<2.4$.
The CMS components taken together provide a nearly hermetic detector,
permitting accurate measurements of the missing transverse momentum~\ptmiss. 

The tracking detector is of special importance for this analysis.
The tracker consists of an inner pixel detector and an outer strip detector.
The tracker used for the 2016 data-taking period,
referred to as the ``Phase-0'' tracker,
measured particles within $\abs{\eta}<2.5$.
At the beginning of 2017,
an upgraded pixel detector was installed~\cite{Phase1Pixel}.
The upgraded tracker is referred to as the ``Phase-1'' tracker.
The Phase-1 tracker,
used for the 2017 and 2018 data-taking periods,
measured particles within $\abs{\eta} < 3.0$. 
Tracks traversing the tracker system encounter 3 (4) pixel layers
within a radius of 102 (160)\mm in the Phase-0 (Phase-1) tracker.
They encounter up to around 10 layers in the strip detector within a radius of 1.2\unit{m}.
Compared to the Phase-0 tracker,
the Phase-1 tracker provides improved tracking and vertex resolution,
enhanced \PQb-tagging performance,
and a better measurement of the \dedx track energy loss~\cite{Phase1Pixel}.

Events of interest are selected using a two-tiered trigger system~\cite{Khachatryan:2016bia}.
The first level (L1), composed of custom hardware processors,
uses information from the calorimeters and muon detectors
to select events at a rate of around 100\unit{kHz}
within a fixed latency of about 4\mus~\cite{CMS:2020cmk}.
The second level, known as the high-level trigger,
consists of a farm of processors running a version of the full event reconstruction software
optimized for fast processing, and reduces the event rate to around 1\unit{kHz}
before data storage~\cite{Khachatryan:2016bia}. 
For the hadronic DTk channel, signal event candidates were recorded
by requiring \ptmiss at the trigger level
to exceed a threshold that varied between 100 and 120\GeV,
depending on the LHC instantaneous luminosity.
For the electron+DTk (muon+DTk) channel, signal event candidates were recorded
by requiring an electron (muon) candidate with $\pt>27$ (24)\GeV at the trigger level,
where \pt is the transverse momentum.
The trigger efficiencies are measured in data using independent cross triggers
and are found to exceed 97\% for events in the hadronic DTk channel
and 80 (90)\% for $\pt>40\GeV$ in the electron+DTk (muon+DTk) channel,
for events that satisfy the selection criteria given below
in Section~\ref{sec:events}.

A detailed description of the CMS detector,
along with definitions of the coordinate system and kinematic variables,
is given in Ref.~\cite{Chatrchyan:2008aa}.

\section{Event reconstruction}
\label{sec:reconstruction}

{\tolerance=800
Individual particles are reconstructed as particle-flow (PF) objects
with the CMS PF algorithm~\cite{CMS:2017yfk},
which identifies them as photons,
charged hadrons, neutral hadrons, electrons, or muons.
To improve the quality of the electron reconstruction,
additional criteria (beyond those of the PF algorithm) are imposed on the
\sieie variable~\cite{CMS:2020uim},
which is a measure of the width of the ECAL
shower shape in the $\eta$ coordinate,
and on the ratio $H/E$ of energies associated with the electron candidate
in the HCAL and ECAL~\cite{CMS:2020uim}.
Specifically,
small values of \sieie and $H/E$ are required,
with the criteria optimized separately
for the barrel and endcaps~\cite{CMS:2020uim}.
For muon candidates,
additional criteria
are imposed on the matching between track segments reconstructed in the
silicon tracker and muon detector~\cite{Sirunyan:2018fpa}.
Electron and muon candidates are required to have $\pt>40\gev$ and $\abs{\eta}<2.4$.
\par}

The primary vertex (PV) is taken to be the vertex corresponding to
the hardest scattering in the event,
evaluated using tracking information alone as described
in Section 9.4.1 of Ref.~\cite{CMS-TDR-15-02}.
Charged-particle tracks associated with vertices other than
the PV are removed from further consideration.
The PV is required to lie within 24\cm of the
center of the detector in the direction along the beam axis
and within 2\cm in the plane transverse to that axis.

Jets are defined by clustering PF candidates
using the anti-\kt jet algorithm~\cite{Cacciari:2008gp,Cacciari:2011ma}
with a distance parameter of~0.4.
Jet quality criteria~\cite{cms-pas-jme-10-003,CMS-PAS-JME-16-003}
are imposed to eliminate jets from
spurious sources such as electronics noise.
The jet energies are corrected for the nonlinear response of the
detector~\cite{Khachatryan:2016kdb}
and to account for the expected contributions of neutral
particles from extraneous $\Pp\Pp$ interactions (pileup)~\cite{Cacciari:2007fd}.
Jets are required to have $\pt>30\GeV$,
$\abs{\eta}<2.4$,
and to not overlap with an identified lepton
within a cone of radius
$\dR\equiv\sqrt{\smash[b]{(\Delta\phi)^2+(\Delta\eta)^2}}=0.4$
around the jet direction,
where $\phi$ is the azimuthal angle.
The number of selected jets in an event is denoted~\njets.

The identification of \PQb jets is performed by applying
a version of the combined secondary vertex algorithm based on
deep neural networks (\textsc{DeepCSV})~\cite{Sirunyan:2017ezt} to the selected jet sample.
The medium working point of this algorithm is used.
The tagging efficiency for \PQb jets with $\pt\approx30\GeV$ is 65\%.
The corresponding misidentification probability for gluon
and light-quark (\PQu, \PQd, \PQs) jets is 1.6\%,
while that for charm quark jets is 13\%.
The number of identified \PQb jets in an event is denoted~\nbjets.

To suppress the contributions of jets erroneously identified as leptons
as well as those due to genuine leptons from hadron decays,
electron and muon candidates are subjected to a lepton-isolation requirement.
This isolation criterion is based on the variable~\imini,
which is the sum of the scalar \pt of charged hadron,
neutral hadron, and photon PF candidates within a cone of radius \dR
around the lepton direction,
divided by the lepton~\pt.
The expected contributions to this sum of neutral particles from pileup
are subtracted~\cite{Cacciari:2007fd}.
The radius \dR of the cone is 0.2 for lepton $\pt<50\GeV$,
$10\GeV/\pt$ for $50<\pt< 200\GeV$,
and 0.05 for $\pt>200\GeV$.
The decrease in \dR with increasing lepton \pt
accounts for the increased collimation of the
decay products from the lepton's parent particle
as the Lorentz boost of the parent particle
increases~\cite{Rehermann:2010vq}.
The isolation requirement is $\imini<0.1$ (0.2) for electrons (muons).

The analysis uses a measure of the missing transverse momentum
called the ``hard~\ptmiss'' \ptmisshard, 
defined as the magnitude of the vector sum of \ptvec over all PF jets
with $\pt>30\GeV$ and $\abs{\eta}<5.0$,
where the PF jets are defined as outlined above but are clustered inclusively,
meaning that isolated leptons and photons are included as ``jets.''
The lepton-isolation and jet-lepton overlap rejection criteria described above
are not applied to jets used to compute \ptmisshard.
Hard~\ptmiss is more robust against pileup and soft radiation
than the \ptmiss variable 
calculated through a sum over individual particles in an event rather than jets.

\section{Simulated event samples}
\label{sec:mc}

Simulated events are used to train boosted decision tree (BDT) multivariate binary classifiers
and to estimate the signal acceptance in the SRs.
The BDTs are used to help identify DTk candidates,
as discussed in Section~\ref{sec:tracks}.

{\tolerance=2000
The SM production of \ttbar, \wjets, and \zjets events,
as well as of multijet events produced through quantum chromodynamics (QCD) processes,
is simulated at leading order (LO) precision
using the {\MGvATNLO}\,2.2.2~\cite{Alwall:2014hca,Alwall:2007fs} event generator.
The \ttbar events are generated with
up to three additional partons in the matrix element calculation.
The \wjets and \zjets events are generated
with up to four additional partons.
Single top quark events produced through the $s$~channel,
diboson events such as those originating from
$\PW\PW$, $\PZ\PZ$, or $\PZ\PH$ production (where $\PH$ is the SM Higgs boson),
and rare events such as those from $\ttbar\PW$,
$\ttbar\PZ$, and $\PW\PW\PZ$ production,
are generated with the {\MGvATNLO} generator
at next-to-leading order (NLO)~\cite{Frederix:2012ps},
with the exception of $\PW\PW$ events when both \PW bosons decay leptonically,
which are generated with the
{\POWHEG}\,v2.0~\cite{Nason:2004rx,Frixione:2007vw,Alioli:2010xd,Alioli:2009je,Re:2010bp}
generator at NLO.
This same \POWHEG generator is used to describe single top quark event
production through the $t$ and $\PQt\PW$ channels.
Normalization of the simulated background samples is based on the cross section
calculations of Refs.~\cite{Alioli:2009je,Re:2010bp,Alwall:2014hca,Melia:2011tj,Beneke:2011mq,
Cacciari:2011hy,Baernreuther:2012ws,Czakon:2012zr,Czakon:2012pz,Czakon:2013goa,
Gavin:2012sy,Gavin:2010az},
which generally correspond to NLO or next-to-NLO (NNLO) precision.
The detector response is based on the \GEANTfour~\cite{Agostinelli:2002hh}
suite of programs.
\par}

With the exception of the pure wino and higgsino models,
simulated signal events are generated at LO using the {\MGvATNLO} generator,
with up to two additional partons included.
The production cross sections are
determined with approximate NNLO plus next-to-next-to-leading logarithmic (NNLL)
accuracy~\cite{bib-nlo-nll-01,bib-nlo-nll-02,bib-nlo-nll-03,bib-nlo-nll-04,bib-nlo-nll-05,
Beenakker:2016lwe,Beenakker:2011sf,Beenakker:2013mva,Beenakker:2014sma,
Beenakker:1997ut,Beenakker:2010nq,Beenakker:2016gmf}.
Events with gluino (squark) pair production are generated
for a range of gluino \mgluino (squark \msquark) and LSP \mlsp mass values,
with $\mlsp<\mgluino$ ($\mlsp<\msquark$).
The ranges of mass considered vary according to the model,
but are generally from around 600--2500\GeV for \mgluino,
200--1700\GeV for \msquark,
and 0--1500\GeV for~\mlsp.
The gluinos and squarks decay according to
the phase space model of Ref.~\cite{Sjostrand:2014zea}.
For the pure wino and higgsino models,
simulated signal events are generated at LO using the
{\PYTHIA}\,8.205 generator~\cite{Sjostrand:2014zea}.

To render the computational requirements manageable,
the detector response for signal events is based on the CMS fast
simulation program~\cite{Abdullin:2011zz,Giammanco:2014bza},
which yields results that are generally
consistent with those from {\GEANTfour}.
To improve the consistency with {\GEANTfour},
we apply a correction of 1\% to the fast simulation samples
to account for differences in the efficiency
of the jet quality requirements~\cite{cms-pas-jme-10-003,CMS-PAS-JME-16-003}
and corrections of 5--12\% to account for differences
in the \PQb jet tagging efficiency.
Additional corrections are applied to account for differences
in the efficiency of tagging charginos with the Phase-0 and -1 trackers.

{\tolerance=2000
Parton showering and hadronization are simulated
with the {\PYTHIA}\,8.205 generator~\cite{Sjostrand:2014zea}.
For background events,
the Phase-0 samples use the CUETP8M1~\cite{Khachatryan:2015pea} tune
while the Phase-1 samples use the CP5~\cite{CMS:2019csb} tune.
For signal events,
the CP2~\cite{CMS:2019csb} tune is used.
Simulated samples generated at LO (NLO) with the CUETP8M1 tune use the
NNPDF3.0LO (NNPDF3.0NLO)~\cite{Ball:2013hta}
parton distribution function (PDF).
Those generated with the CP2 or CP5 tune use the NNPDF3.1LO
(NNPDF3.1NNLO)~\cite{Ball:2017nwa} PDF.
\par}

To improve the {\MGvATNLO} modeling of the jet multiplicity from initial-state radiation (ISR),
we follow the procedure introduced in Refs.~\cite{CMS:2017abv,CMS:2017okm}.
A control sample enriched in \ttbar events is selected in both data and simulation
by requiring two lepton candidates ($\Pe\Pe$, $\Pgm\Pgm$, or $\Pe\Pgm$)
and two tagged {\cPqb} jets.
The number of all other jets in an event is denoted~\njisr.
Reweighting factors are applied to the simulated \ttbar events
so that the \njisr distribution in simulation agrees with that in data.
The same reweighting factors are also applied to the simulated signal events.
The reweighting factors are 0.920, 0.821, 0.715, 0.662, 0.561, and 0.511 for 
$\njisr= 1,2,3,4,5,$ and $\geq 6$, respectively.

\section{Selection of disappearing track candidates}
\label{sec:tracks}

Two categories of DTk are defined: short tracks and long tracks.
Short tracks have hits recorded only in the pixel detector.
Long tracks have hits in both the pixel and strip detectors.
The track \pt is required to exceed 25 (40)\GeV for the short (long) category.
The pseudorapidity is required to satisfy $\abs{\eta}<2.0$.
This latter requirement reduces the background from spurious tracks
(Section~\ref{sec:background}),
which appears preferentially in the forward region.
To ensure that the tracks are isolated,
we calculate the relative isolation \isorel,
defined by the sum of the scalar \pt of other tracks within a cone of radius $\dR=0.3$
around the track direction,
divided by the track~\pt.
We require $\isorel<0.2$.
The impact parameter $d_{xy}$ of the track in the transverse plane
and the impact parameter $d_{z}$ in the direction
along the beam axis must both be less than 0.1\cm,
where $d_{xy}$ and $d_{z}$ are measured relative to the~PV.
The track is required to have at least three hits,
one of which must be on the innermost pixel layer.
Studies of simulated data show that charginos that decay within the tracker
are rarely reconstructed as PF candidates.
We therefore require the track to lie at least $\dR=0.01$
away from any reconstructed particle and at least $\dR=0.4$
away from any reconstructed jet with $\pt>15$\GeV.

To select tracks that are consistent with a~DTk,
long tracks are required to not have hits in the outermost two layers of the strip detector,
while short tracks are required to not have any hits whatsoever in the strip detector.
In addition, tracks must not be identified as an electron, muon, or hadron.
To ensure that DTk candidates are not associated with a large
deposit of calorimetric energy,
we calculate the quantity \edep,
which is the sum of ECAL and HCAL cluster energies
deposited within $\dR=0.4$ of the candidate track.
Different criteria on $\edep$ and $\edep/p$
(with $p$ the track momentum) are used to define SRs
and control regions (CRs) for short and long tracks,
respectively, and for the Phase-0 and -1 detectors,
as indicated in Table~\ref{tab:sr-cr}.
We use $\edep/p$ for long tracks because the resulting signal distribution peaks near~1,
which is convenient for the selection process.
However, because the uncertainty in $p$ is larger for short tracks,
the $\edep/p$ distribution for short tracks does not exhibit a clear peak,
making it simpler to use~\edep.
(The momentum resolution for long tracks is 10\% or better,
while for short tracks it is 20\% or worse,
depending on the number of hits associated with the track.)
To further reduce background from electrons,
additional criteria are applied to reject tracks that enter detector regions
with an ECAL crystal with anomalously low light yield,
a disabled crystal, or a noisy channel.
These additional criteria reduce the signal efficiency by less than~2\%.

\begin{table*}[t]
\topcaption{
Selection criteria on the BDT classifier score
and on the calorimetric energy \edep
associated with a disappearing track candidate
for the search region (SR) and control region (CR) samples
discussed in Section~\ref{sec:background}.
}
\centering
\cmsTable{
\begin{scotch}{lllll}
Phase/category            & Phase 0/short   & Phase 0/long   & Phase 1/short   & Phase 1/long  \\ \hline
BDT for SR samples        & ${>}0.1$        & ${>}0.12$      & ${>}0.15$       & ${>}0.08$     \\ 
\edep for SR samples      & ${<}15\GeV$     & ${<}0.2p$      & ${<}15\GeV$     & ${<}0.2p$     \\[\cmsTabSkip] 
\edep for \crreal samples & $[30, 300]$\GeV & $[0.3,1.2]p$   & $[30,300]$\GeV  & $[0.3,1.2]p$  \\ 
BDT for \crreal samples   & ${>}0.1$        & ${>}0.05$      & ${>}0.05$       & ${>}0.08$     \\[\cmsTabSkip]
BDT for \crfake samples   & $[-0.10,-0.05]$ & $[-0.1,0.0]$   & $[-0.10,0.05]$  & $[-0.1,0.0]$  \\
\end{scotch}
}
\label{tab:sr-cr}
\end{table*}

\begin{figure*}[t]
\centering
\includegraphics[width=.45\linewidth]{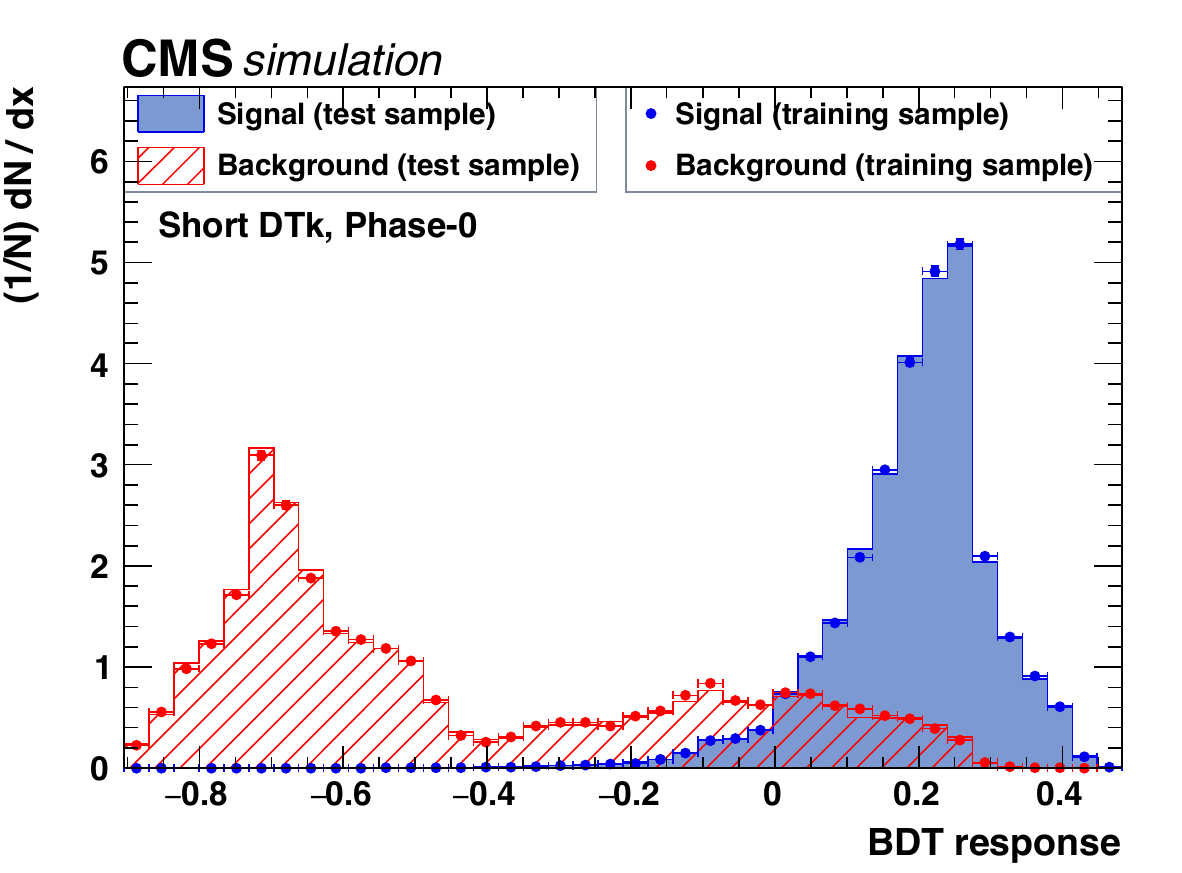}
\includegraphics[width=.45\linewidth]{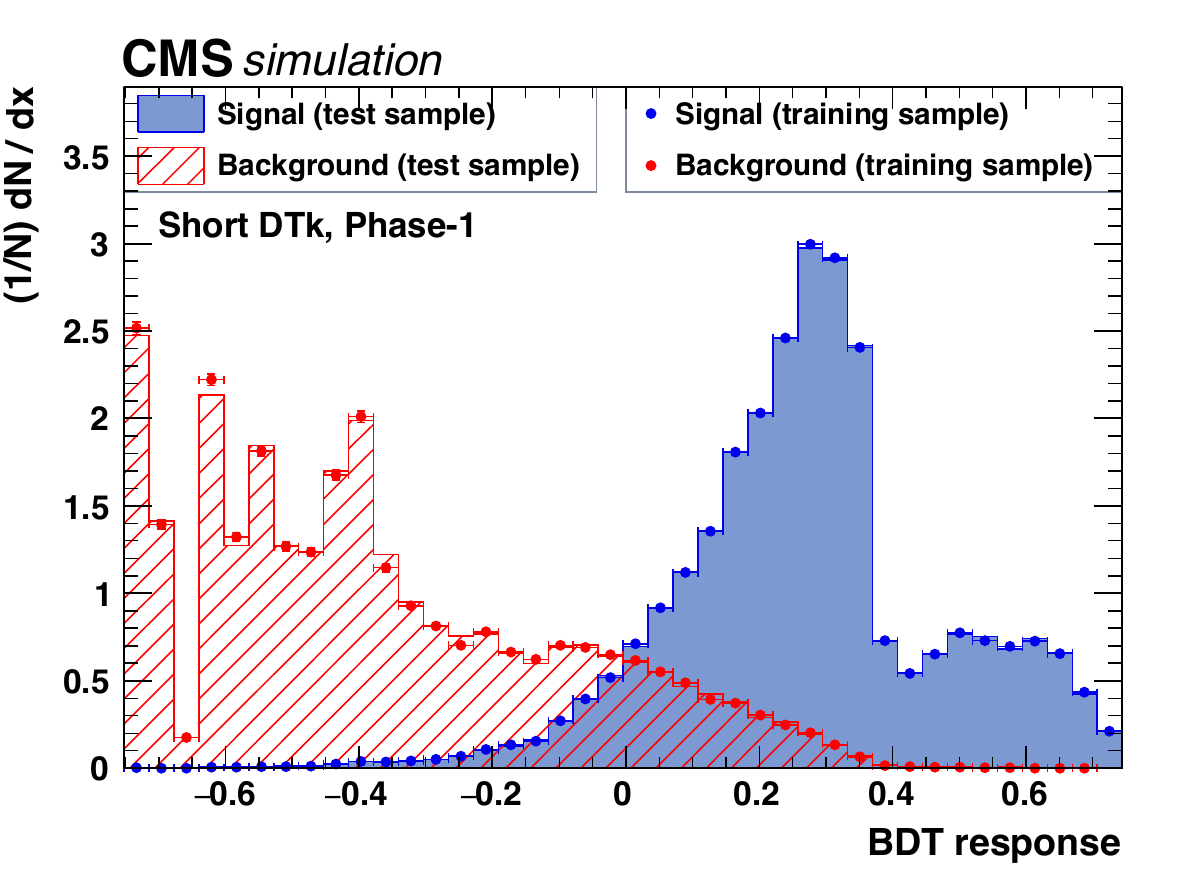}\\
\includegraphics[width=.45\linewidth]{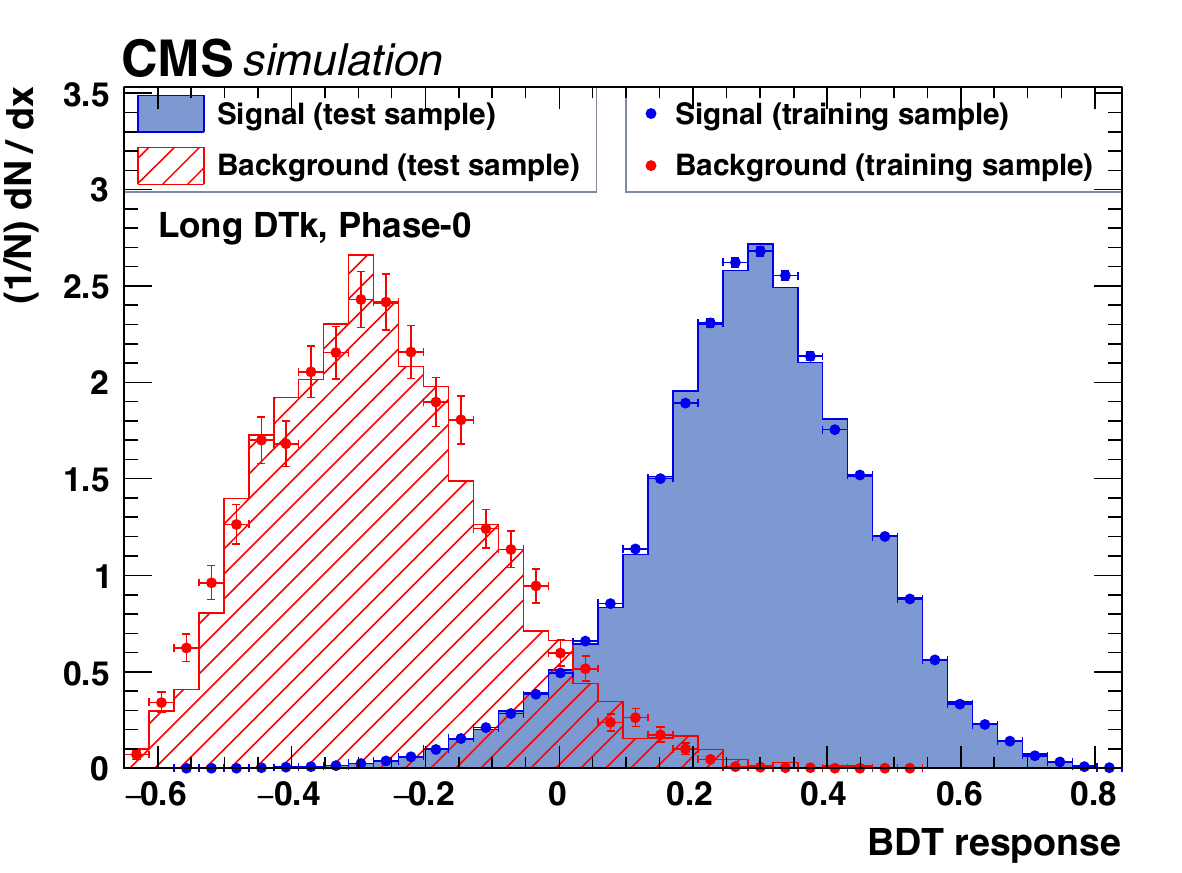}
\includegraphics[width=.45\linewidth]{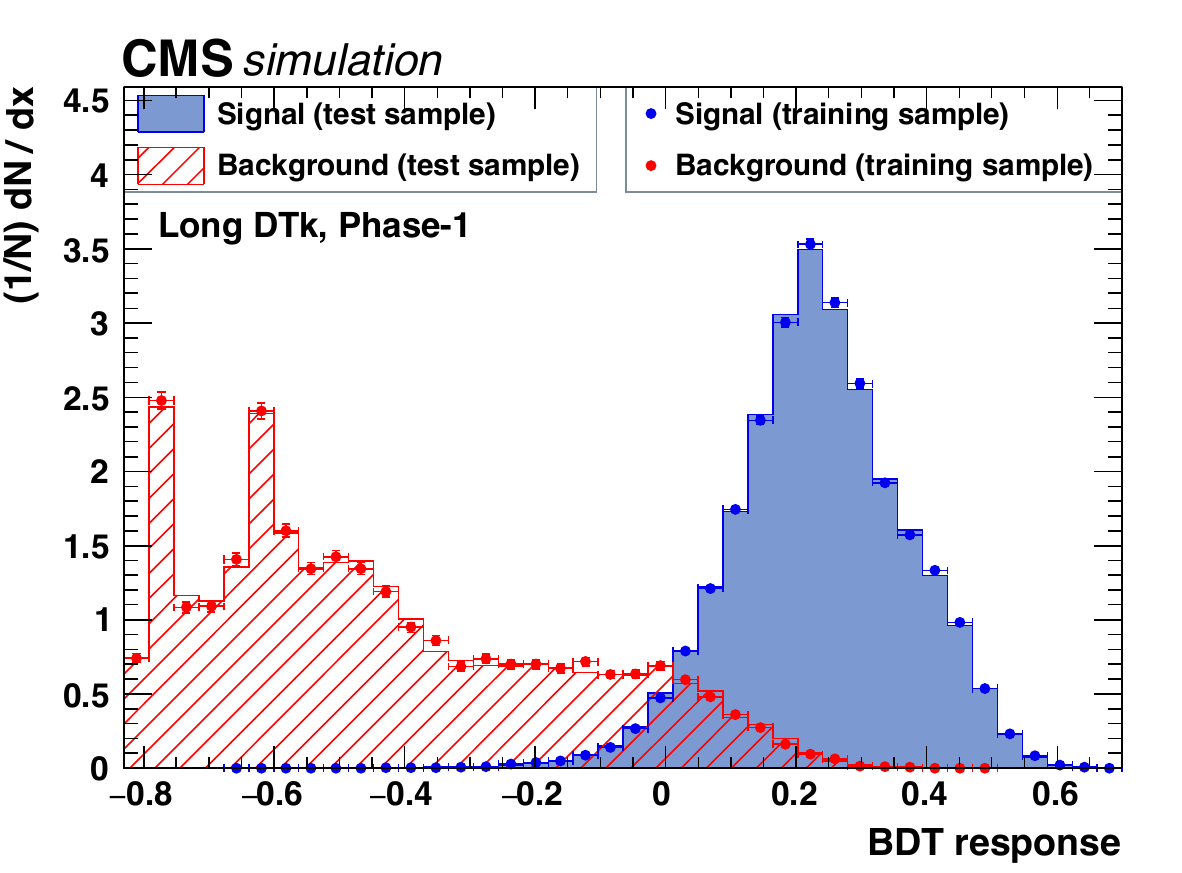}
\caption{
The distributions of simulated events
used to train and validate the BDT classifiers.
The left (right) column corresponds to the Phase-0 (Phase-1) detector
and the upper (lower) row to the short (long) track category.
The uncertainty bars shown for the training samples
indicate the Poisson uncertainties.
No events appear outside of the regions shown.
}
\label{fig:bdts}
\end{figure*}

A BDT classifier is used to improve the purity of the DTk candidate sample.
For the signal training,
chargino-matched tracks from simulated gluino and top squark pair-production events are used.
The matching criterion is $\dR(\mathrm{track},\charginogen)<0.01$,
where \charginogen represents a generator-level chargino.
For the background training,
tracks from all simulated SM processes are used.
The input variables to the classifier are
$d_{xy}$, $d_{z}$, \isorel, $\Delta\pt/\pt^{2}$, $\chi^{2}/n_{\mathrm{d}}$
(with $\Delta\pt$ the uncertainty in the reconstructed \pt value
and $n_\mathrm{d}$ the number of degrees of freedom),
the number of pixel hits,
and,
for long tracks,
the number of strip hits and the number of missing outer strip hits,
where this latter quantity is the number of strip layers
between the outermost strip hit and the outermost strip layer
for which no hit is recorded.
Separate BDTs are trained for the short- and long-track categories
and for the Phase-0 and -1 detectors.
Distributions of the BDT output scores
for these categories are shown in Fig.~\ref{fig:bdts}.
Thresholds on the classifier output values are applied in defining SRs and CRs 
for the Phase-0 and -1 detectors,
as indicated in Table~\ref{tab:sr-cr}.
The classifier significantly improves the selection performance
compared to so-called ``cut-based'' approaches,
increasing the signal efficiency for short tracks by a factor ranging from 2 to 4,
depending on the data-taking period,
and of long tracks by around 50\%,
for an equivalent background rejection.

These requirements select 45 to 55\% of charginos that decay after the second tracker layer
and before the penultimate one,
depending on the length of the track.

A novel technique is used to validate the DTk selection efficiency.
First, well reconstructed, isolated muons are selected from both data and simulation.
Tracker hits associated with the muon are iteratively removed from the event one-by-one,
starting with the outermost hit and working inward,
until only three tracker hits remain.
These shortened track segments are used as a proxy for long-lived charged particles.
After each step in the process,
the complete event reconstruction procedure is reapplied
and the efficiency for the tracking algorithm and DTk classifier to identify the
shortened segments as a DTk candidate is determined.
This efficiency ranges from 45 to 55\%
depending on the length of the shortened muon track,
similar to the reconstruction efficiency for charginos mentioned above.
The procedure is applied to samples corresponding to the various data-taking periods.
Comparisons of the results between data and simulation are used to derive scale factors
that are applied to the DTk selection efficiencies from simulation.
The scale factors are consistent with unity with uncertainties,
of statistical origin,
that vary between 10 and 17\%.

The DTk candidates are further classified
on the basis of their \dedx energy loss value in the pixel detector.
As the measure of the \dedx,
we use the so-called ``harmonic-2'' hit-weighted
average \harmonic~\cite{CMS:2011arq,CMS:2016kce},
which is an estimate of the linear stopping power of a particle
that is robust with respect to outlier measurements.
The \harmonic variable is defined by
\begin{linenomath}
  \begin{equation}
    \harmonic  = \left(\frac{1}{N} \sum_{i=1}^{N} {\left(\Delta E/\Delta x\right)_i^{-2}}\right)^{-1/2},
    \label{eq:harmonic2}
  \end{equation}
\end{linenomath}
where
$\Delta E$ is the energy loss estimated from the charge collected by a pixel hit,
$\Delta x$ is the length of the track segment corresponding to that hit,
and $N$ is the number of pixel hits on the track.
We define two intervals in~\harmonic:
$\harmonic<4.0$ and $\harmonic>4.0$\MeV/cm.
In the following, the term ``\dedx'' refers to the \harmonic value.

{\tolerance=5000
Time-dependent calibration of the \dedx measurement is performed in data using the position
of the center of a Gaussian fit to the minimum-ionizing-particle peak of well-reconstructed muons,
separately for the barrel and endcap regions.
The \dedx values from all data-taking periods are scaled to be consistent
with the result from simulation,
with a fit mean of 2.87\MeV/cm.
\par}

The \dedx value can further be used to
estimate the mass of the DTk,
through the relation
(with $m$ the charged-particle mass)
\begin{linenomath}
  \begin{equation}
    \dedx = K \frac{m^{2}}{p^{2}} + C,
  \end{equation}
\end{linenomath}
which is valid for singly charged particles~\cite{CMS:2011arq}.
The constants $K$ and $C$ are determined
as described in Ref.~\cite{CMS:2016kce}
to be $2.684 \pm 0.001$
and $3.375 \pm 0.001\MeV$/cm, respectively.
We refer to the DTk mass value determined in this manner as~\mdedx.
We do not use the \mdedx variable for quantitative purposes
but merely examine its distribution in Section~\ref{sec:results}.

\section{Event selection and search regions}
\label{sec:events}

The number of DTk candidates in an event is denoted~\nDTk.
Signal events are required to have \mbox{$\nDTk\geq 1$}, 
$\njets\geq 1$, and $\ptmisshard>30$\GeV.
At least one DTk candidate must satisfy $\mT(\mathrm{DTk}, \mathrm{\ptmisshard})>20\GeV$,
where \mT is the transverse mass~\cite{Arnison:1983rp}
formed from the two quantities in parentheses.
For $\nDTk=1$, the search is performed inclusively in the hadronic DTk,
electron+DTk, and muon+DTk channels.
We also search in a channel with $\nDTk\geq 2$.
For this latter channel,
events are required to satisfy the hadronic trigger requirements
in the case of zero leptons
or the trigger requirements for the electron (muon) channel
in the case of $\nelec\geq1$ ($\nmuon\geq1$),
where \nelec (\nmuon) is the number of electron (muon) candidates in an event.
The different search channels are defined as follows.

\begin{itemize}
\item Hadronic channel:
\begin{itemize}
  \item $\nDTk=1$,
  \item $\nelec=\nmuon=0$,
  \item $\ptmisshard>150$\GeV.
\end{itemize}
\item Muon channel:
\begin{itemize}
  \item $\nDTk=1$,
  \item $\nelec=0$,
  \item $\nmuon\geq1$,
  \item $m_{\text{DTk},\PGm}>120$\GeV,
  \item $\mT(\Pgm,\ptmisshard)>110$\GeV.
\end{itemize}
\item Electron channel:
\begin{itemize}
  \item $\nDTk=1$,
  \item $\nelec\geq1$,
  \item $m_{\text{DTk},\Pe}>120$\GeV,
  \item $\mT(\Pe,\ptmisshard)>110$\GeV.
\end{itemize}
\item $\nDTk\geq 2$ channel:
\begin{itemize}
  \item Satisfies the criteria given above
  for either the hadronic, muon, or electron channel,
  except for the \nDTk requirement,
  \item $\nDTk\geq2$.
\end{itemize}
\end{itemize}
The invariant mass \mDTkell is formed from the DTk
and the highest \pt electron or muon candidate in the event.
The \mDTkell and \mT requirements suppress the
Drell-Yan (DY) and \wjets backgrounds, respectively.
For the \mDTkell calculation,
the mass of the DTk is assumed to be zero.
Note that we require $\nelec=0$ for the muon channel in order to render it orthogonal to the electron channel,
but do not require $\nmuon=0$ for the electron channel because it is unnecessary for the analysis. 

The above requirements define the global SR,
referred to as the ``baseline'' region.
Events in the baseline region are divided into nonoverlapping SRs as follows.
Events with  $\nDTk=1$ are sorted into 48 SRs
based on the DTk length category,
the \dedx interval,
and the values of \njets, \nbjets, and \ptmisshard.
The choice of binning is designed to provide sensitivity to a wide range of scenarios,
covering strong and electroweak production
and the presence of all three generations of fermions.
Events with $\nDTk\geq2$ are grouped into a single~SR,
bringing the total number of SRs to~49.
The definition of the SRs is given in \mbox{Tables~\ref{tab:SR1} and~\ref{tab:SR2}.}

\begin{table*}[t!]
\centering
\topcaption{
Definition of the search regions (SRs) for the hadronic channel.
}
\begin{tabular}{|c|c|c|c|c|c|c|}
\hline
\multicolumn{7}{|c|}{Hadronic channel ($\nDTk=1$, $\nmuon=0$, $\nelec=0$)} \\ \hline
\ptmisshard (\GeVns)  & \multicolumn{1}{l|}{\nbjets} & \multicolumn{1}{l|}{\njets} & \nshort & \nlong & \dedx (\MeVns/cm) & SR \\ \hline
\multirow{16}{*}{150--300}     & \multirow{8}{*}{0}       & \multirow{4}{*}{1--2}  & \multirow{2}{*}{0}  & \multirow{2}{*}{1}  & $<$4.0 & 1  \\ \cline{6-7}
                               &                          &                         &                     &                     & $>$4.0 & 2  \\ \cline{4-7} 
                               &                          &                         & \multirow{2}{*}{1}  & \multirow{2}{*}{0}  & $<$4.0 & 3  \\ \cline{6-7} 
                               &                          &                         &                     &                     & $>$4.0 & 4  \\ \cline{3-7} 
                               &                          & \multirow{4}{*}{$\geq$3}& \multirow{2}{*}{0} & \multirow{2}{*}{1}  & $<$4.0 & 5  \\ \cline{6-7} 
                               &                          &                         &                     &                     & $>$4.0 & 6  \\ \cline{4-7} 
                               &                          &                         & \multirow{2}{*}{1}  & \multirow{2}{*}{0}  & $<$4.0 & 7  \\ \cline{6-7} 
                               &                          &                         &                     &                     & $>$4.0 & 8  \\ \cline{2-7} 
                               & \multirow{8}{*}{$\geq$1} & \multirow{4}{*}{1--2}  & \multirow{2}{*}{0}  & \multirow{2}{*}{1}  & $<$4.0 & 9  \\ \cline{6-7} 
                               &                          &                         &                     &                     & $>$4.0 & 10 \\ \cline{4-7} 
                               &                          &                         & \multirow{2}{*}{1}  & \multirow{2}{*}{0}  & $<$4.0 & 11 \\ \cline{6-7} 
                               &                          &                         &                     &                     & $>$4.0 & 12 \\ \cline{3-7}
                               &                          & \multirow{4}{*}{$\geq$3}& \multirow{2}{*}{0}  & \multirow{2}{*}{1}  & $<$4.0 & 13 \\ \cline{6-7}
                               &                          &                         &                     &                     & $>$4.0 & 14 \\ \cline{4-7} 
                               &                          &                         & \multirow{2}{*}{1}  & \multirow{2}{*}{0}  & $<$4.0 & 15 \\ \cline{6-7} 
                               &                          &                         &                     &                     & $>$4.0 & 16 \\ \hline
\multirow{8}{*}{$>$300}        & \multirow{8}{*}{Any}     & \multirow{4}{*}{1--2}  & \multirow{2}{*}{0}  & \multirow{2}{*}{1}  & $<$4.0 & 17 \\ \cline{6-7} 
                               &                          &                         &                     &                     & $>$4.0 & 18 \\ \cline{4-7} 
                               &                          &                         & \multirow{2}{*}{1}  & \multirow{2}{*}{0}  & $<$4.0 & 19 \\ \cline{6-7} 
                               &                          &                         &                     &                     & $>$4.0 & 20 \\ \cline{3-7} 
                               &                          & \multirow{4}{*}{$\geq$3}& \multirow{2}{*}{0}  & \multirow{2}{*}{1}  & $<$4.0 & 21 \\ \cline{6-7} 
                               &                          &                         &                     &                     & $>$4.0 & 22 \\ \cline{4-7} 
                               &                          &                         & \multirow{2}{*}{1}  & \multirow{2}{*}{0}  & $<$4.0 & 23 \\ \cline{6-7} 
                               &                          &                         &                     &                     & $>$4.0 & 24 \\ \hline
\end{tabular}
\label{tab:SR1}
\end{table*}

\begin{table*}[t!]
\centering
\topcaption{
Definition of the search regions (SRs) for the muon, electron, and $\nDTk\geq 2$ channels.
}
\begin{tabular}{|c|c|c|c|c|c|c|}
\hline
\multicolumn{7}{|c|}{Muon channel ($\nDTk=1$, $\nmuon\geq 1$, $\nelec=0$)} \\ \hline
\ptmisshard (\GeVns) & \multicolumn{1}{l|}{\nbjets}      & \multicolumn{1}{l|}{\njets} & \nshort & \nlong & \dedx (\MeVns/cm) & SR \\ \hline
\multirow{8}{*}{30--100}      & \multirow{4}{*}{0}       & \multirow{12}{*}{$\geq$1}   & \multirow{2}{*}{0} & \multirow{2}{*}{1} & $<$4.0 & 25 \\ \cline{6-7} 
                              &                          &                             &                    &                    & $>$4.0 & 26 \\ \cline{4-7} 
                              &                          &                             & \multirow{2}{*}{1} & \multirow{2}{*}{0} & $<$4.0 & 27 \\ \cline{6-7} 
                              &                          &                             &                    &                    & $>$4.0 & 28 \\ \cline{2-2} \cline{4-7} 
                              & \multirow{4}{*}{$\geq$1} &                             & \multirow{2}{*}{0} & \multirow{2}{*}{1} & $<$4.0 & 29 \\ \cline{6-7} 
                              &                          &                             &                    &                    & $>$4.0 & 30 \\ \cline{4-7} 
                              &                          &                             & \multirow{2}{*}{1} & \multirow{2}{*}{0} & $<$4.0 & 31 \\ \cline{6-7} 
                              &                          &                             &                    &                    & $>$4.0 & 32 \\ \cline{1-2} \cline{4-7} 
\multirow{4}{*}{$>$100}       & \multirow{4}{*}{Any}     &                             & \multirow{2}{*}{0} & \multirow{2}{*}{1} & $<$4.0 & 33 \\ \cline{6-7} 
                              &                          &                             &                    &                    & $>$4.0 & 34 \\ \cline{4-7} 
                              &                          &                             & \multirow{2}{*}{1} & \multirow{2}{*}{0} & $<$4.0 & 35 \\ \cline{6-7}
                              &                          &                             &                    &                    & $>$4.0 & 36 \\ \hline
\multicolumn{7}{|c|}{Electron channel ($\nDTk=1$, $\nelec\geq 1$)} \\ \hline
\multirow{8}{*}{30--100}      & \multirow{4}{*}{0}       & \multirow{12}{*}{$\geq$1}   & \multirow{2}{*}{0} & \multirow{2}{*}{1} & $<$4.0 & 37 \\ \cline{6-7} 
                              &                          &                             &                    &                    & $>$4.0 & 38 \\ \cline{4-7} 
                              &                          &                             & \multirow{2}{*}{1} & \multirow{2}{*}{0} & $<$4.0 & 39 \\ \cline{6-7} 
                              &                          &                             &                    &                    & $>$4.0 & 40 \\ \cline{2-2} \cline{4-7} 
                              & \multirow{4}{*}{$\geq$1} &                             & \multirow{2}{*}{0} & \multirow{2}{*}{1} & $<$4.0 & 41 \\ \cline{6-7} 
                              &                          &                             &                    &                    & $>$4.0 & 42 \\ \cline{4-7} 
                              &                          &                             & \multirow{2}{*}{1} & \multirow{2}{*}{0} & $<$4.0 & 43 \\ \cline{6-7} 
                              &                          &                             &                    &                    & $>$4.0 & 44 \\ \cline{1-2} \cline{4-7} 
\multirow{4}{*}{$>$100}       & \multirow{4}{*}{Any}     &                             & \multirow{2}{*}{0} & \multirow{2}{*}{1} & $<$4.0 & 45 \\ \cline{6-7} 
                              &                          &                             &                    &                    & $>$4.0 & 46 \\ \cline{4-7} 
                              &                          &                             & \multirow{2}{*}{1} & \multirow{2}{*}{0} & $<$4.0 & 47 \\ \cline{6-7} 
                              &                          &                             &                    &                    & $>$4.0 & 48 \\ \hline
\multicolumn{7}{|c|}{$\nDTk\geq 2$ channel}\\ \hline
$>$30                         & Any                      & $\geq$1                     & \multicolumn{3}{|c|}{Any}                   & 49 \\ \hline
\end{tabular}
\label{tab:SR2}
\end{table*}

\section{Background estimation}
\label{sec:background}

Disappearing track signatures are rare in the SM but can occasionally
arise because of an instrumental effect,
either from the misreconstruction of a charged particle
or from the coincidental alignment of hits from different tracks.
We refer to these backgrounds as the ``genuine-particle'' background
and the ``spurious-particle'' background, respectively.
We use methods based entirely on data to evaluate these backgrounds,
as described below.
These methods employ the so-called ``ABCD'' technique,
for which a basic description is given in Ref.~\cite{Kasieczka:2020pil}.

\subsection{Genuine-particle background}
\label{sec:prompt_tracks}

In a typical case,
genuine-particle background might arise from a track that
showers in a crack between ECAL crystals or into a crystal with anomalously low light yield,
yielding a significantly undermeasured deposited energy that degrades the consistency
between the measured track \pt and associated calorimetric energy,
leading to a failure in the PF particle reconstruction.
In other cases, a particle might emit a highly energetic photon,
causing a recoil that similarly degrades this consistency.
Concomitantly,
the two outermost layers of the silicon tracker may not have hits associated with the track,
which can arise through rare inefficiencies or from misassociation of hits in the strip detector. 
We refer to this type of background as the ``genuine-particle showering'' background.
Less often, genuine-particle background can arise from a muon
if there is a large or otherwise anomalous occupancy in the muon system
or as a result of a rare misalignment between the track segment
in the tracker with that in the muon system,
again in conjunction with missing hits in the two outermost layers of the silicon tracker.
We refer to this background as the ``genuine-particle muon'' background.
These genuine-particle backgrounds represent the dominant
background for the long-track category of~DTks.

{\tolerance=2000
The genuine-particle showering background is evaluated
using sideband control regions \crreal, one for each SR,
and a separate ``DY measurement'' CR of Drell-Yan events.
The \crreal samples are defined by the
intervals in \edep and the BDT classifier
listed in the third and fourth rows of Table~\ref{tab:sr-cr}.
These criteria yield a high purity for genuine-particle showering events
and thus negligible contamination.
The \crreal samples are selected using a loosened isolation requirement
on the DTks compared to the SRs.
Specifically,
the DTk candidate must only lie at least $\dR=0.1$
away from any reconstructed jet with $\pt>15$\GeV,
rather than at least $\dR=0.4$ as in Section~\ref{sec:tracks}.
The reason for this loosened restriction is to increase
the statistical precision of the background estimate.
Other than these differences,
the criteria used to select the \crreal samples are the same as
for the SR samples.
\par}

The DY measurement CR is selected by requiring events to
contain an electron and a DTk candidate of opposite charge.
The DTk candidate must satisfy
the selection criteria for either the SR or \crreal samples.
The invariant mass $m(\mathrm{DTk},\Pe)$ formed from the DTk and electron
must be consistent with a \PZ boson,
namely lie in a range from 65 to 110\GeV.
For this calculation, the DTk is assigned a mass of zero.
To improve the purity,
we require $\mT(\Pe,\ptmisshard)<100\GeV$
and that there be a relatively small difference $\Delta\phi$
in the azimuthal angle between the DTk and \ptmisshard:
$\Delta\phi<\pi/2$ for long tracks and $\Delta\phi<\pi/4$ for short tracks.
The $\Delta\phi$ requirement is imposed because genuine-particle DTks
correspond to particles that are not reconstructed,
which can lead to significant \ptmisshard along the direction of the DTk.

\begin{table*}[tb]
\centering
\topcaption{
The transfer factors \klowhi and \kmuon used for the evaluation of the genuine-particle backgrounds.
The ``Genuine shower'' columns refer to the \klowhi factors
while the ``Genuine muon'' columns refer to the \kmuon factors.
The genuine-particle muon background is negligible for the short category of DTks.
The uncertainties are statistical only.
}
\label{tab:kappa}
\cmsTable{
\begin{scotch}{ccccccc}
\multirow{2}{*}{\klowhi, \kmuon:} & \multicolumn{2}{c}{Phase 0} & \multicolumn{2}{c}{Phase 1} & \multicolumn{2}{c}{Combined Phase 0 and 1}   \\ 
    & \multicolumn{1}{c}{Genuine shower} & Genuine muon & \multicolumn{1}{c}{Genuine shower} & Genuine muon & \multicolumn{1}{c}{Genuine shower} & Genuine muon  \\
\hline
Short & \multicolumn{1}{c}{$0.65 \pm 0.11$} & \NA                    & \multicolumn{1}{c}{$0.435 \pm 0.049$} & \NA                   & \multicolumn{1}{c}{$0.492 \pm 0.046$} & \NA \\ 
Long  & \multicolumn{1}{c}{$0.247 \pm 0.020$}  & $0.00099 \pm 0.00017$  & \multicolumn{1}{c}{$0.389 \pm 0.032$} & $0.00047 \pm 0.00012$ & \multicolumn{1}{c}{$0.307 \pm 0.018$} & $0.00074 \pm 0.00011$ \\
\end{scotch}
}
\end{table*}

A transfer factor \klowhi is derived from the DY measurement CR,
given by the ratio of the number of events with small and large values of the \fdep variable,
\begin{linenomath}
\begin{equation}
  \klowhi=N^{\text{DY}}_{\fdeplow} / N^{\text{DY}}_{\fdephi}, 
\end{equation}
\end{linenomath}
where $\fdep=\edep/p$ for long tracks
and $\fdep = \edep$ for short tracks.
The \edep requirements for the \fdeplow (\fdephi) region are the same as those listed
for the SR (\crreal) samples in Table~\ref{tab:sr-cr}.
The transfer factor is determined separately for short and long tracks
and for the Phase-0 and -1 detectors.
The DY measurement CR is nearly 100\% pure in genuine-particle showering events
except for the short-track category at low \fdep,
where approximately 50\% of the events arise from spurious tracks.
The spurious-track contamination in this sample is evaluated
using the method described in Section~\ref{sec:combinatoric}
and is then subtracted before the \klowhi factors are calculated,
with an associated systematic uncertainty evaluated
as described in Section~\ref{sec:systematic}.
The values of the \klowhi factors
are reported in the ``Genuine shower'' columns of \mbox{Table~\ref{tab:kappa}}.

The estimate $N_{\mathrm{SR}_i}^{\text{genuine}}$ of the number of
genuine-particle showering background events
in the $i$th SR is
\begin{linenomath}
\begin{equation}
  N_{\mathrm{SR}_i}^{\text{genuine}} = \klowhi \,\nCRreal,
\end{equation}
\end{linenomath}
with \nCRreal the number of events in the $i$th \crreal sample.  

An analogous procedure is employed to evaluate the much smaller genuine-particle muon background,
making use of sideband control regions \crrealmu defined analogously to the SRs 
but inverting the requirement from Section~\ref{sec:tracks}
that the DTk not be identified as a PF muon,
namely the track \textit{must} be identified as a muon.
We call this track a ``DTk-proxy'' track.
A DY measurement CR is defined in the same manner as described above
for the genuine-particle showering background
except with the \PZ boson candidate formed from a muon and the DTk-proxy track.
The \crrealmu and this DY measurement CR are essentially 100\% pure in
genuine-particle muon events and thus have negligible contamination.
A transfer factor \kmuon is derived from the DY measurement CR
as the ratio of the number of events with the muon veto of Section~\ref{sec:tracks} applied
to the corresponding number with the veto inverted.
The values of the transfer factors \kmuon
are listed in the ``Genuine muon'' columns of Table~\ref{tab:kappa}. 
Note that this background is negligible for short tracks
because of the high efficiency of the
combined strip tracker and muon systems. 

The estimate $N_{\mathrm{SR}_i}^{\text{genuine-}\mu}$
of the number of genuine-particle muon background events
in the $i$th SR is
\begin{linenomath}
\begin{equation}
  N_{\mathrm{SR}_i}^{\text{genuine-}\mu} = \kmuon \,{\nCRrealmu},
\end{equation}
\end{linenomath}
with \nCRrealmu the number of events in the $i$th \crrealmu sample.

\begin{figure*}[tp]
\centering
\includegraphics[width=.45\linewidth]{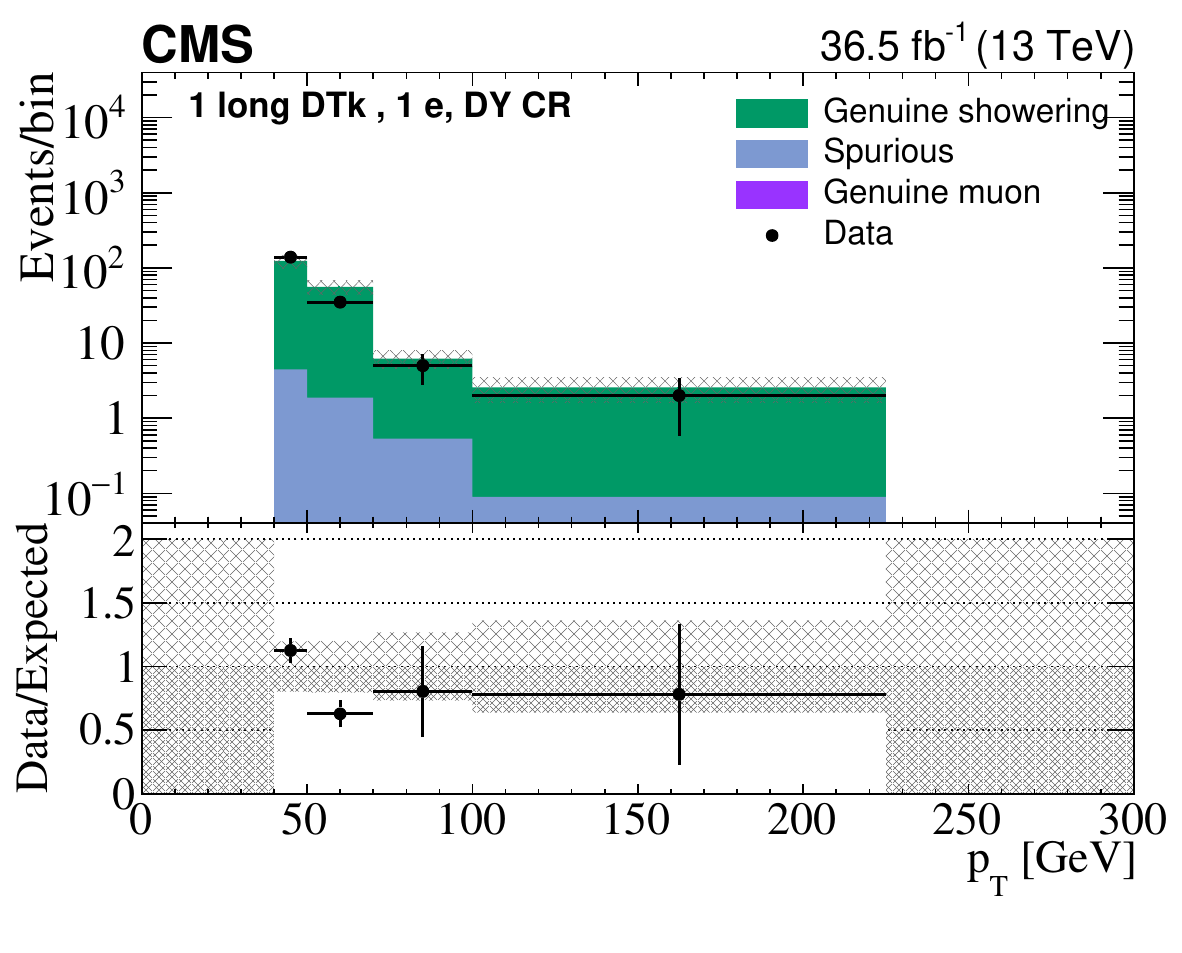}
\includegraphics[width=.45\linewidth]{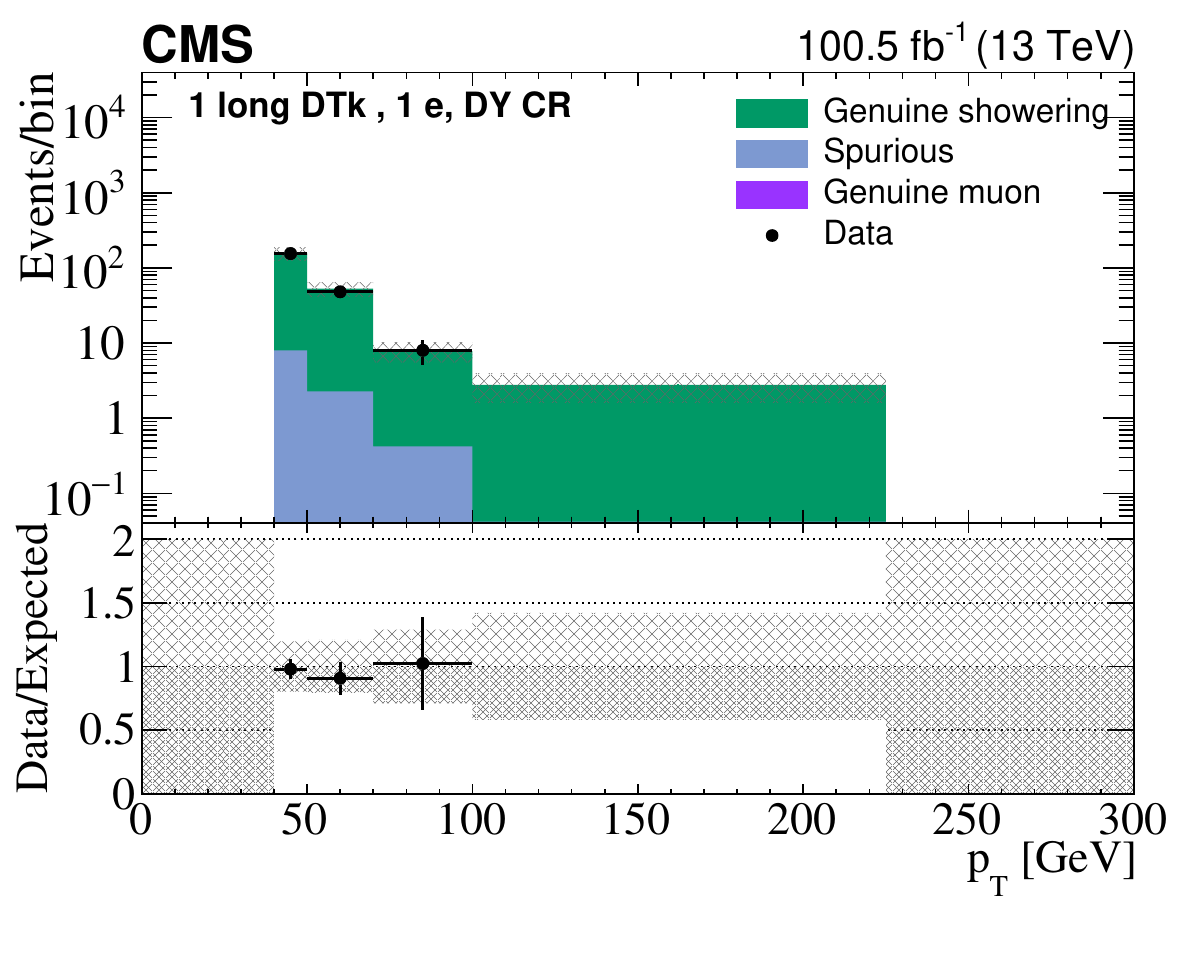}\\
\includegraphics[width=.45\linewidth]{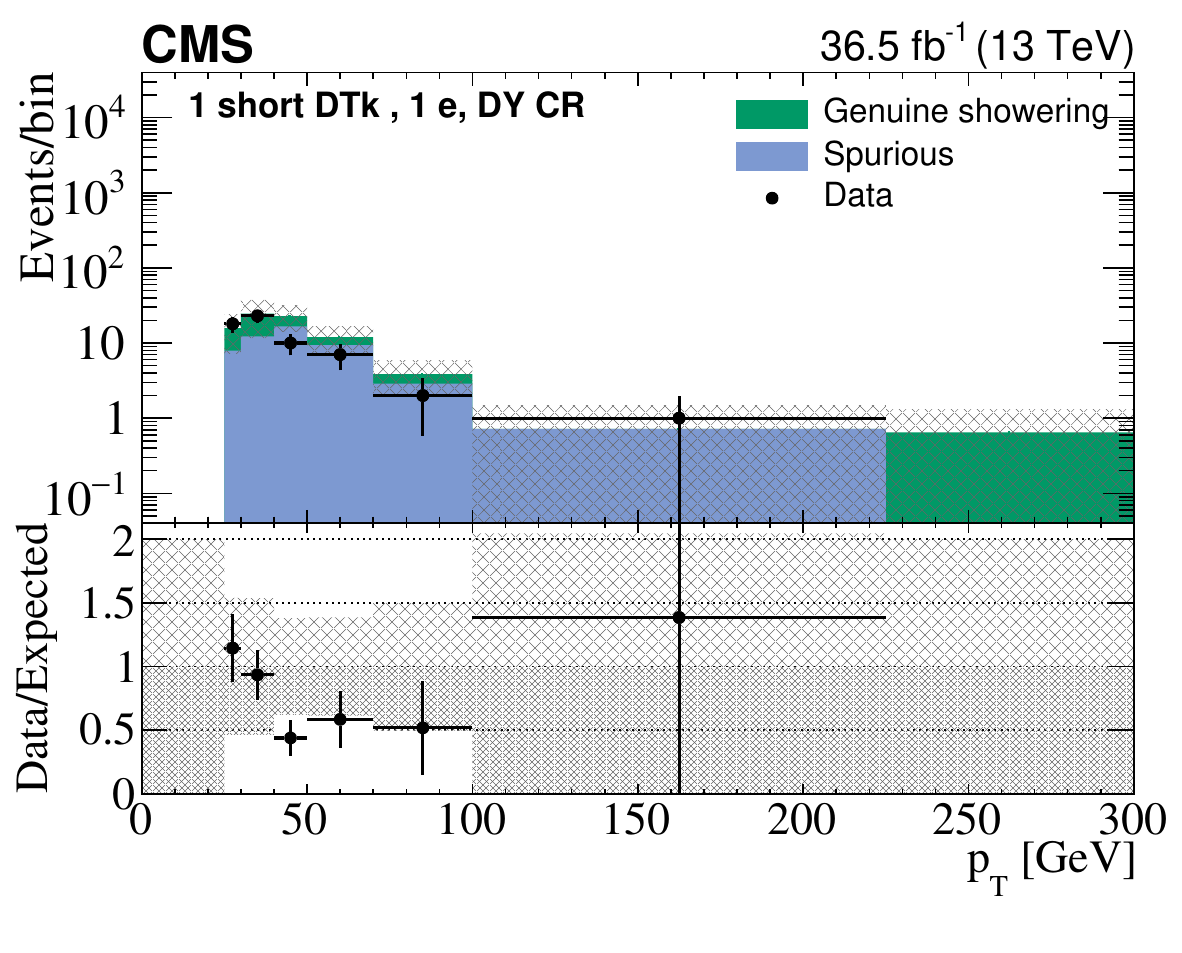}
\includegraphics[width=.45\linewidth]{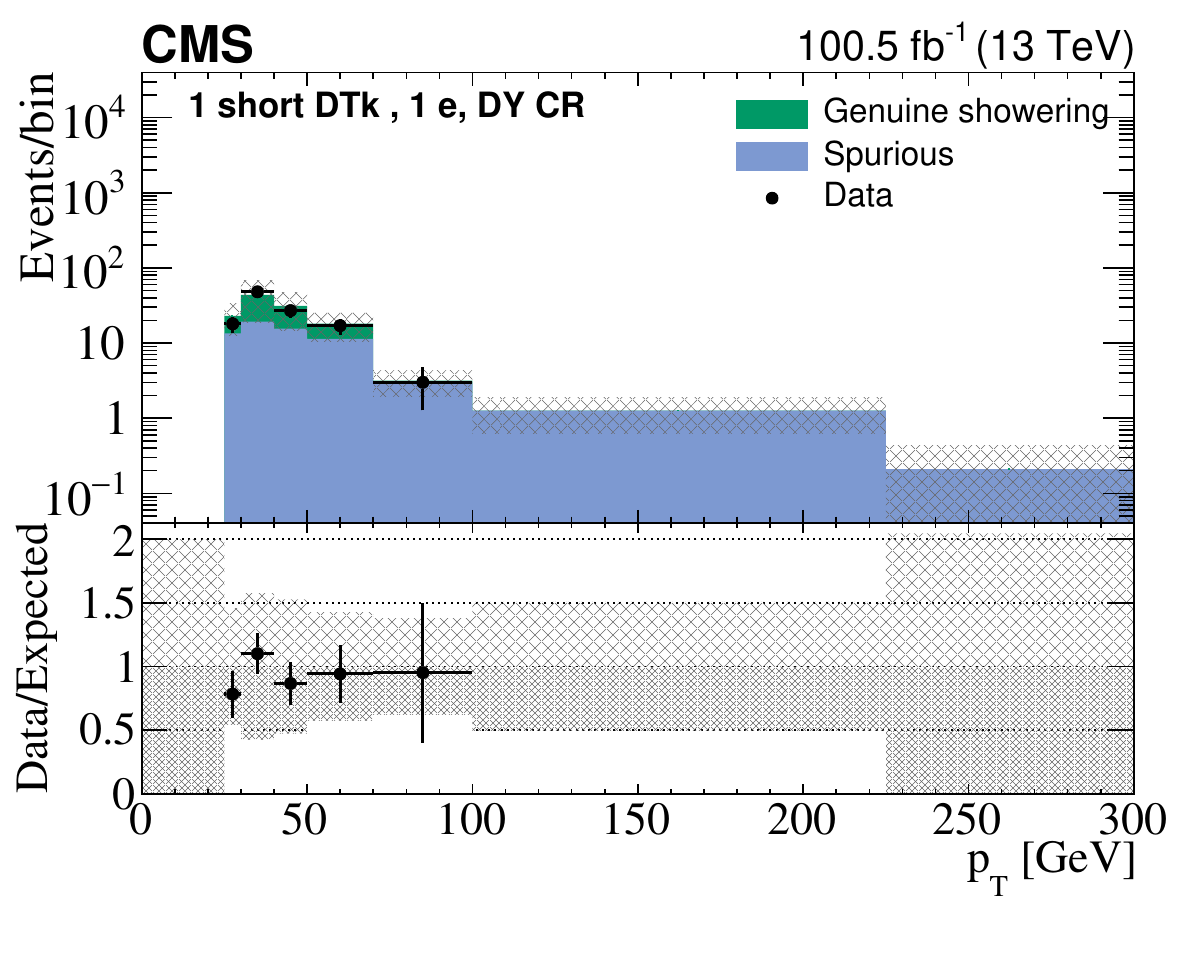}\\
\includegraphics[width=.45\linewidth]{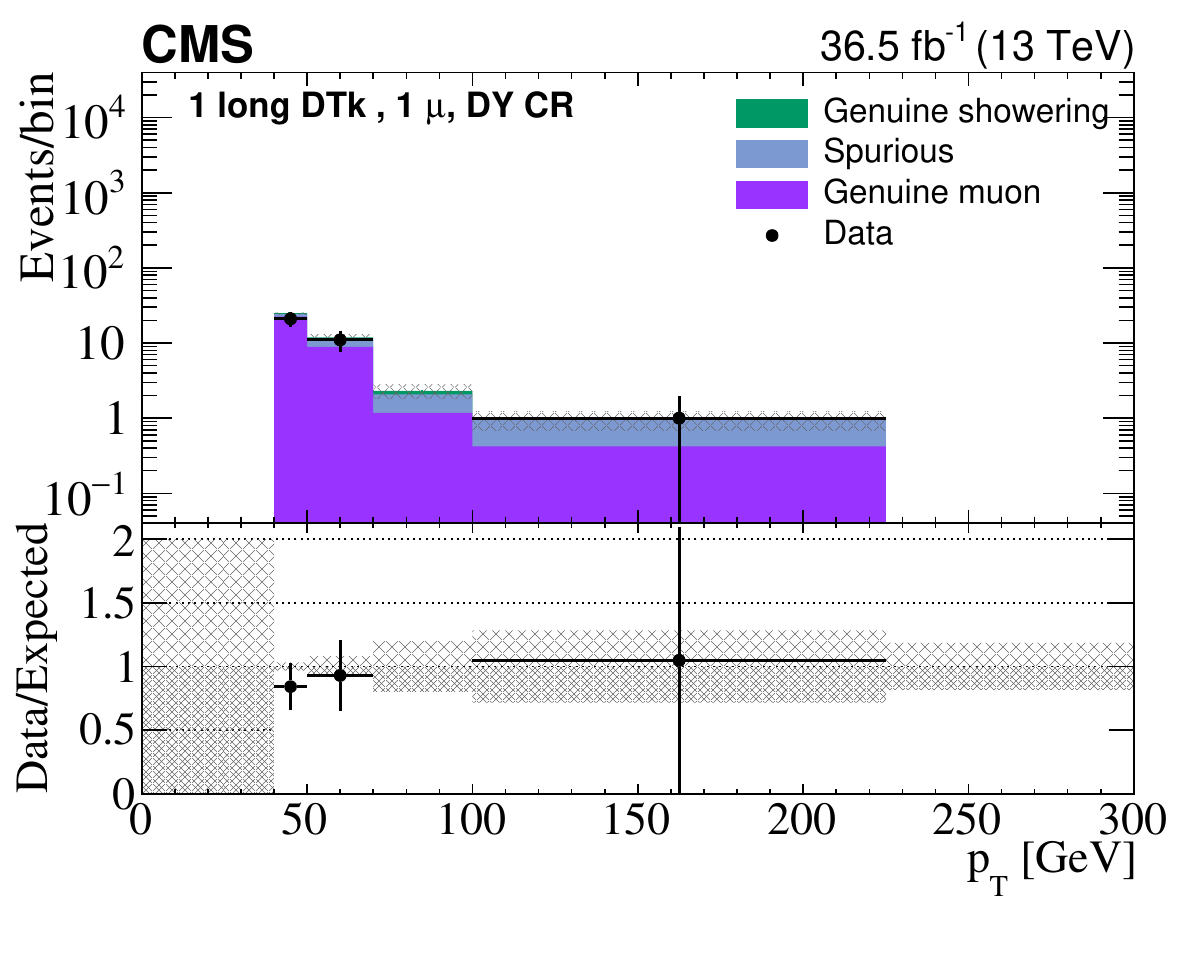}
\includegraphics[width=.45\linewidth]{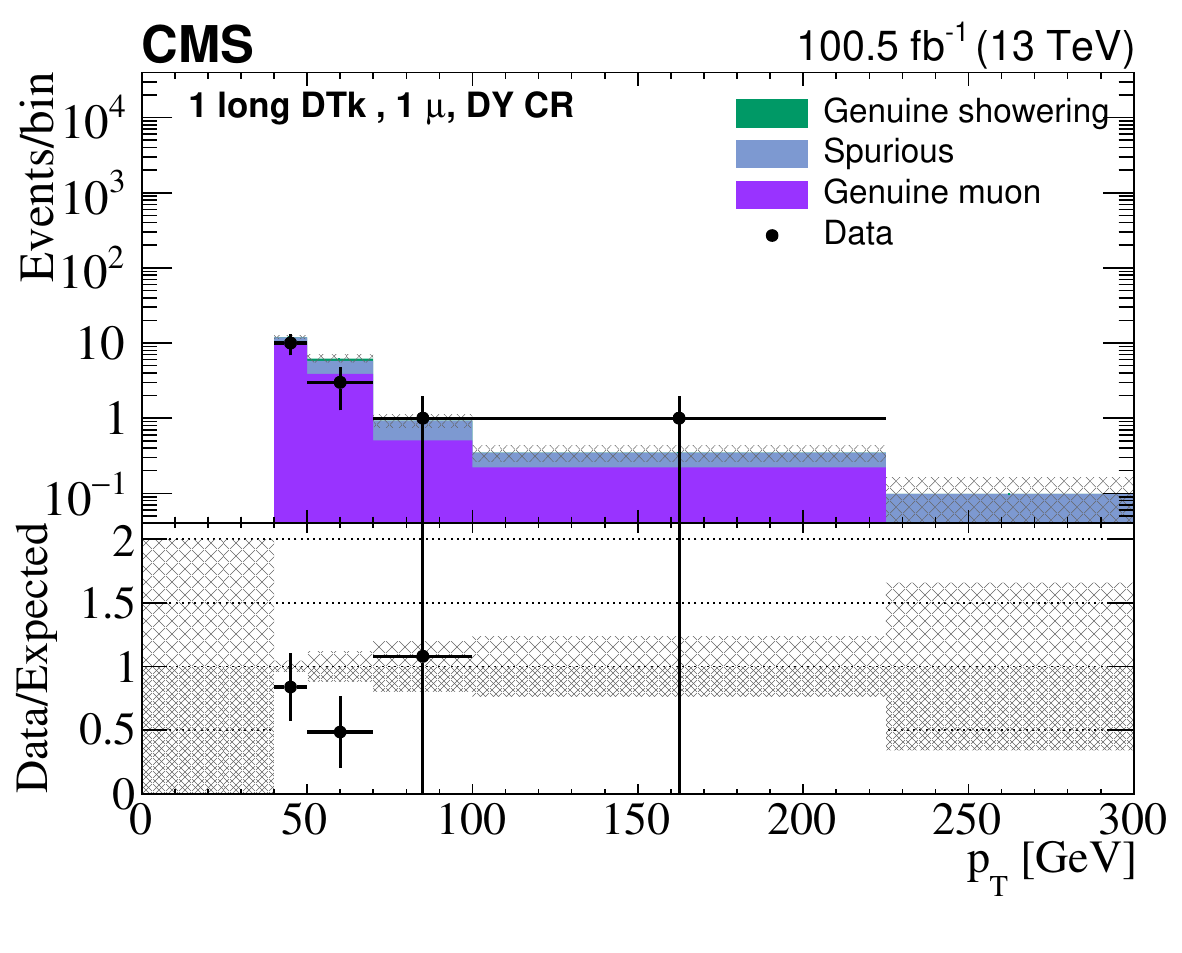}
\caption{
Comparison of the \pt distributions of DTks in the \klowhi DY measurement control region 
for the data and background prediction
for long (upper) and short (middle) showering tracks
and in the \kmuon DY measurement control region for long muon tracks (lower).
The left (right) column corresponds to the Phase-0 (Phase-1) detector.
The uncertainty bars on the ratios in the lower panels indicate the fractional Poisson uncertainties
in the observed counts.
The gray bands show the fractional Poisson uncertainties in the sideband region counts,
added in quadrature with the systematic uncertainties.
The leftmost (rightmost) bin includes underflow (overflow).
}
\label{fig:prompt-meas}
\end{figure*}

\begin{figure*}[tp]
\centering
\includegraphics[width=.45\linewidth]{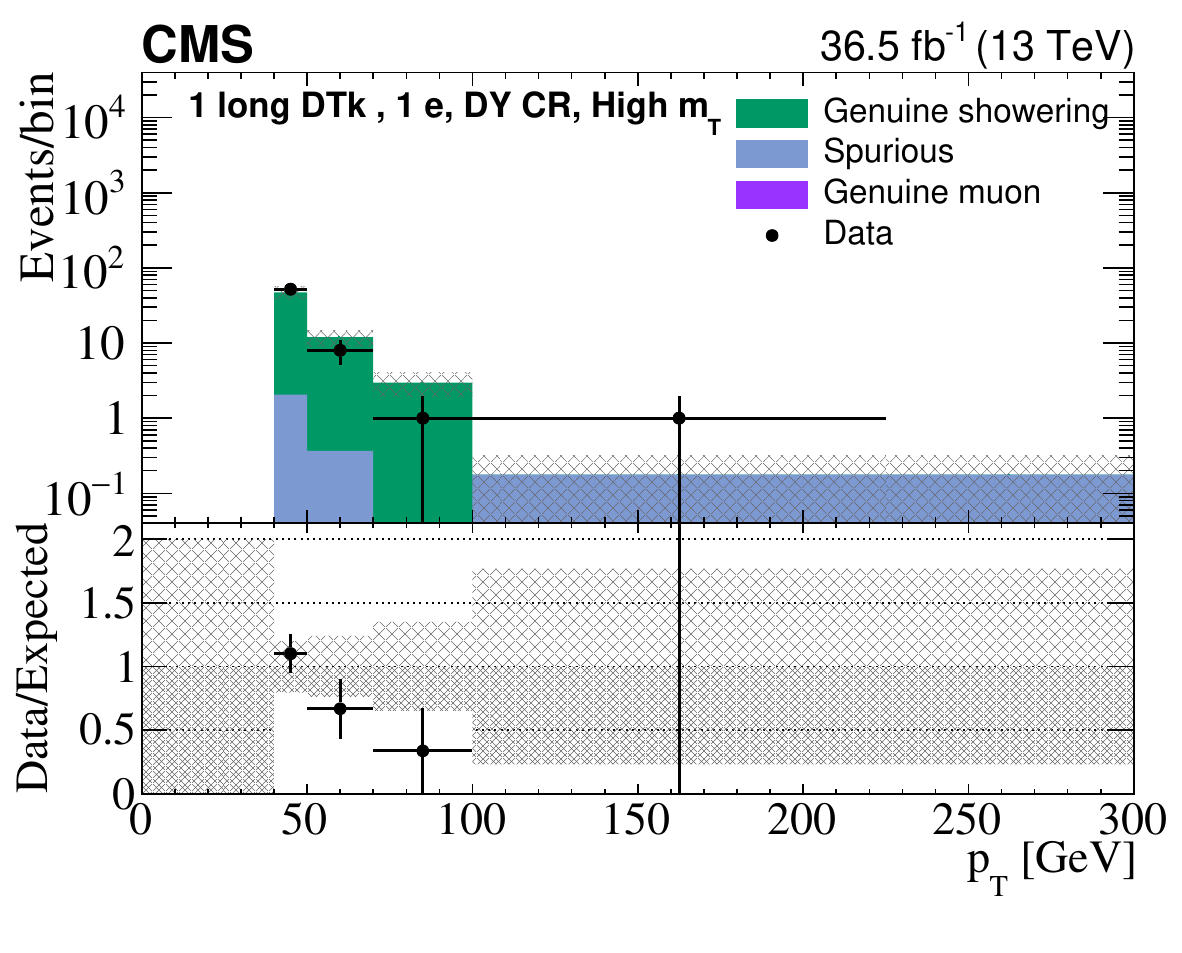}
\includegraphics[width=.45\linewidth]{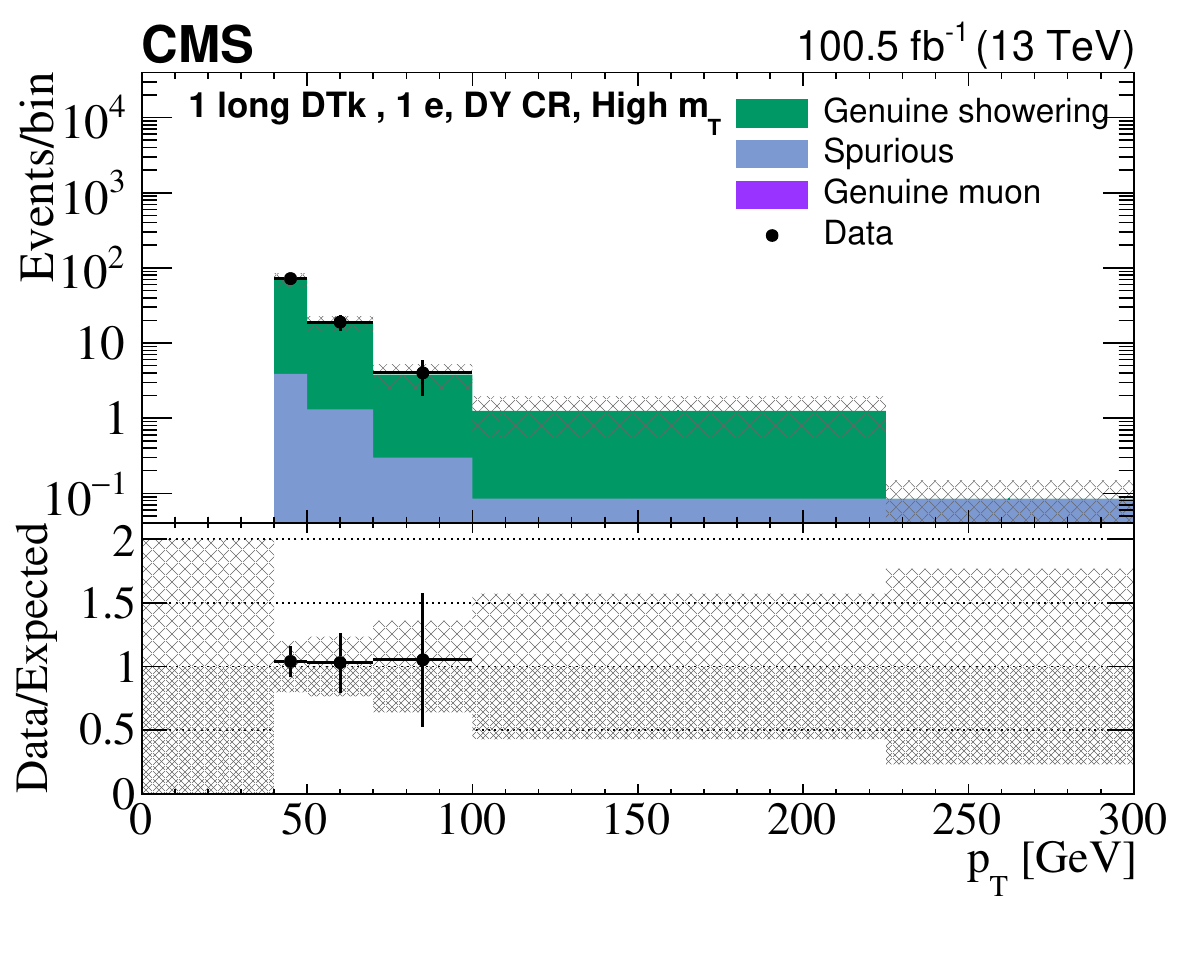}\\
\includegraphics[width=.45\linewidth]{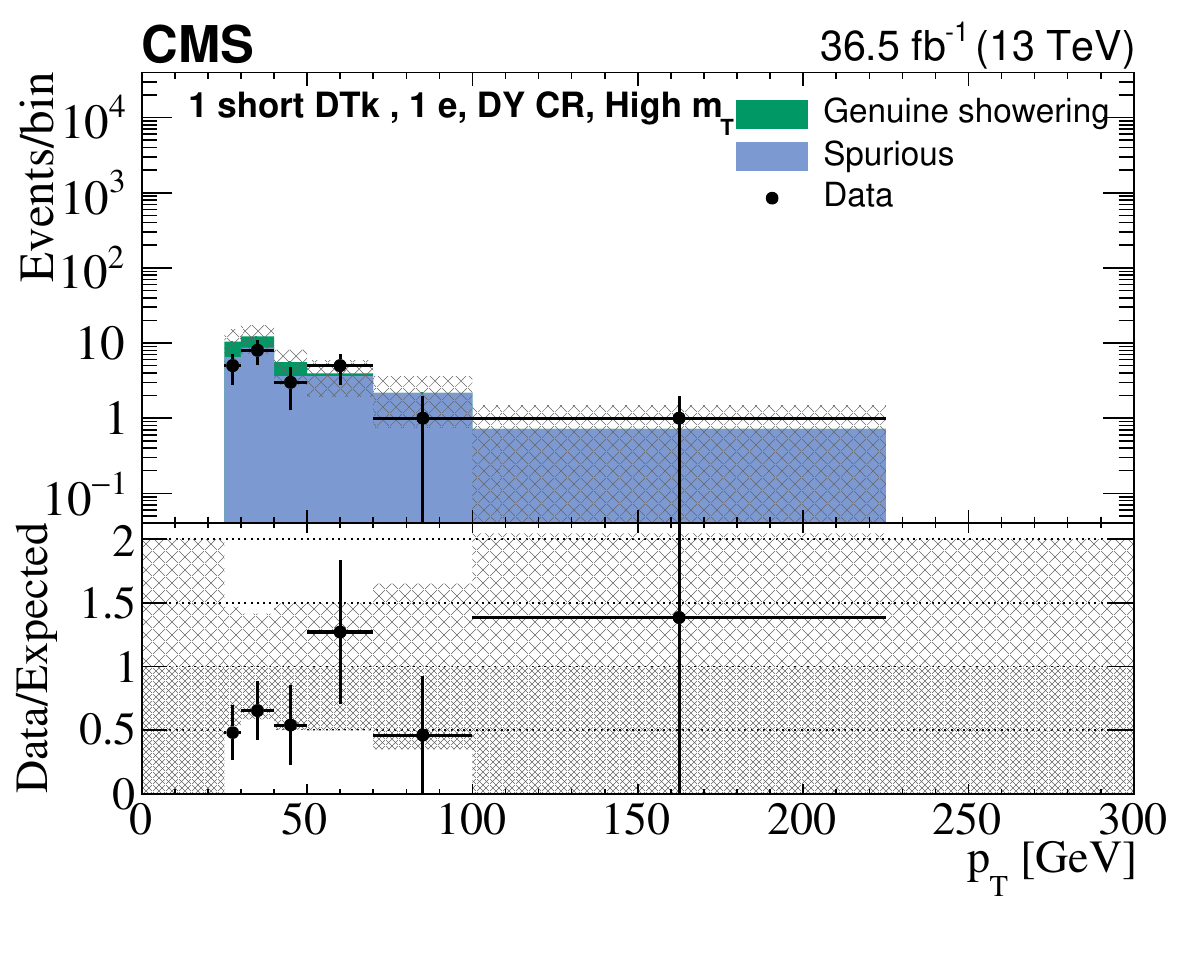}
\includegraphics[width=.45\linewidth]{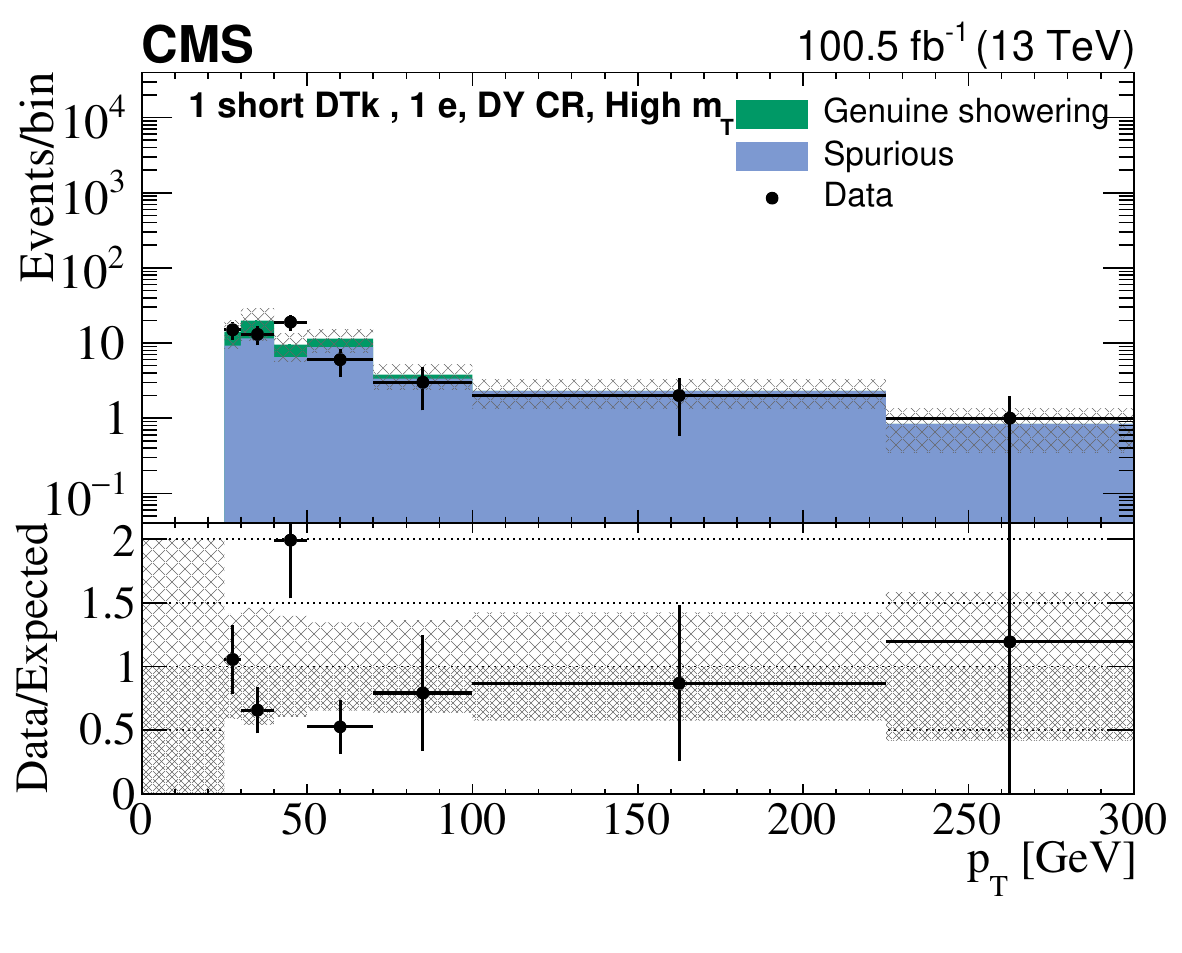}\\
\includegraphics[width=.45\linewidth]{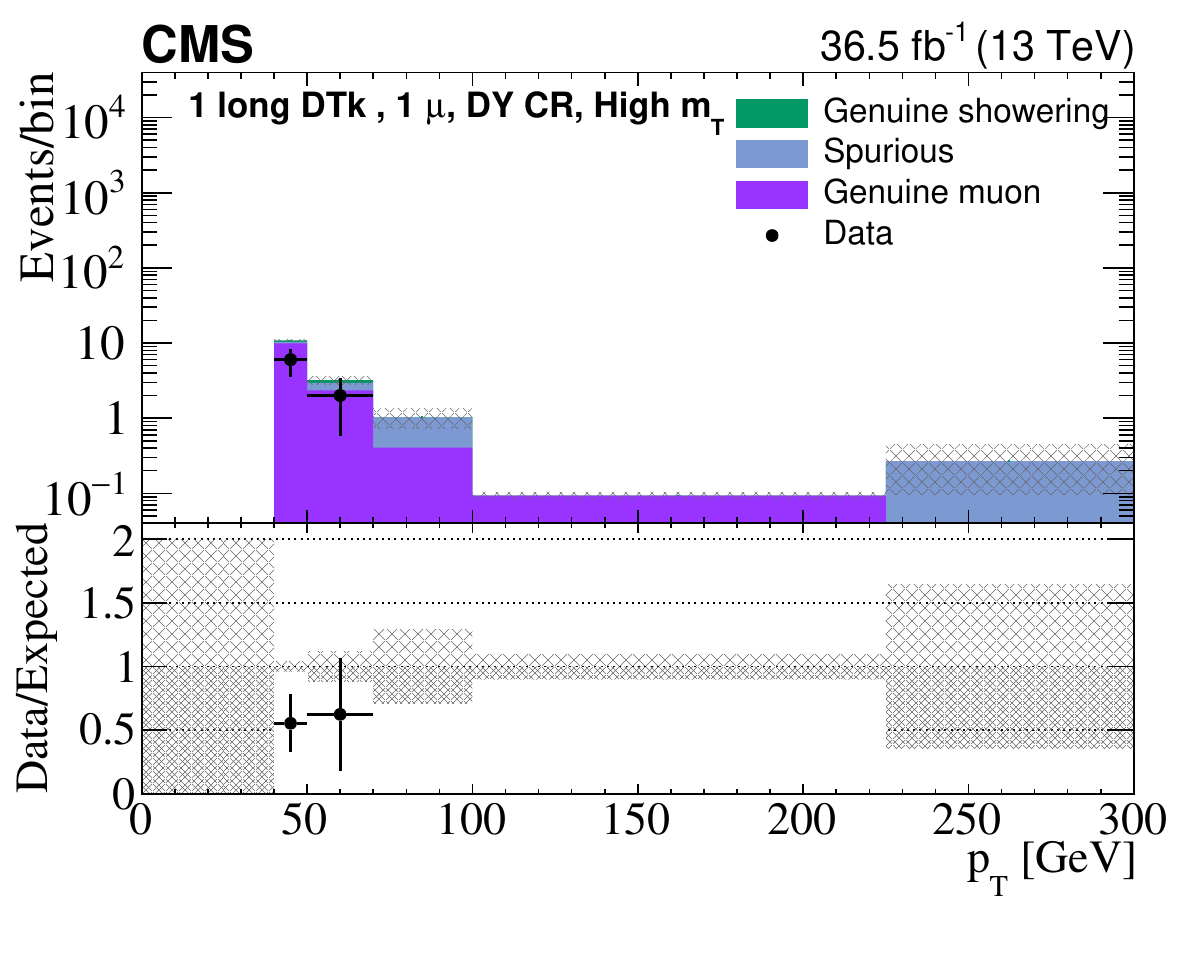}
\includegraphics[width=.45\linewidth]{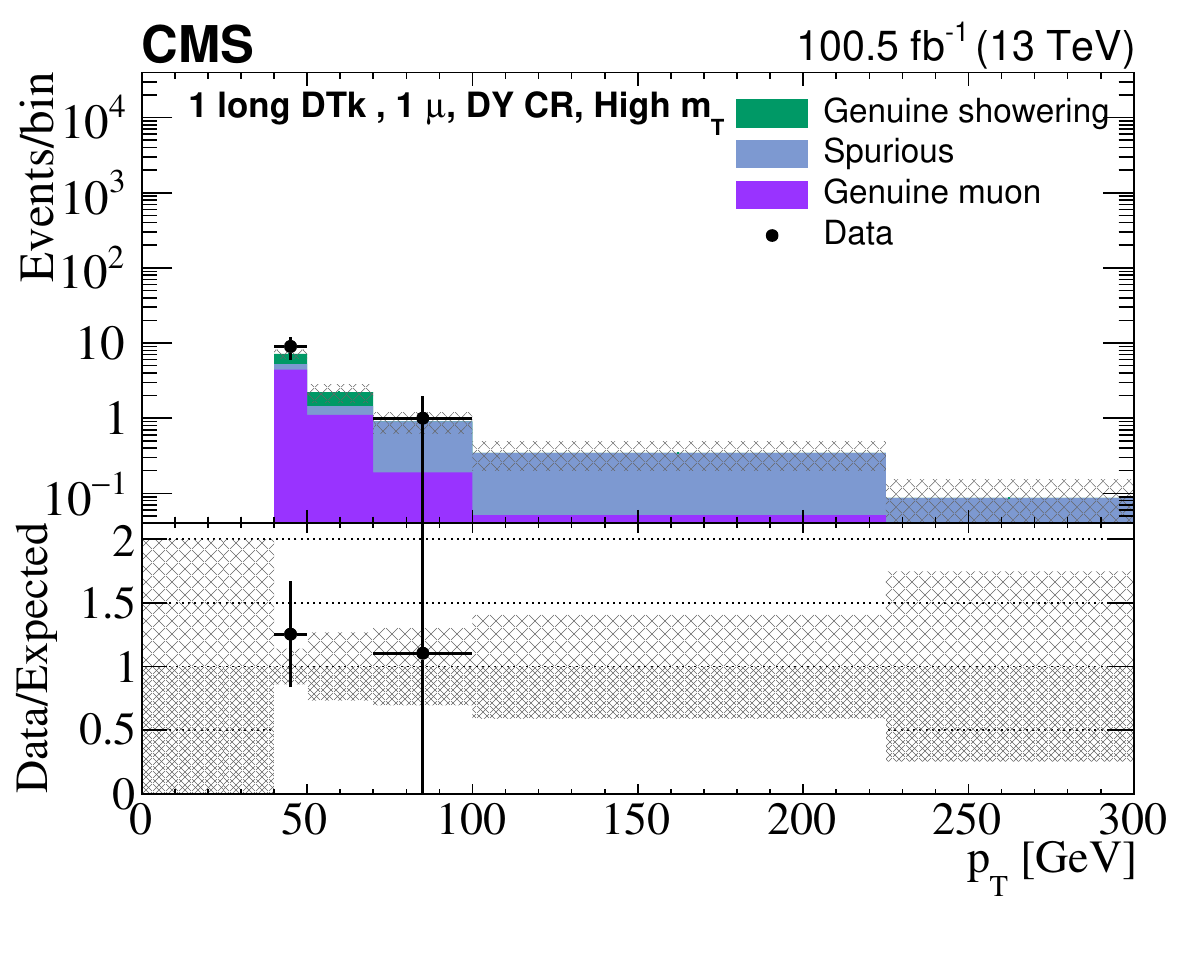}
\caption{
Comparison of the \pt distributions of DTks in the high-\mT validation region 
for the data and background prediction
for long (upper) and short (middle) showering tracks
and for long muon tracks (lower).
The left (right) column corresponds to the Phase-0 (Phase-1) detector.
The uncertainty bars on the ratios in the lower panels indicate the fractional Poisson uncertainties
in the observed counts.
The gray bands show the fractional Poisson uncertainties in the sideband region counts,
added in quadrature with the systematic uncertainties.
The leftmost (rightmost) bin includes underflow (overflow).
}
\label{fig:prompt-test}
\end{figure*}

We study the dependence of the transfer factors \klowhi and \kmuon
on the kinematic properties of the tracks.
It is found that the \pt spectra of DTks in the DY measurement CRs are predicted reasonably well,
with statistically significant deviations within 20\% for the showering background tracks.
Comparisons between the predicted and observed \pt spectra of the tracks
in the two DY measurement CRs are presented in Fig.~\ref{fig:prompt-meas}.
Tests of the method are performed in high-\mT validation regions
defined in the same manner as the corresponding DY measurement CR
but with the inverted requirement $\mT>110\GeV$.
The results in the validation regions are shown in Fig.~\ref{fig:prompt-test}
and indicate agreement between the predicted and observed results 
to within the statistical and systematic uncertainties. 

\subsection{Spurious-particle background}
\label{sec:combinatoric}

The spurious-particle background arises when 
hits in the silicon tracker from two or more particles
align to form a pattern that mimics the signature of a single particle.
Such combinatoric patterns
typically do not have associated calorimetric or muon system activity.
They thereby can satisfy the selection criteria for~DTks and appear as background in our search.
The spurious-particle background depends primarily on the level of activity in an event,
particularly the occupancy of the tracker.
It forms the dominant background
for the short-track category of~DTks. 

The spurious-particle background is evaluated using sideband control regions \crfake,
one for each SR,
and a ``QCD measurement CR'' enhanced in QCD multijet events.
The \crfake samples are selected
using the same criteria as for the corresponding SRs
except with lower ranges in the BDT variable.
These ranges are listed in the last row of Table~\ref{tab:sr-cr}.
The ranges are chosen so as to yield sufficiently populated samples,
a high purity of spurious tracks,
and a phase space similar to the SRs.

The QCD measurement CR is selected by requiring
$\nelec=\nmuon=\nbjets=0$ and $30<\ptmisshard<60\GeV$.
It is essentially 100\% pure in spurious-track events,
with negligible contamination.
A transfer factor \thilow is derived from the QCD measurement CR sample as
\begin{linenomath}
\begin{equation}
  \thilow = \nbdthi/\nbdtlo,
\end{equation}
\end{linenomath}
with \nbdthi and \nbdtlo the number of events satisfying the
BDT selection criteria in Table~\ref{tab:sr-cr} for the SR and \crfake selections,
respectively.
The values of the transfer factors \thilow are listed in Table~\ref{tab:theta}.

The estimate \nfake of the number of spurious-particle background events in the $i$th SR is
\begin{linenomath}
\begin{equation}
  \nfake = \thilow\nCRfake,
\end{equation}
\end{linenomath}
with \nCRfake the number of events in the $i$th \crfake event sample.

Because of the limited size of the data sample,
six SRs in the large \dedx category have no
events in the corresponding \crfake sample:
these are bins 18, 22, 34, 42, 44, and 46
(see Tables~\ref{tab:SR1} and~\ref{tab:SR2}).
To obtain a prediction for these bins,
an extrapolation is made based on the predicted spurious-track background 
in the adjacent SR, namely,
the equivalent SR in the low \dedx category.
The prediction is made as
\begin{linenomath}
\begin{equation}
  \nfakehi = \dedxlohi\nfakelo,
\end{equation}
\end{linenomath}
where the transfer factor \dedxlohi is
studied in various CRs and is found to be $0.18\pm0.05$
for both long and short tracks,
independent of any analysis variable.
Relative systematic uncertainties associated with the extrapolated counts
are taken from the bin from which the value is extrapolated,
in addition to the uncertainty in~\dedxlohi.

\begin{table}[t]
\centering
\topcaption{
The transfer factor \thilow used for the evaluation of the spurious-particle background.
The  uncertainties are statistical only.
}
\begin{scotch}{cccc}
\multicolumn{4}{c}{Spurious particle} \\
\thilow: & \multicolumn{1}{c}{Phase 0}   & \multicolumn{1}{c}{Phase 1}     & \multicolumn{1}{c}{Combined Phase 0 and 1}   \\[2pt]
\hline
Short             & $0.722 \pm 0.079$ & $0.210 \pm 0.020$           & $0.346 \pm 0.025$  \\
Long              & $0.089 \pm 0.015$ & $0.042 \pm 0.021$           & $0.080 \pm 0.013$  \\
\end{scotch}
\label{tab:theta}
\end{table}

\begin{figure*}[t]
\centering
\includegraphics[width=.45\linewidth]{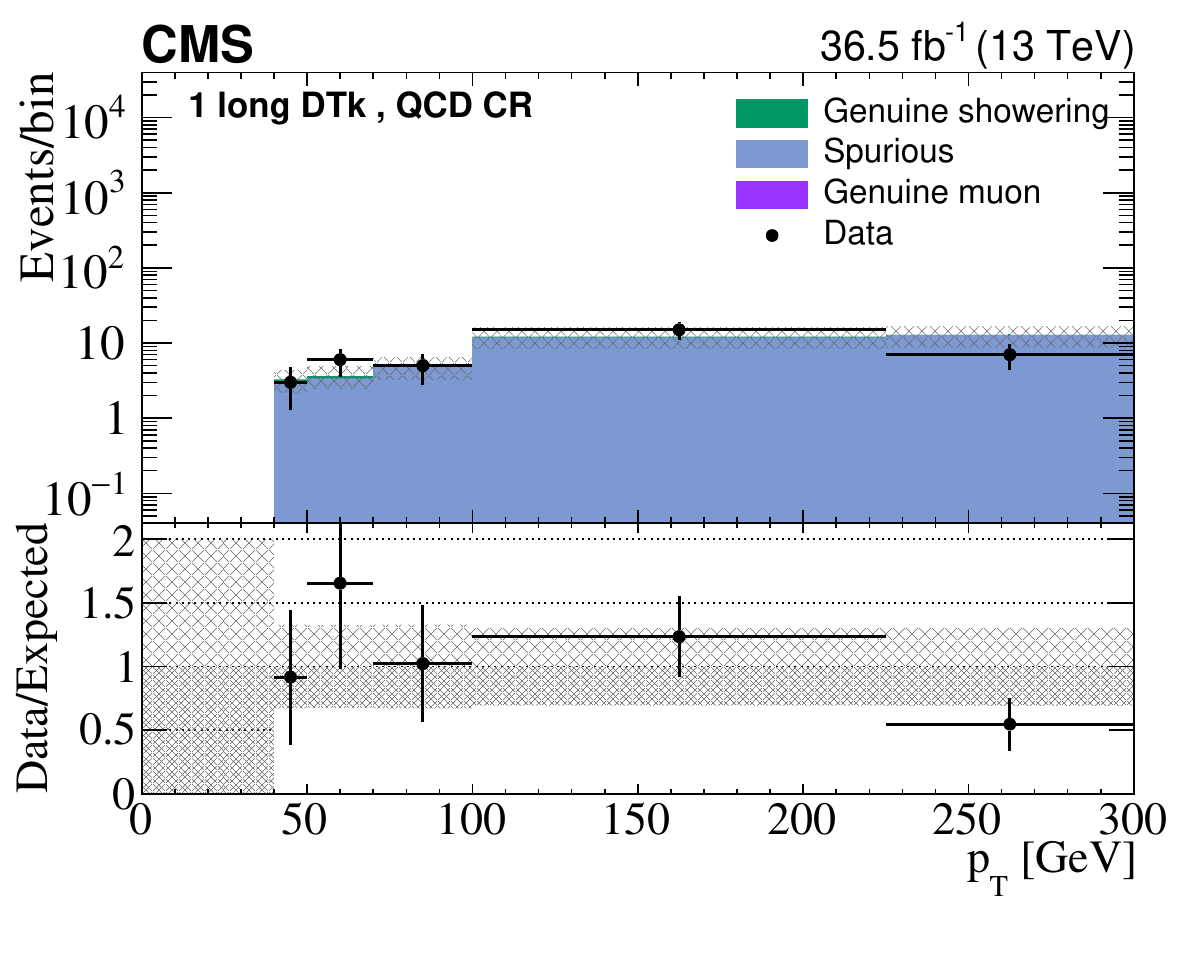}
\includegraphics[width=.45\linewidth]{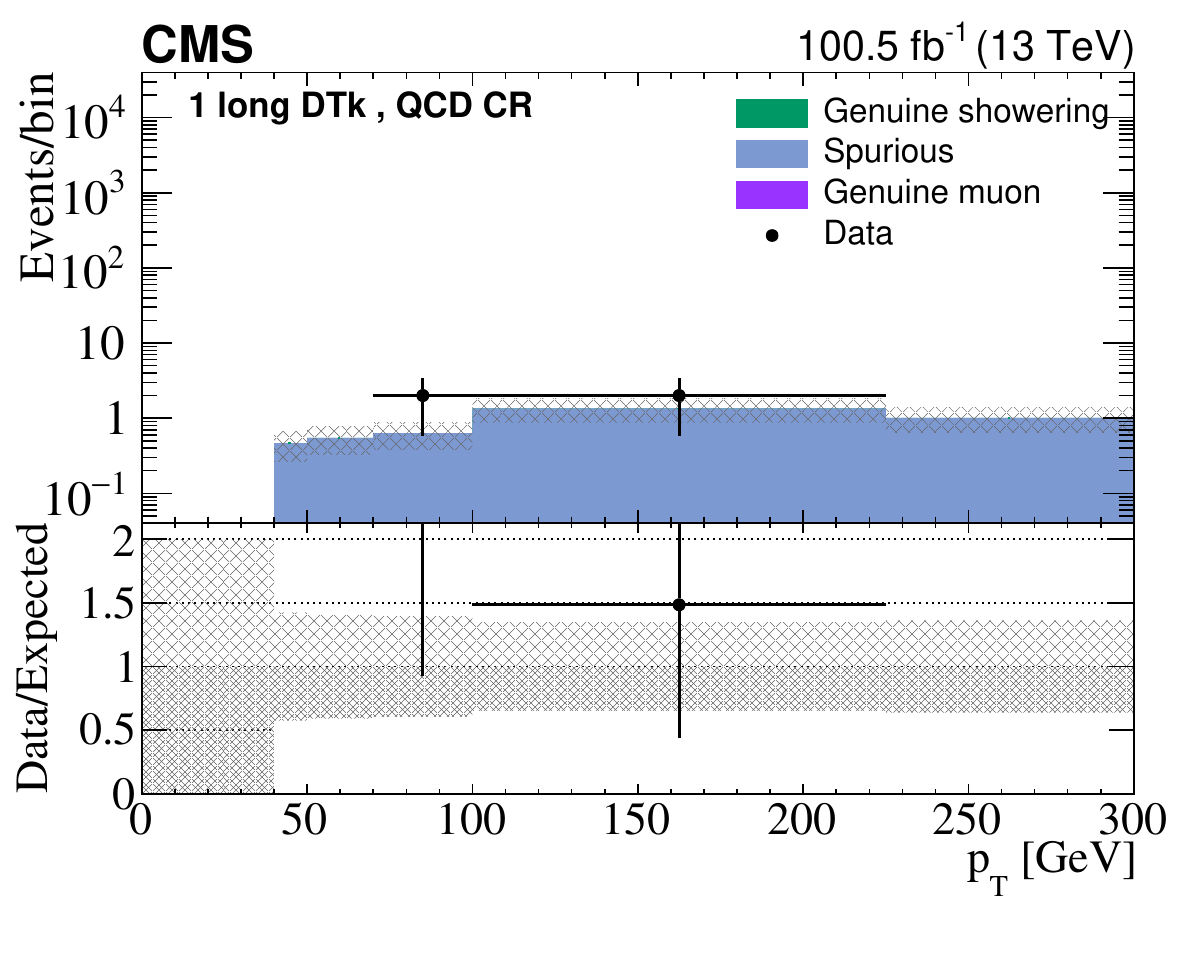}\\
\includegraphics[width=.45\linewidth]{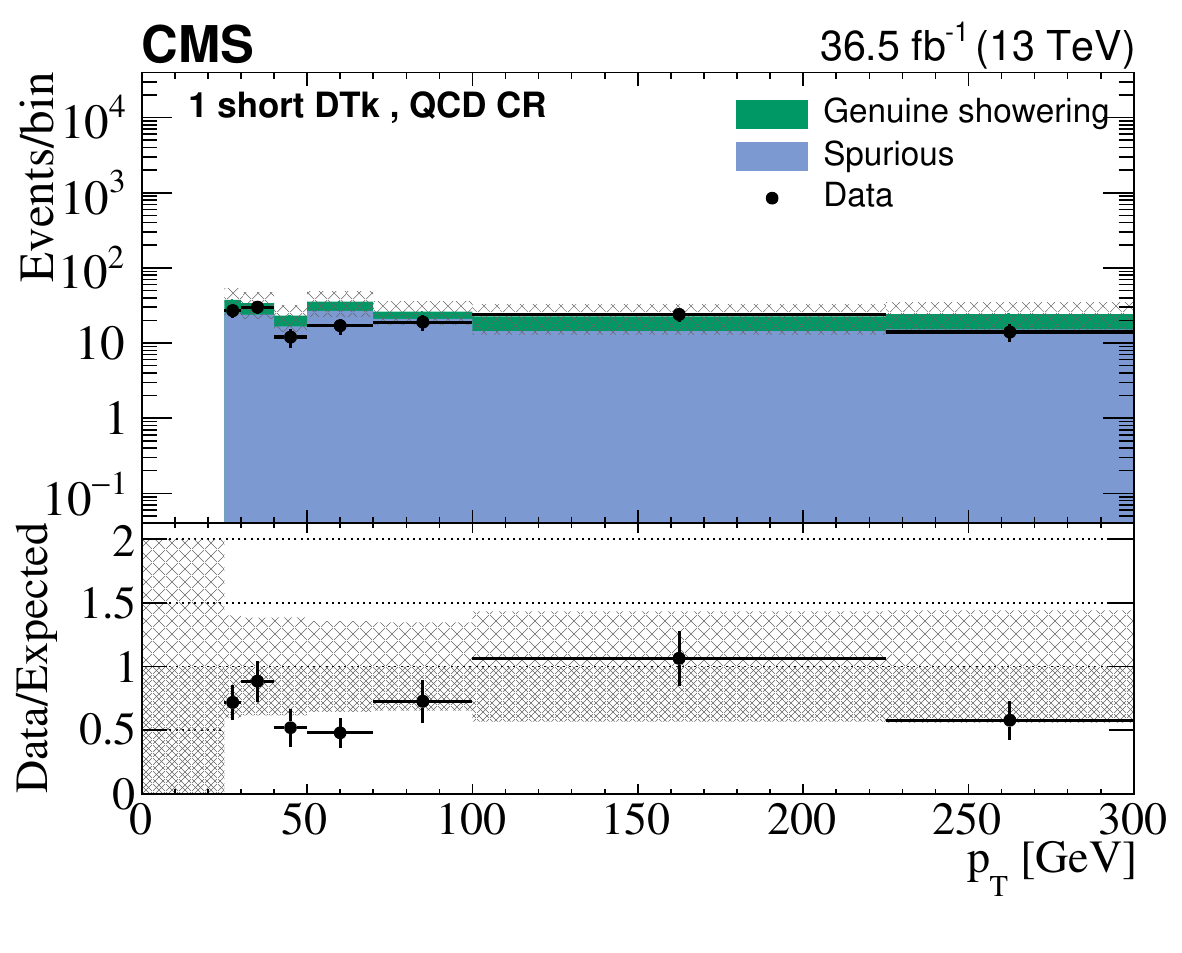}
\includegraphics[width=.45\linewidth]{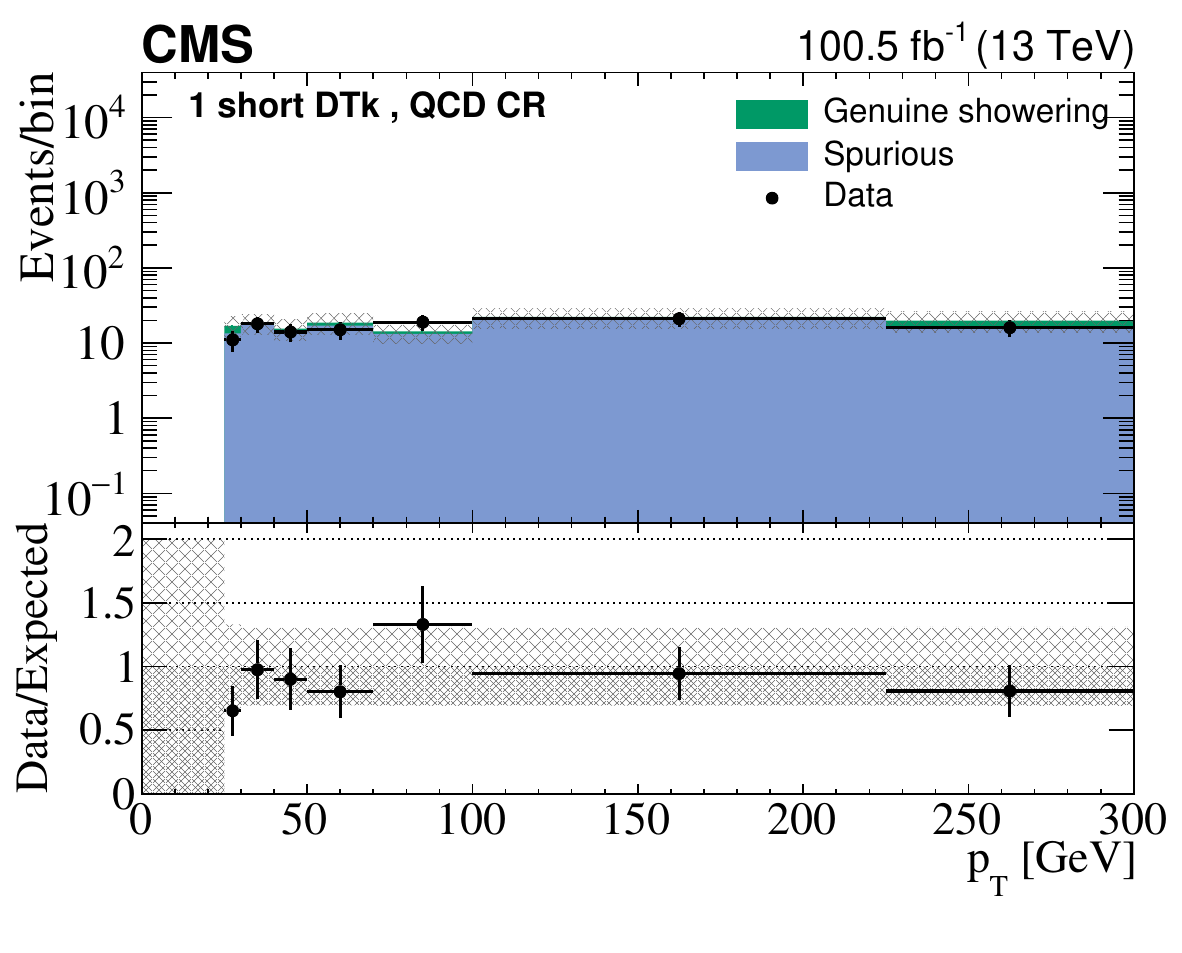}
\caption{
Comparison of the \pt distributions of DTks in the \thilow QCD measurement control region 
for the data and background prediction
for long (upper) and short (lower) tracks.
The left (right) column corresponds to the Phase-0 (Phase-1) detector.
The uncertainty bars on the ratios in the lower panels indicate the fractional Poisson uncertainties
in the observed counts.
The gray bands show the fractional Poisson uncertainties in the control region counts,
added in quadrature with the systematic uncertainties.
The leftmost (rightmost) bin includes underflow (overflow).
}
\label{fig:fake-meas}
\end{figure*}

\begin{figure*}[t]
\centering
\includegraphics[width=.45\linewidth]{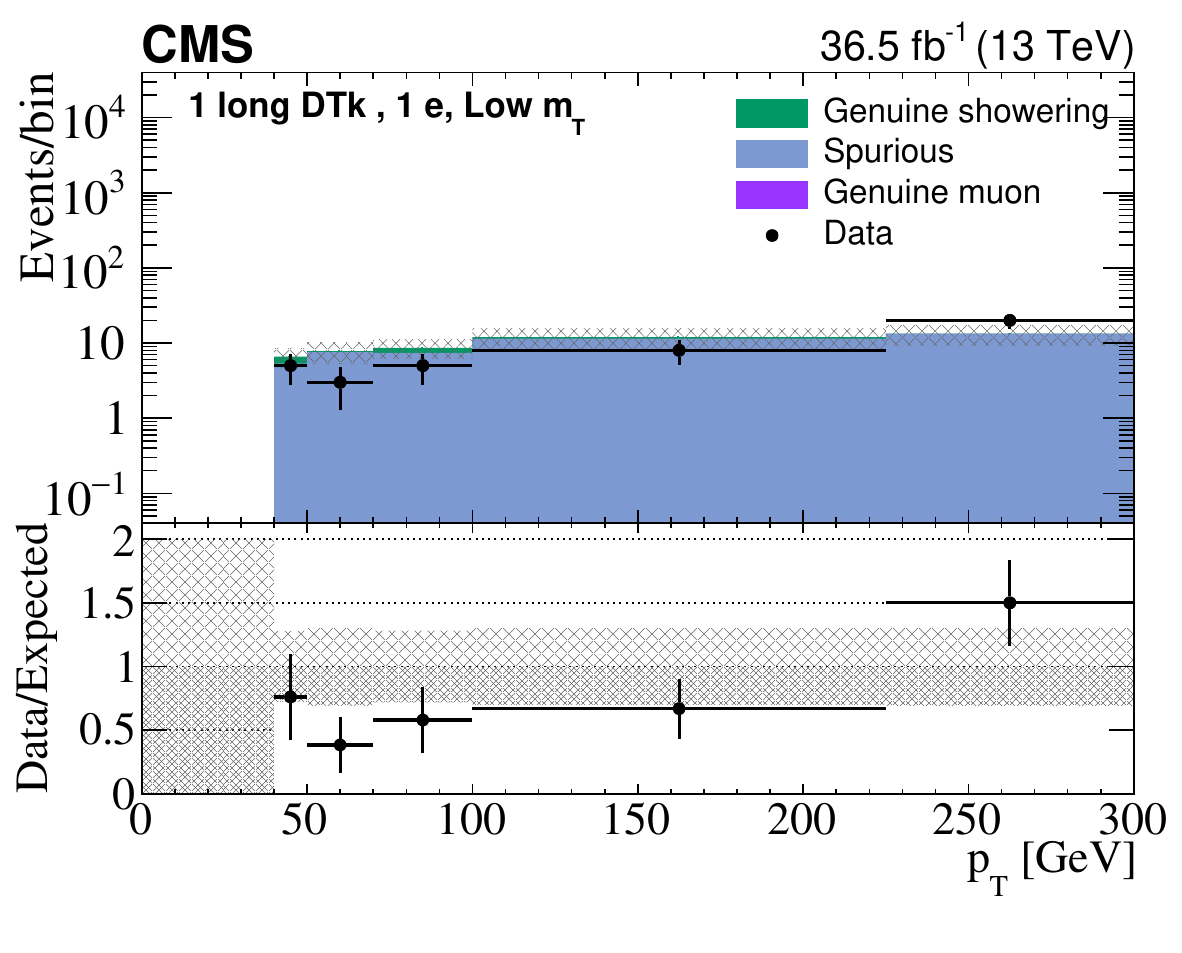}
\includegraphics[width=.45\linewidth]{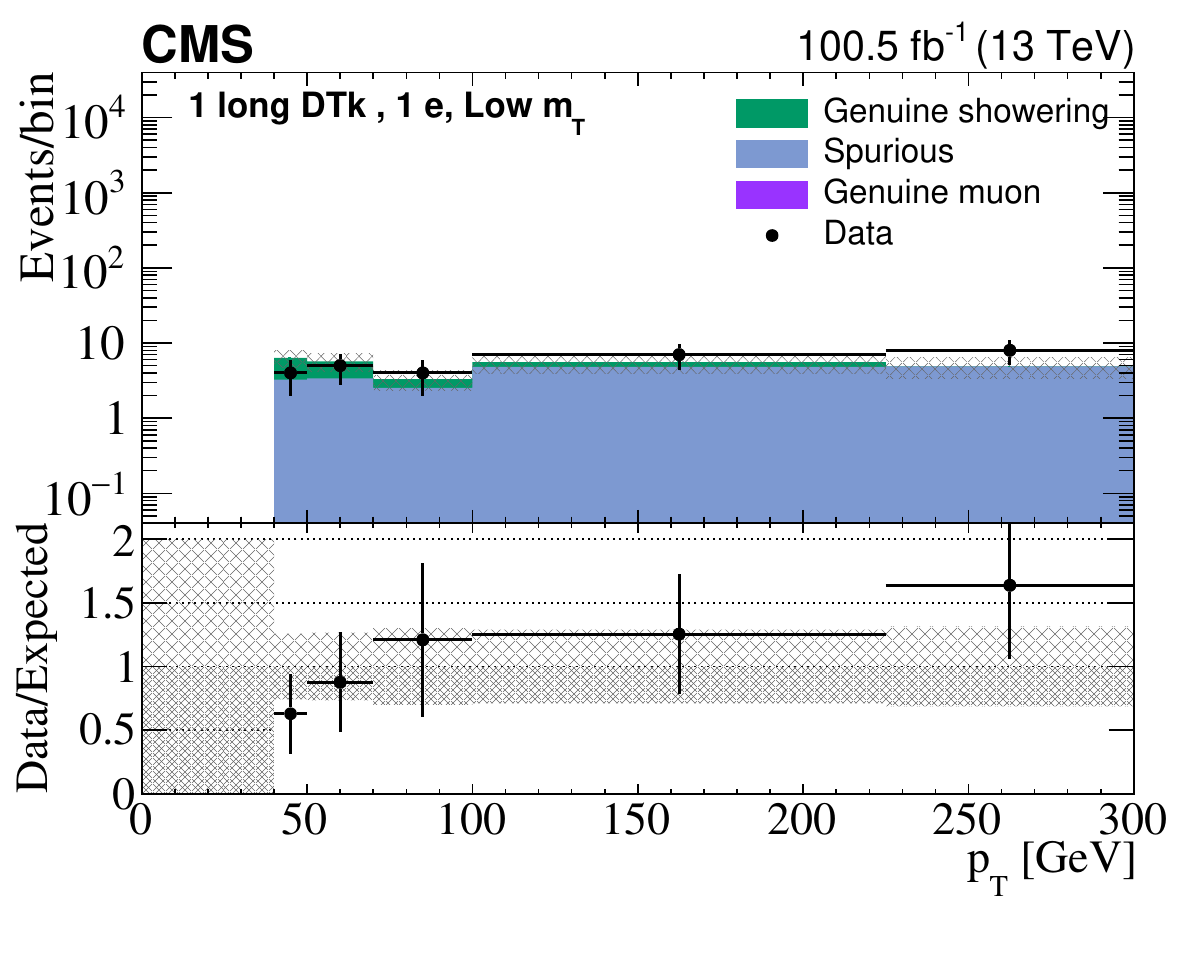}\\
\includegraphics[width=.45\linewidth]{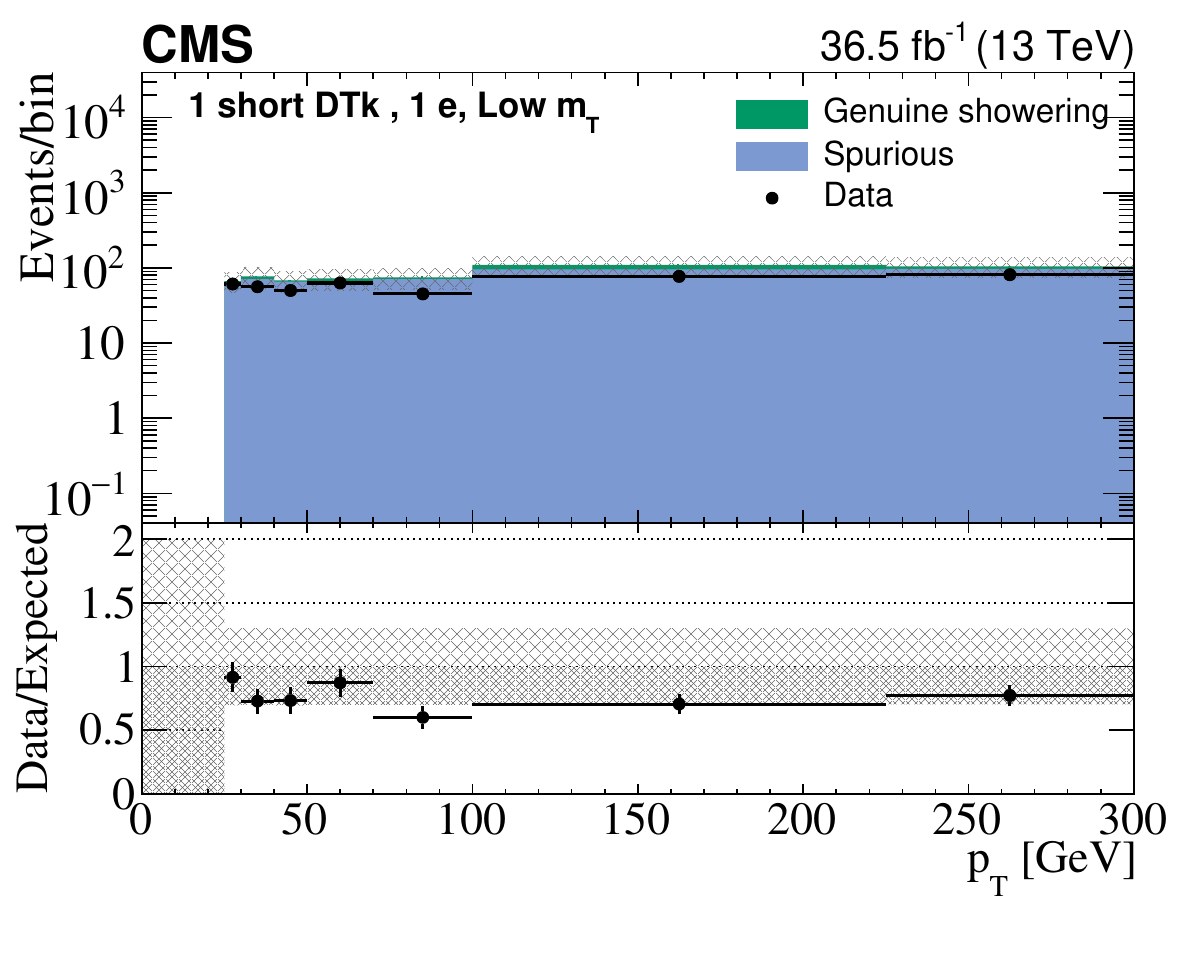}
\includegraphics[width=.45\linewidth]{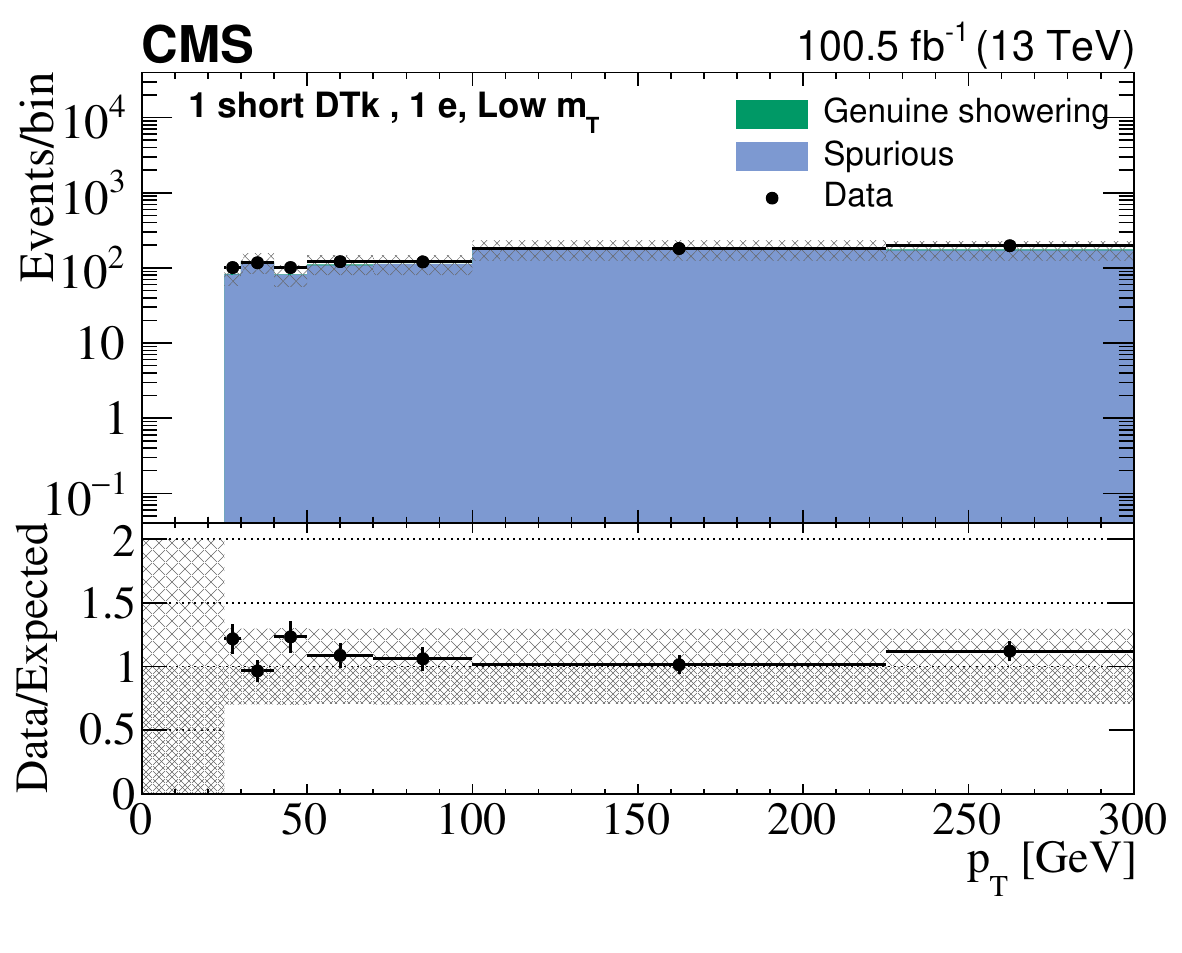}
\caption{
Comparison of the \pt distributions of DTks in events with one electron and one DTk,
in a validation region with $\mT<110$\GeV and $m_{\text{DTk},\ell} \not\in [65,110]$\GeV,
for the data and background prediction
for long (upper) and short (lower) tracks.
The left (right) column corresponds to the Phase-0 (Phase-1) detector.
The uncertainty bars on the ratios in the lower panels indicate the fractional Poisson uncertainties
in the observed counts.
The gray bands show the fractional Poisson uncertainties in the control region counts,
added in quadrature with the systematic uncertainties.
The leftmost (rightmost) bin includes underflow (overflow).
}
\label{fig:fake-test}
\end{figure*}

Comparisons between the predicted and observed \pt spectra of DTk candidates in the
QCD measurement CR are shown in Fig.~\ref{fig:fake-meas}.
The spectra are seen to be reasonably well modeled,
with statistically significant deviations below around~30\%,
within the total uncertainty.
Tests of the spurious-particle background evaluation procedure are
performed in a validation region defined by requiring the presence of one electron and one DTk,
with the restrictions $\mT<110\GeV$ and $m_{\text{DTk},\ell} \not\in [65,110]$\GeV.
Note that this validation region also accounts for genuine-particle background
that might be present in the spurious-particle sample.
The results,
shown in Fig.~\ref{fig:fake-test},
again demonstrate general agreement within the uncertainties.  

\section{Systematic uncertainties}
\label{sec:systematic}

We consider the following sources of systematic uncertainty in the signal event yields.
Nuisance parameters associated with systematic uncertainties
are treated as correlated across all search bins.

\begin{itemize}
\item \emph{DTk selection efficiency:}
  The uncertainty in the DTk selection efficiency is taken to be the uncertainty in the
  DTk selection efficiency scale factors mentioned in Section~\ref{sec:tracks}.
\item \emph{Integrated luminosity:}
 The systematic uncertainty in the determination of the integrated luminosity
 is 1.6\% \cite{CMS:LUM-17-003,CMS-PAS-LUM-17-004,CMS-PAS-LUM-18-002,Giraldi:2022mwf}.
\item \emph{Jet energy scale and resolution:}
  Uncertainties in the jet energy scale and resolution are evaluated
  as a function of jet \pt and $\eta$,
  and propagated to higher level quantities such as \njets
  and \ptmisshard~\cite{Khachatryan:2016kdb,CMS-PAS-JME-16-003,CMS:2019ctu}.
\item \emph{\PQb jet tagging efficiency and mistagging:}
  \pt, $\eta$, and flavor-dependent uncertainties
  in the \PQb jet tagging
  and light-quark jet mistagging scale factors,
  accounting both for data-versus-{\GEANTfour}-simulation differences
  and for {\GEANTfour}-versus-fast-simulation differences,
  are derived from control samples of \ttbar and QCD multijet events.
 \item \emph{Renormalization and factorization scales:}
  Uncertainties associated with the renormalization (\rscale) and
  factorization (\fscale) scales are evaluated by independently varying \rscale and \fscale
  by a factor of 2.0 and 0.5~\cite{Kalogeropoulos:2018cke,Catani:2003zt,Cacciari:2003fi}.
\item \emph{Initial-state radiation:}
  The uncertainty in the modeling of initial-state radiation
  is taken to be one half of the deviation from unity in the ISR reweighting factors
  presented in Section~\ref{sec:mc}.
\item \emph{Pileup modeling:}
  The uncertainty associated with pileup reweighting is evaluated by varying 
  the value of the total inelastic cross section by 4.6\%~\cite{CMS:2020ebo} from its
  nominal value of 69.2\unit{mb}.
\item \emph{Trigger efficiency:}
  The uncertainty in the trigger efficiency of the hadronic channel
  is assessed from its observed dependence on jet multiplicity
  and of the leptonic channels from
  the observed difference relative to an alternative trigger.
  In addition, the 2016 and 2017 data-taking periods were affected by
  an erroneous ``prefiring'' of the L1 trigger 
  on the previous bunch crossing,
  prohibiting the triggering of some fraction of potential signal events.
  We evaluate the resulting uncertainty
  in the trigger efficiency using a preliminary L1 prefiring efficiency map based on 
  the end of the 2017 data-taking period,
  representing a worst case scenario. 
\item \emph{\dedx calibration:}
  A time-dependent calibration of the \dedx measurement
  is performed using
  tracks associated with a muon,
  with the \dedx value extracted from a Gaussian fit around the peak of the \dedx distribution.
  A similar set of calibration factors is derived using
  low-\pt protons from \PGL baryon decays.
  The difference in the calibration factors derived from the muons and protons
  is used to determine a systematic uncertainty in the \dedx calibration.
\end{itemize}

\begin{table}[t]
\centering
\topcaption{
Upper: The considered sources of systematic uncertainty in the predicted signal yield
and the corresponding range of values over the 49 search regions.
A value of 0 is reported when the relative uncertainty is determined to be less than~0.5\%.
Lower: The ranges for the total pre-fit uncertainty in the predicted background counts
with respect to the respective background contribution.
}
\begin{scotch}{lc}
\multirow{2}{*}{Uncertainty source} & Relative uncertainty \\ 
                                    & in yield (\%) \\ \hline
Signal & \\
\quad DTk selection efficiency & 10\textendash17 \\
\quad Integrated luminosity & 1.6 \\
\quad Jet energy scale and resolution & 0\textendash24 \\
\quad \PQb jet tagging & 0\textendash4 \\
\quad Renormalization and factorization scales & 0\textendash2 \\
\quad Initial-state radiation & 0\textendash3 \\
\quad Pileup modeling & 0\textendash2 \\
\quad Trigger efficiency & 0\textendash4 \\
\quad \dedx calibration & 3\textendash8 \\ [\cmsTabSkip]
Background & \\
\quad Genuine particle showering long & 20\textendash28 \\
\quad Genuine particle showering short & 100\textendash104 \\
\quad Genuine particle muon long & 25\textendash38 \\
\quad Spurious particle long & 5\textendash52 \\
\quad Spurious particle short & 6\textendash28 \\
\end{scotch}
\label{tab:syst}
\end{table}

The following systematic uncertainties are evaluated for the estimates of the
number of background events.
\begin{itemize}
  \item \emph{Spurious-particle contamination:} An uncertainty of 100\% is assigned
  to the estimated number of genuine-particle short-track background events to account for
  the presence of spurious-particle contamination in the DY CR used to calculate
  the \klowhi transfer factors for this category.
  A 100\% uncertainty is assigned because the measurement region for
  short tracks is dominated by spurious tracks and, after the subtraction,
  this is the order of how well the transfer factor is known. 
  This uncertainty does not impact the sensitivity because this background
  is nearly negligible in the short track SRs.
\item \emph{Residual \pt dependence:}
  Differences in the predicted and observed \pt spectra of DTks
  in the transfer factor CRs are used to assess uncertainties of
  30 and 20\% in the number of spurious-particle long-track and genuine-particle long-track
  background events, respectively.
\end{itemize}
Statistical uncertainties in the SRs and CRs are accounted for in the fit
described in Section~\ref{sec:results}.

The sources of systematic uncertainty
and their assessed ranges across the 49 SRs
are summarized in Table~\ref{tab:syst}.

\section{Results and interpretation}
\label{sec:results}

\begin{figure*}[]
\centering
\includegraphics[width=0.49\linewidth]{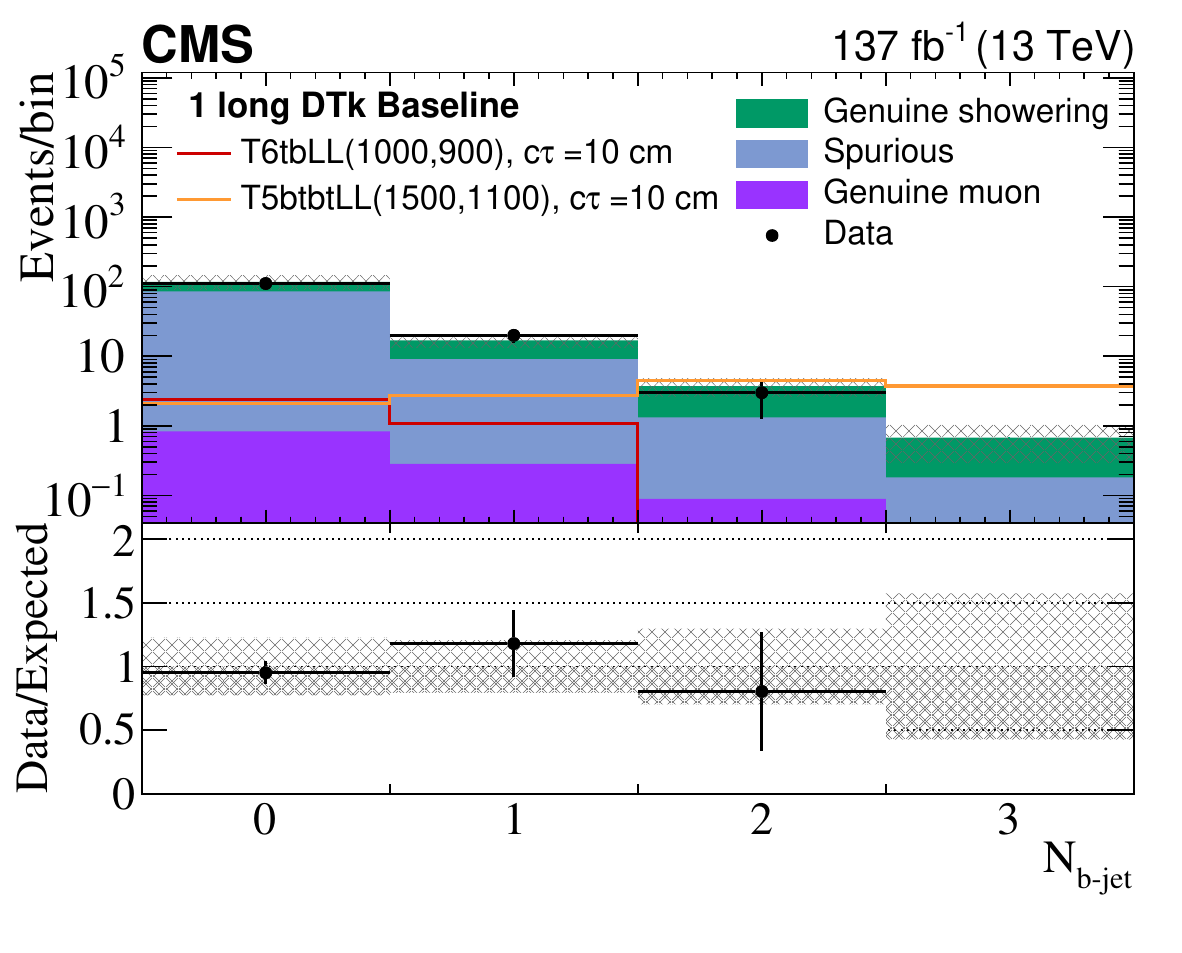}
\includegraphics[width=0.49\linewidth]{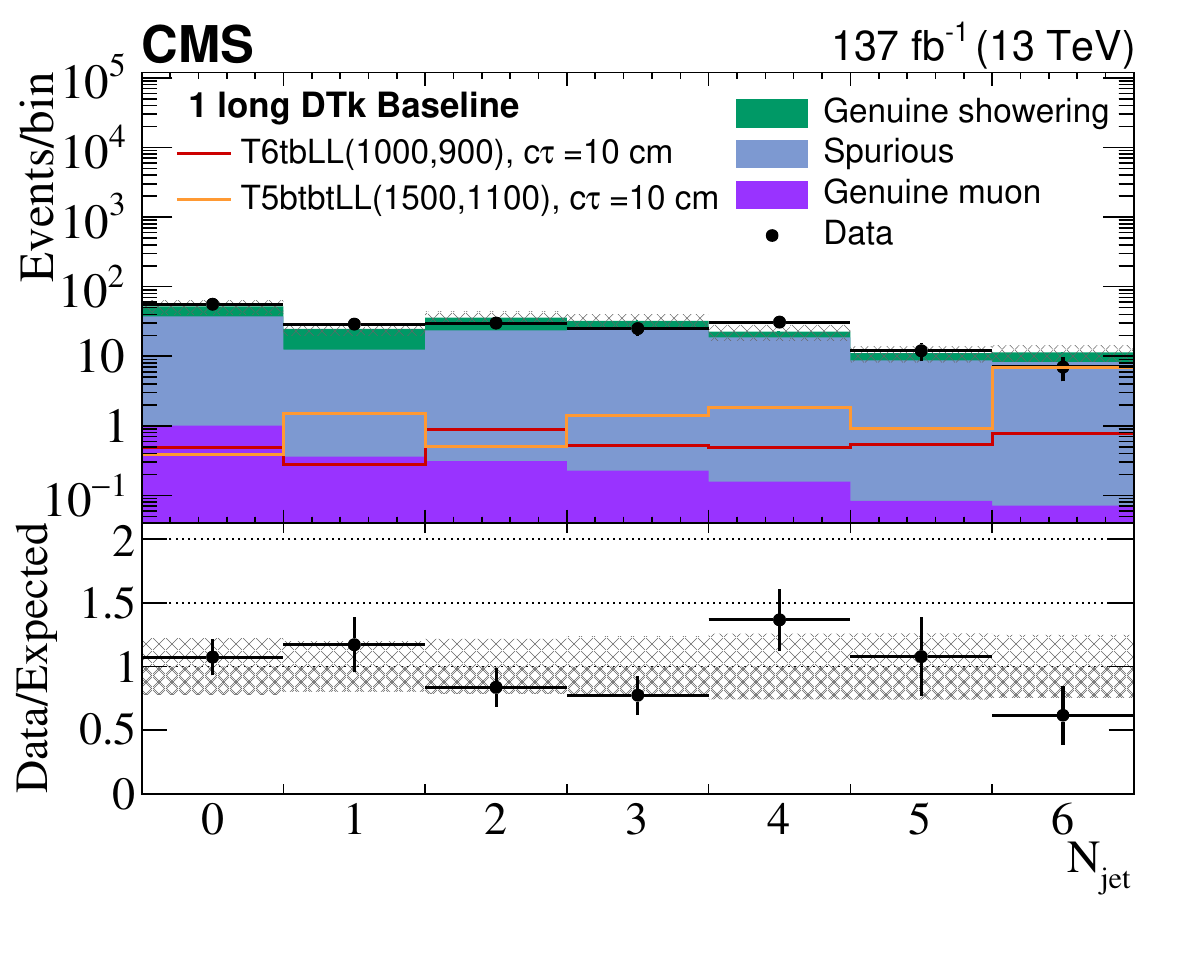} \\
\includegraphics[width=0.49\linewidth]{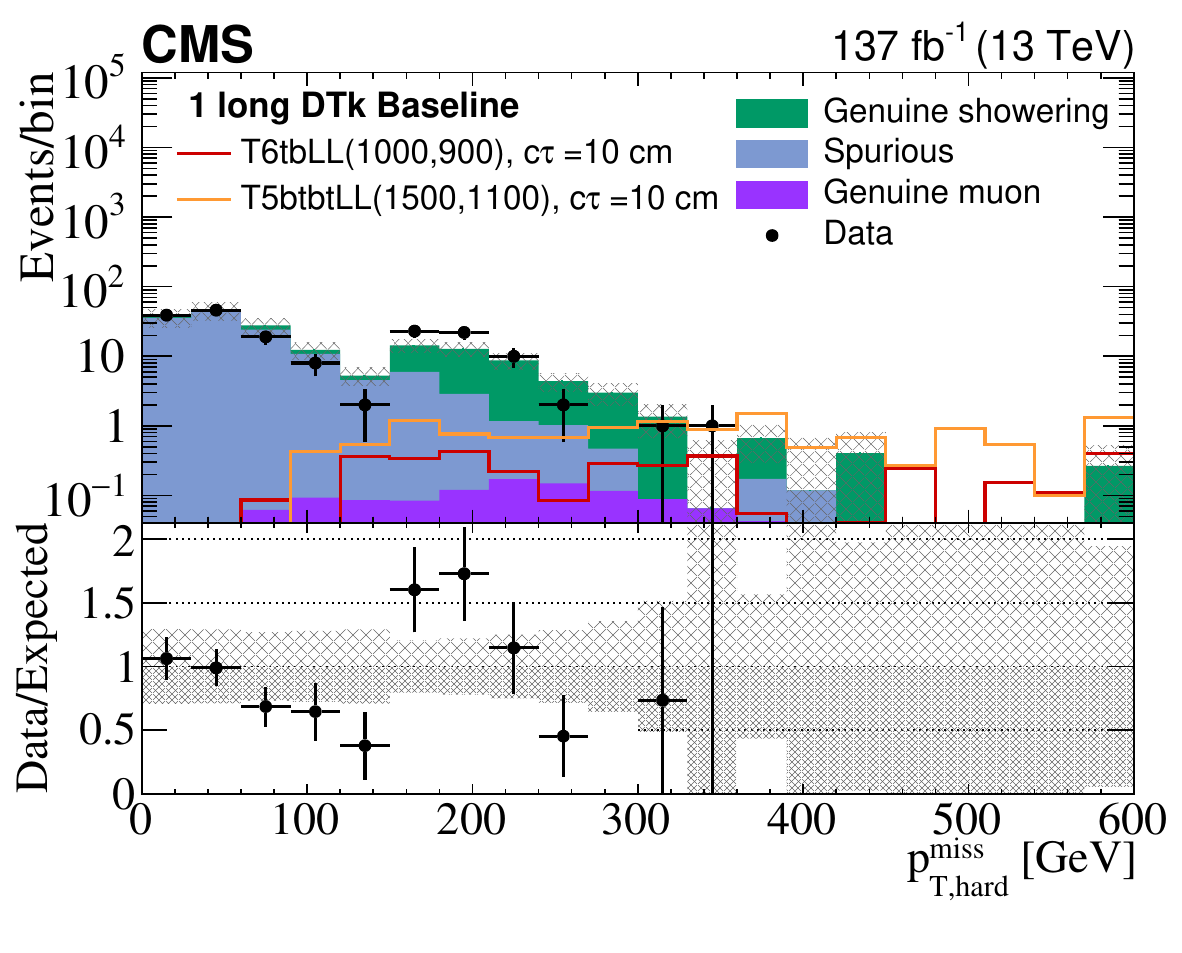}
\includegraphics[width=0.49\linewidth]{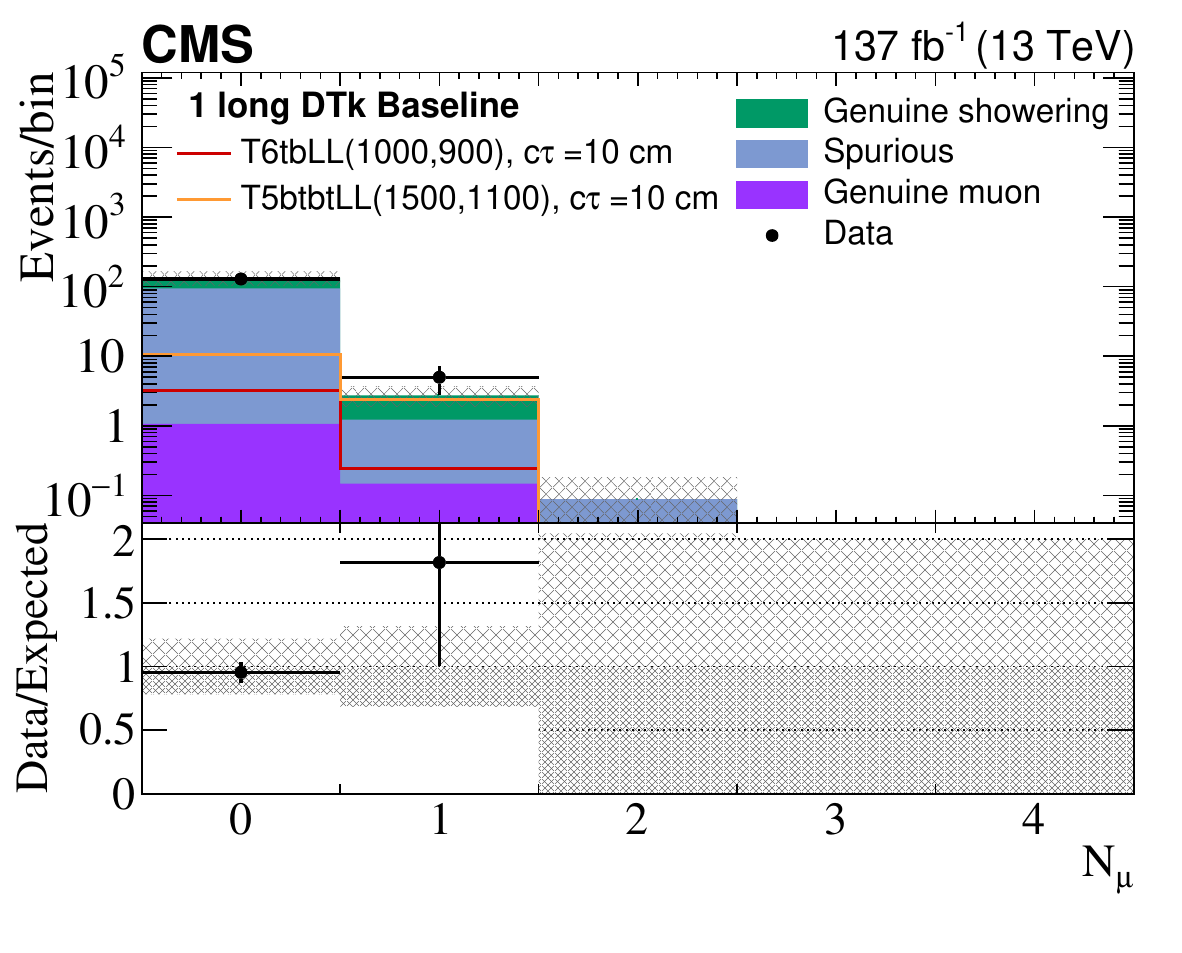} \\
\includegraphics[width=0.49\linewidth]{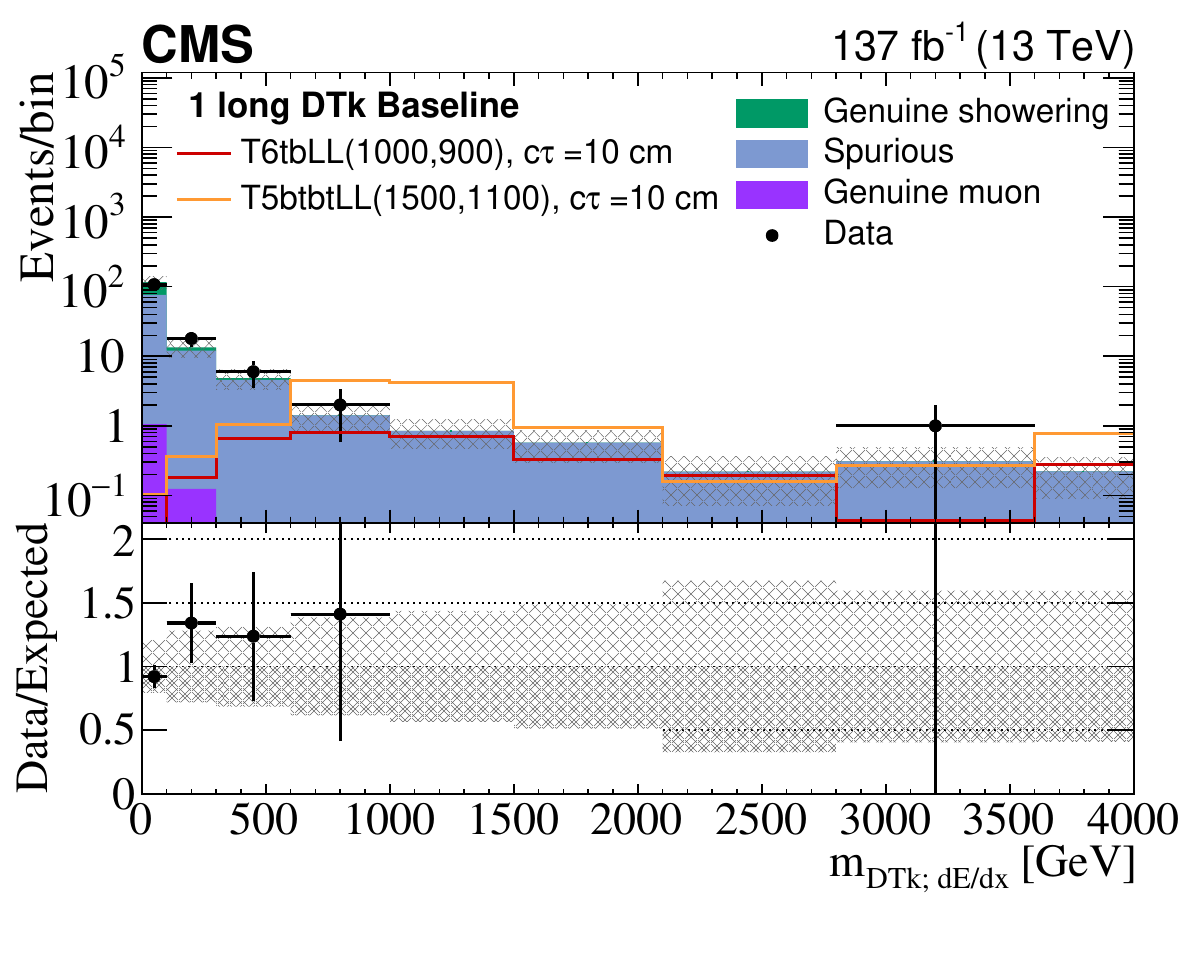}
\includegraphics[width=0.49\linewidth]{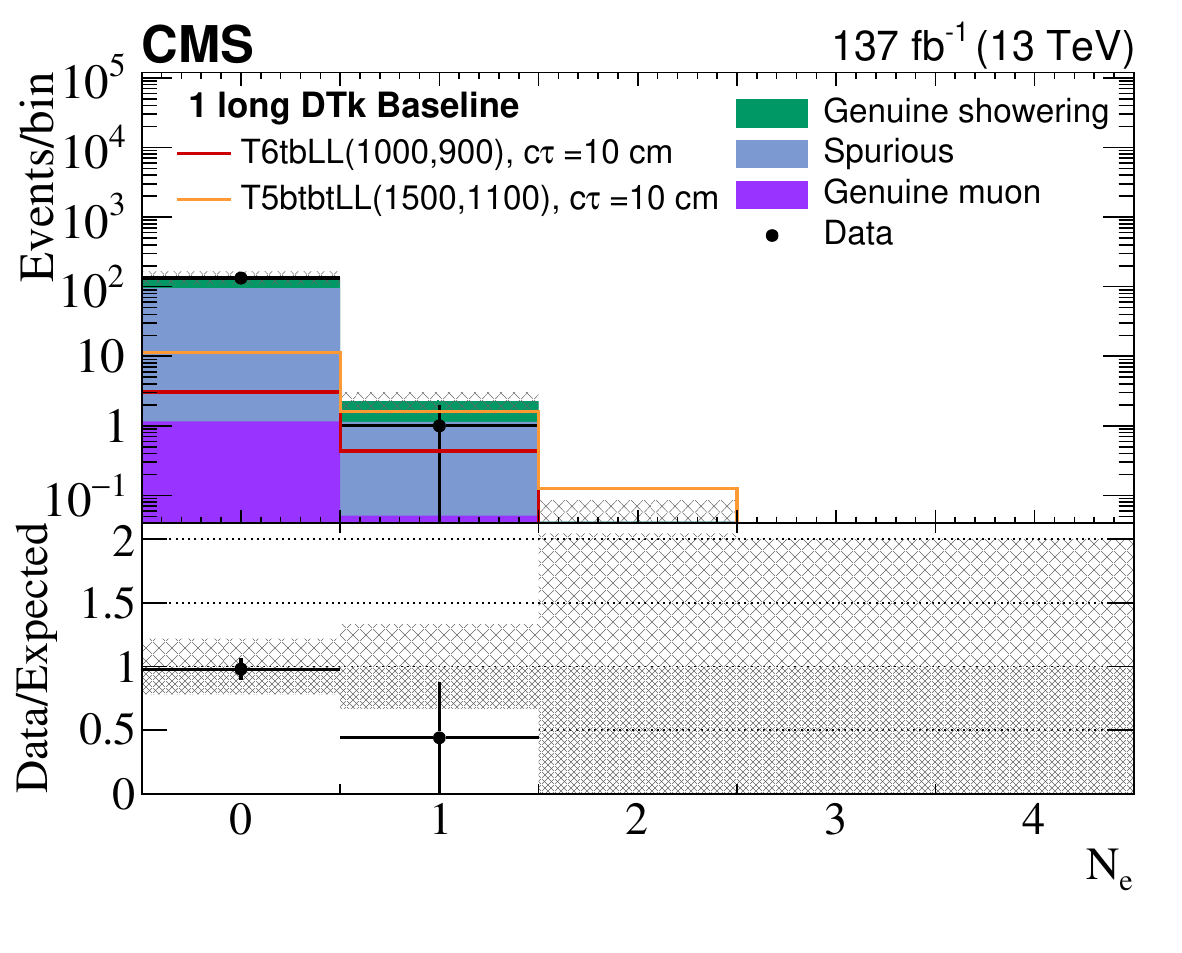}
\caption{
Comparison in the baseline region for the long-track DTk category
between the data and pre-fit predicted SM background
for the \njets (upper left), \nbjets (upper right), \ptmisshard (middle left),
\nelec (middle right), \nmuon (lower left), and \mdedx (lower right) distributions.
The uncertainty bars on the ratios in the lower panels indicate the fractional Poisson uncertainties
in the observed counts.
The gray bands show the fractional Poisson uncertainties in the control region counts,
added in quadrature with the systematic uncertainties.
The leftmost (rightmost) bin includes underflow (overflow).
For purposes of illustration,
results from the T6tbLL and T5btbtLL models are shown,
where the first and second numbers in parentheses indicate the squark (or gluino) mass
and the LSP mass, respectively, in \GeV.
}
\label{fig:baseline1}
\end{figure*}

\begin{figure*}[]
\centering
\includegraphics[width=0.49\linewidth]{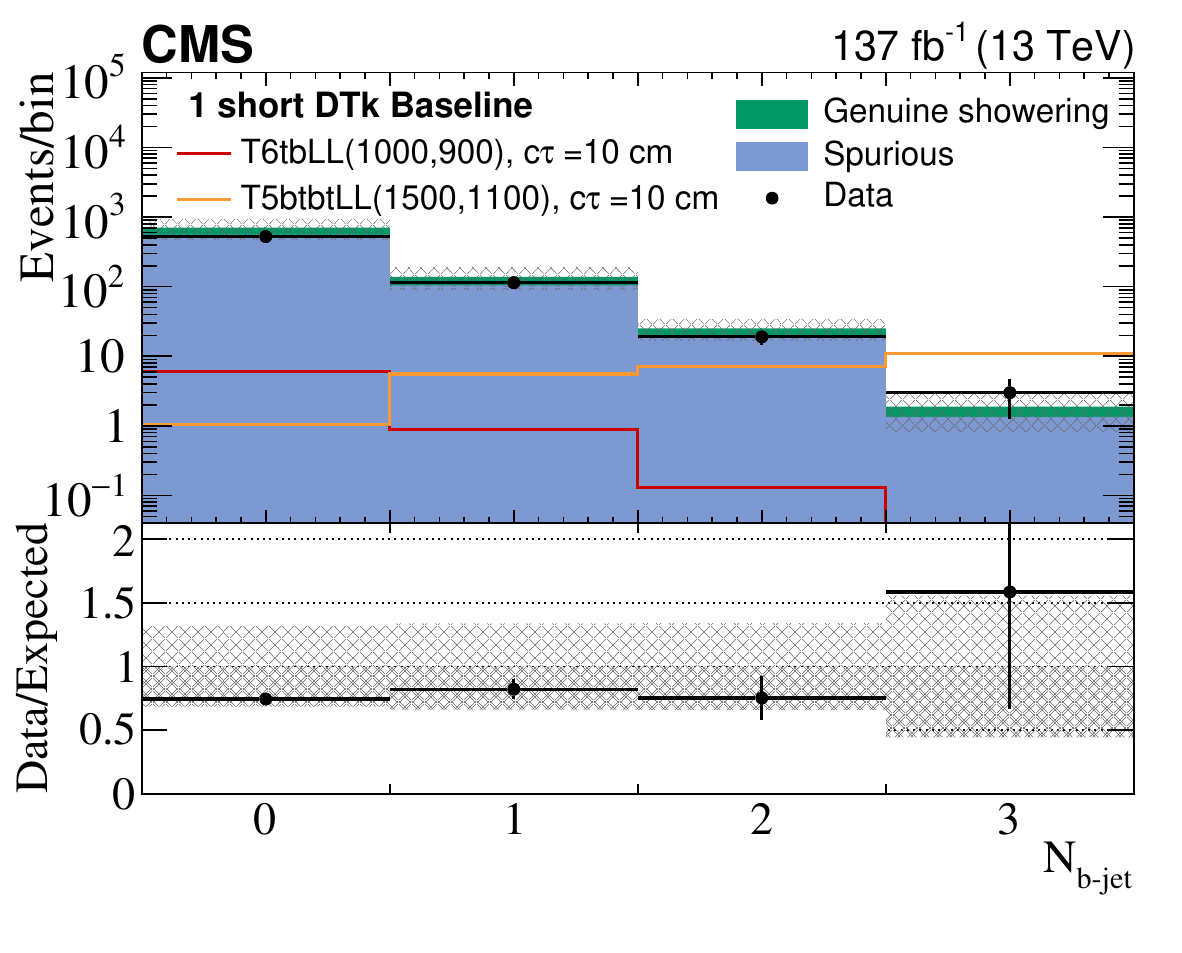} 
\includegraphics[width=0.49\linewidth]{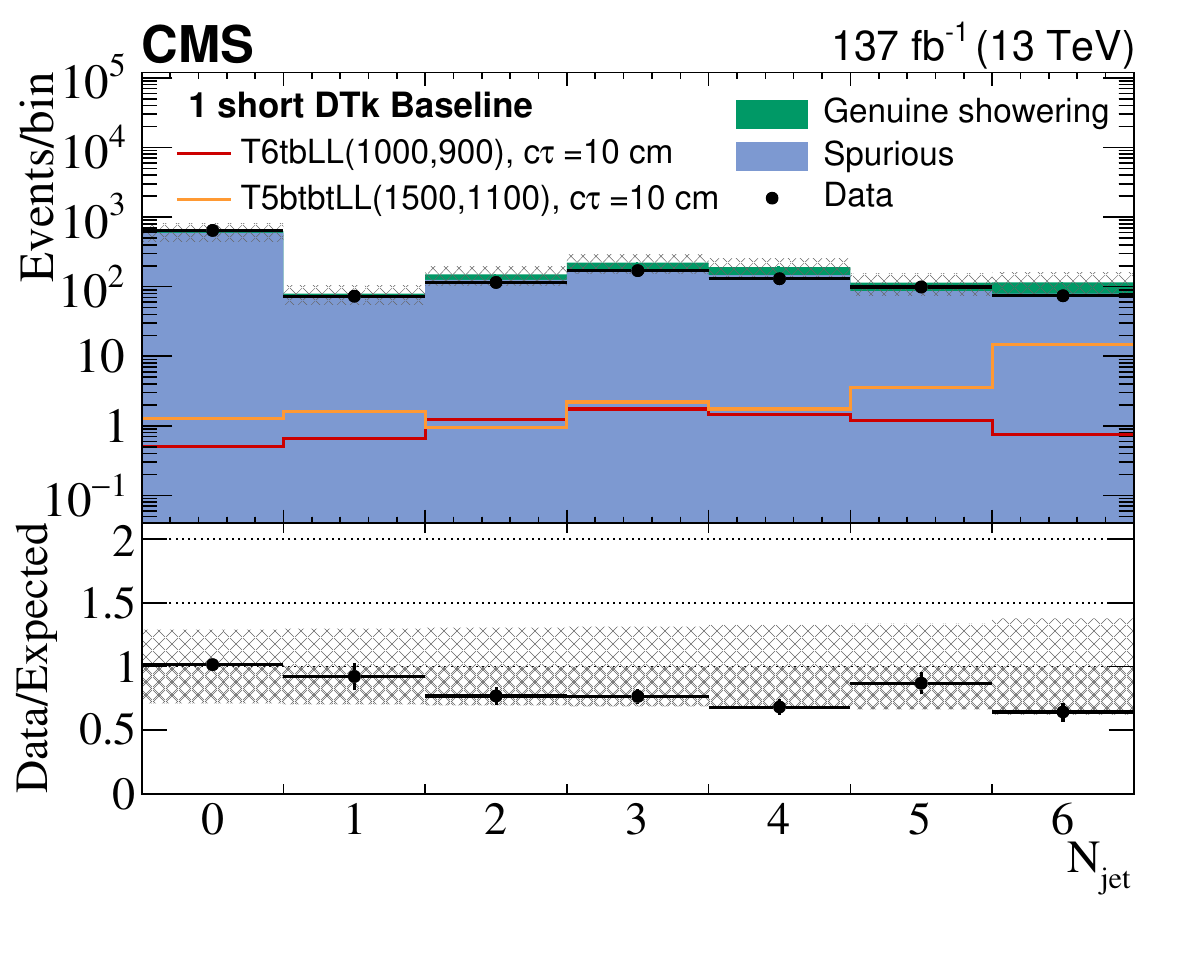} \\
\includegraphics[width=0.49\linewidth]{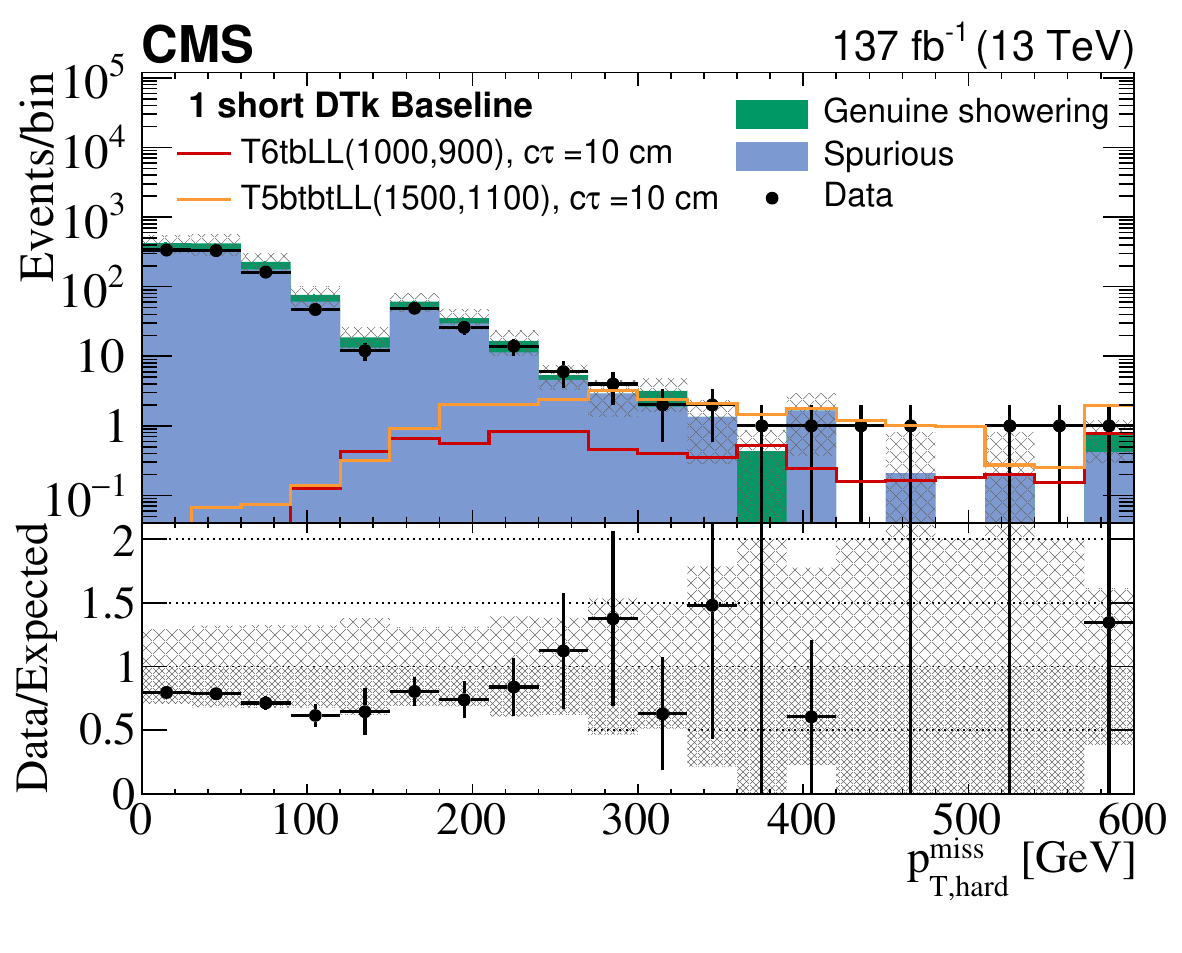}
\includegraphics[width=0.49\linewidth]{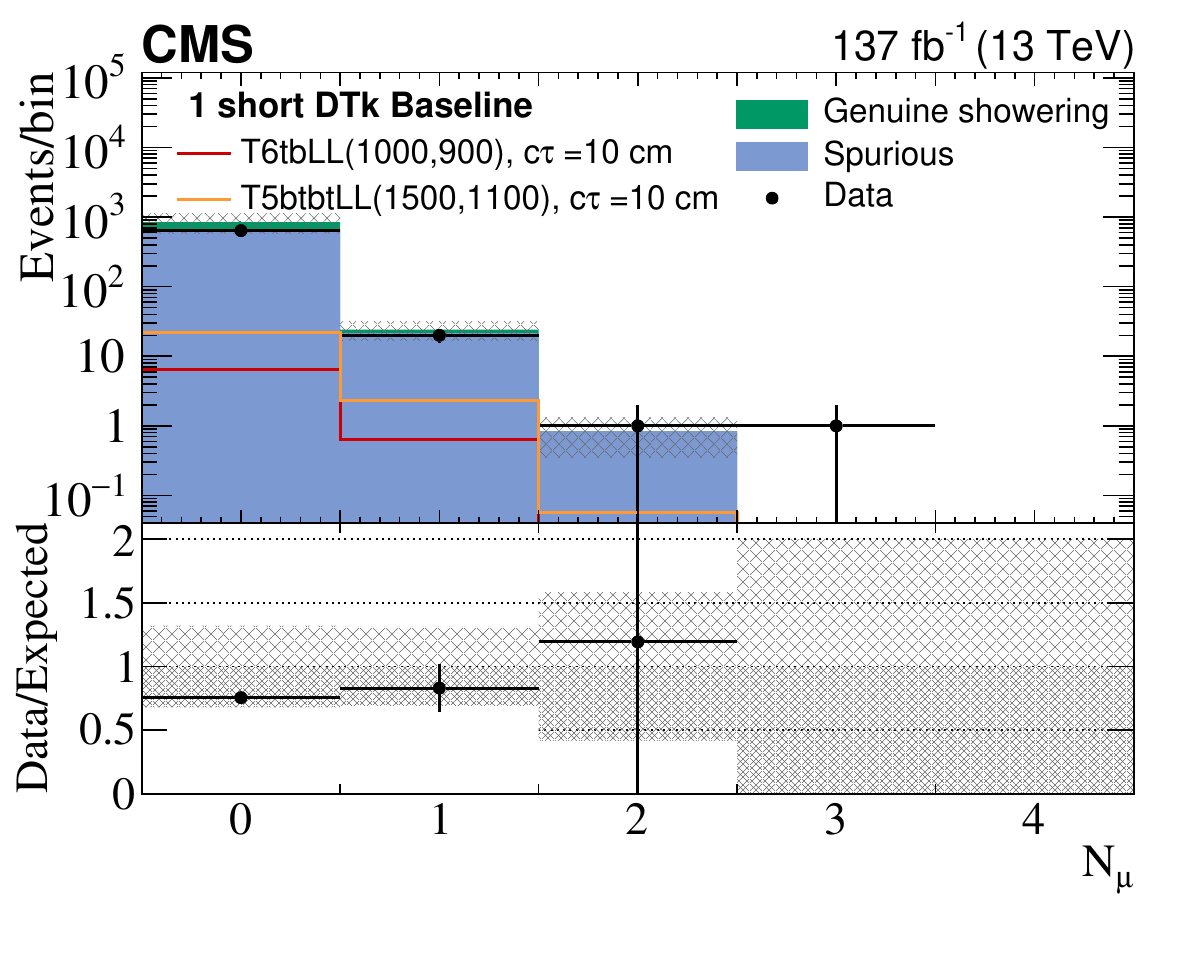} \\
\includegraphics[width=0.49\linewidth]{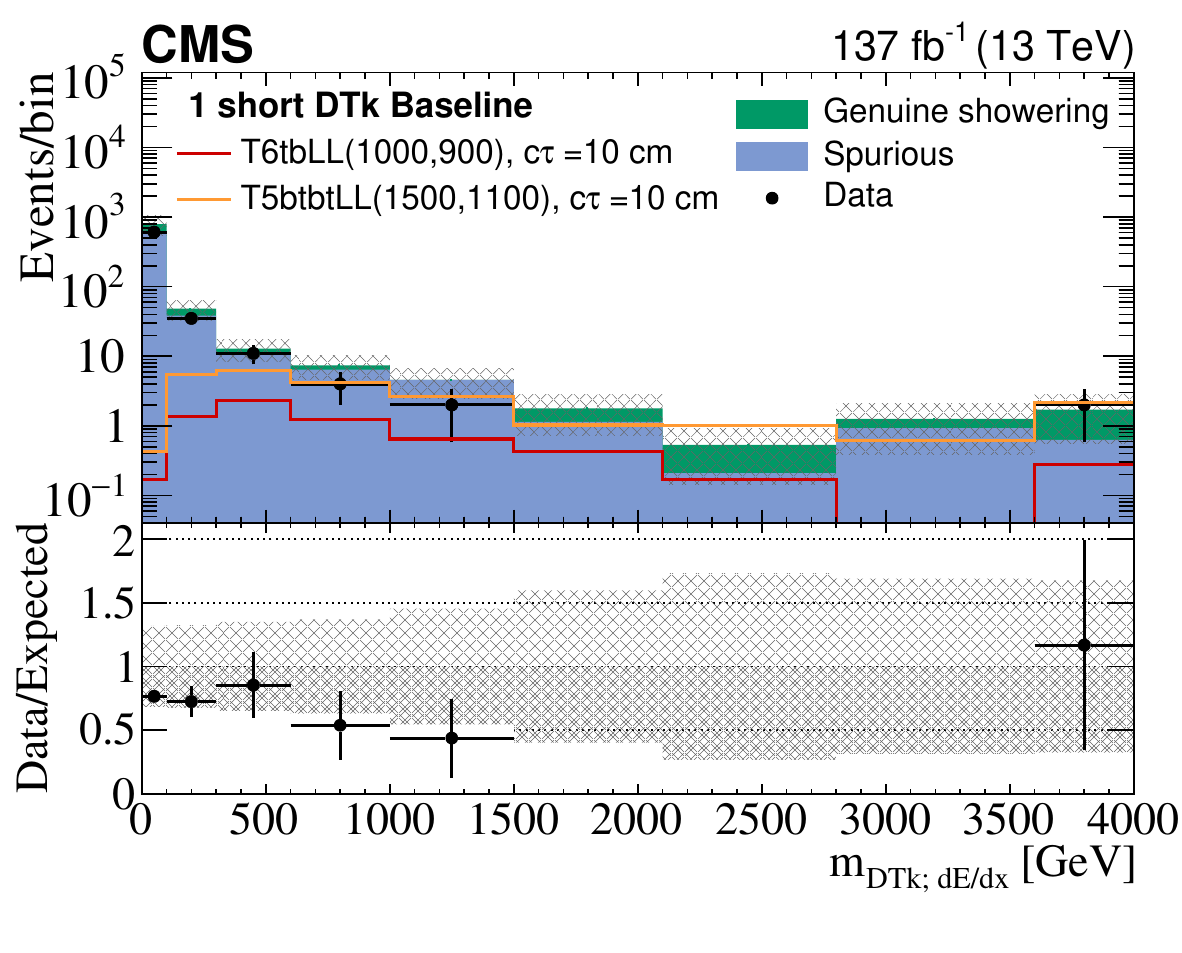} 
\includegraphics[width=0.49\linewidth]{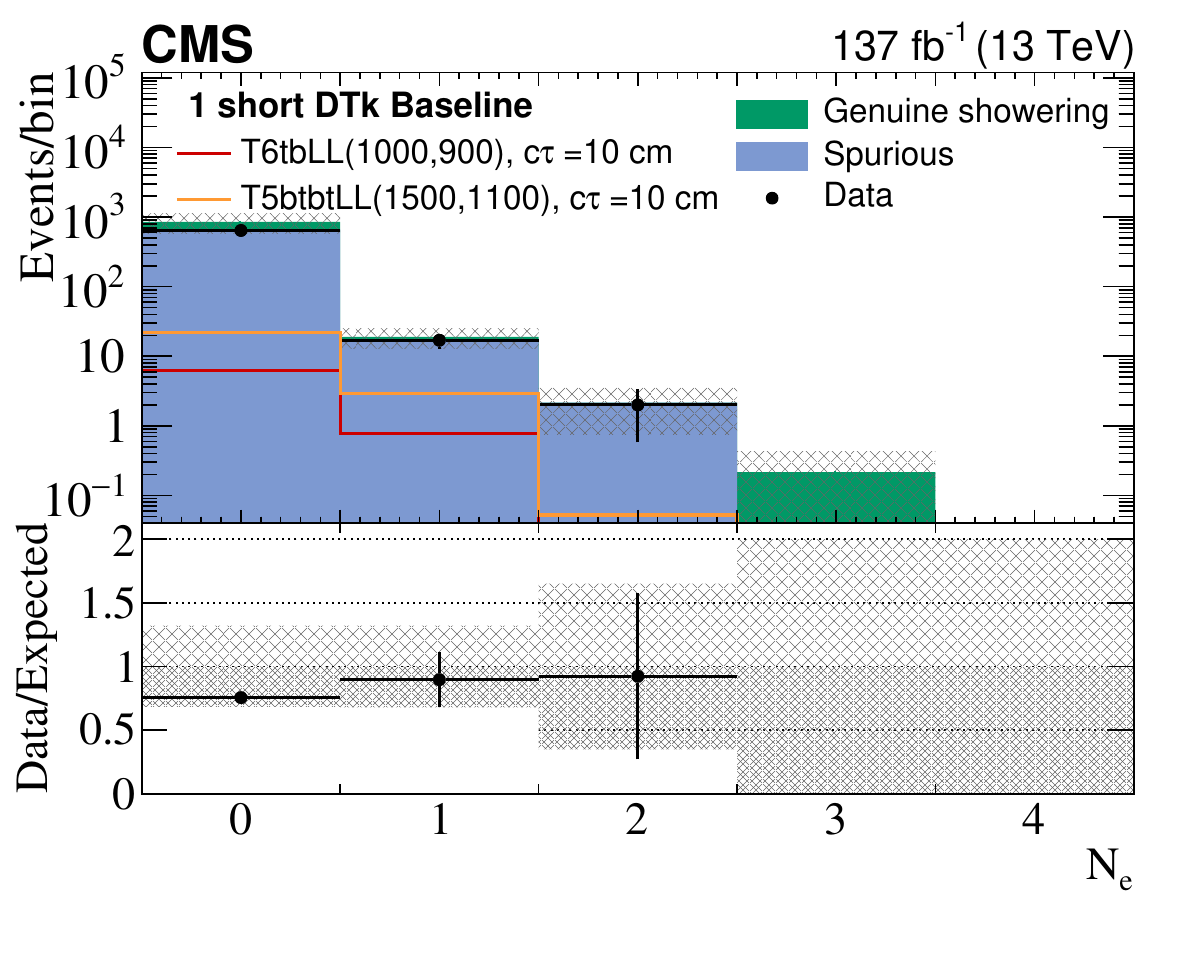}
\caption{
Comparison in the baseline region for the short-track DTk category
between the data and pre-fit predicted SM background
for the \njets (upper left), \nbjets (upper right), \ptmisshard (middle left),
\nelec (middle right), \nmuon (lower left), and \mdedx (lower right) distributions.
The uncertainty bars on the ratios in the lower panels indicate the fractional Poisson uncertainties
in the observed counts.
The gray bands show the fractional Poisson uncertainties in the control region counts,
added in quadrature with the systematic uncertainties.
The leftmost (rightmost) bin includes underflow (overflow).
For purposes of illustration,
results from the T6tbLL and T5btbtLL models are shown,
where the first and second numbers in parentheses indicate the squark (or gluino) mass
and the LSP mass, respectively, in \GeV.
}
\label{fig:baseline2}
\end{figure*}

\begin{figure*}[t]
\centering
\includegraphics[width=.49\linewidth]{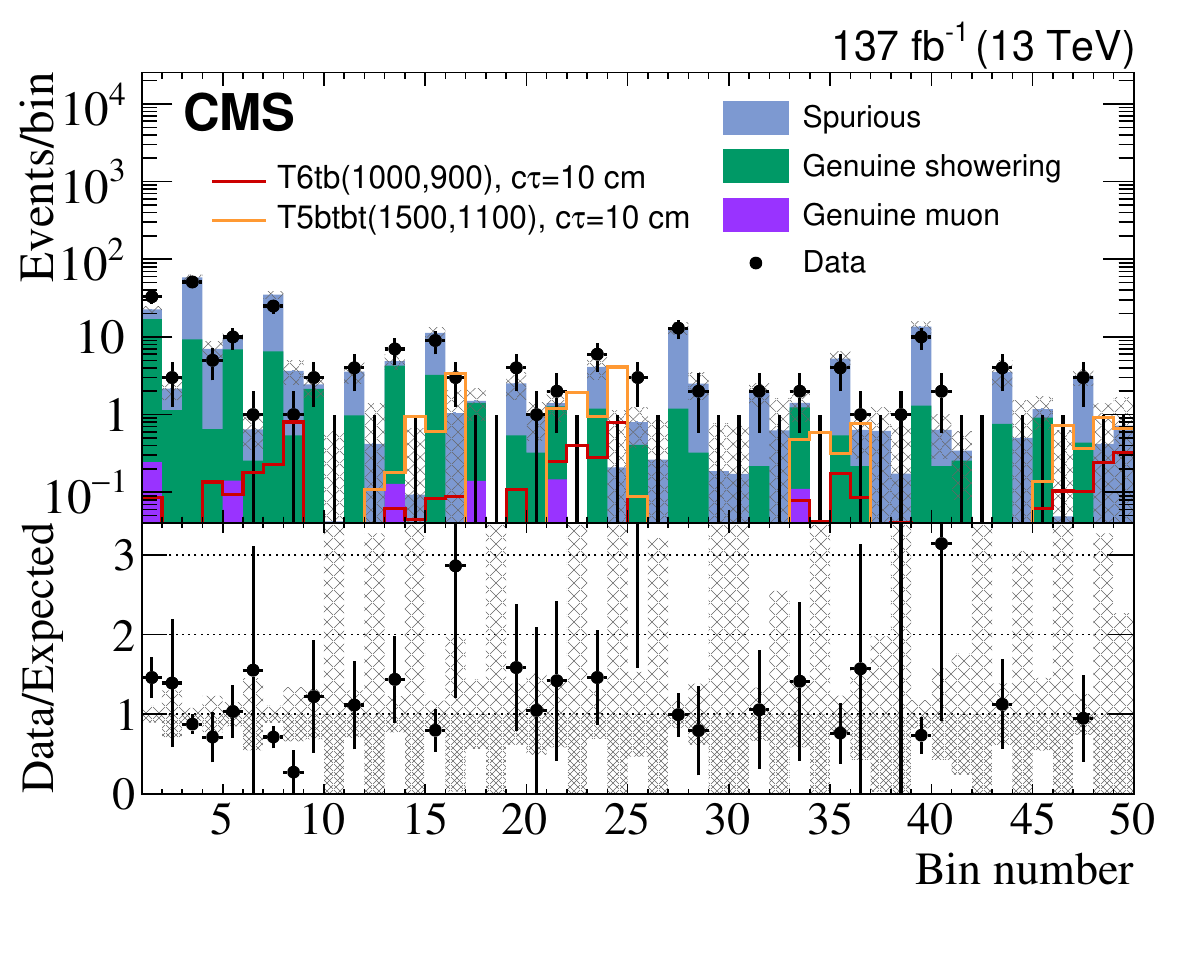}
\includegraphics[width=.49\linewidth]{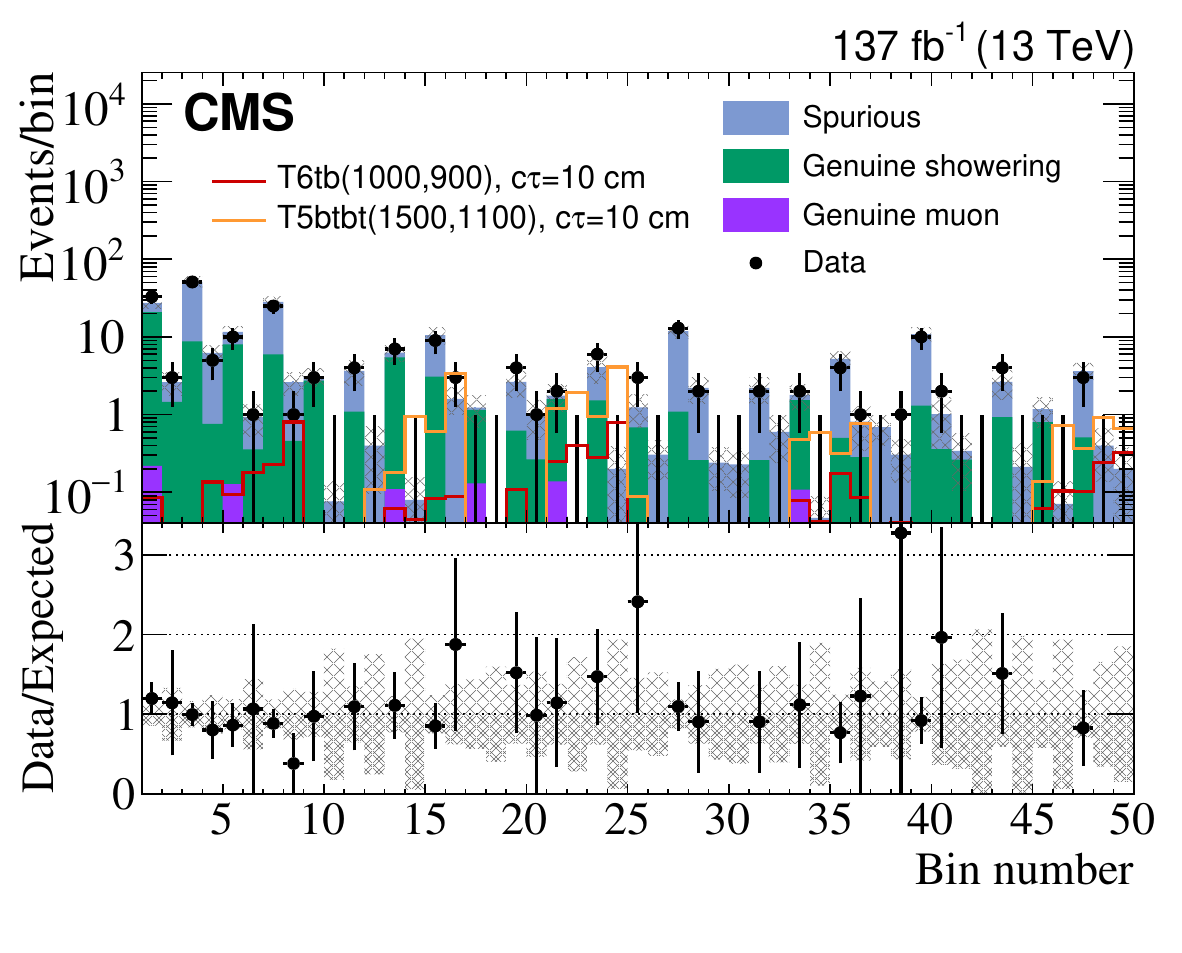}
\caption{
Comparison between the data and SM background predictions for the 49 search regions.
The left (right) plot shows the pre-fit (post-fit) background predictions.
The uncertainty bars on the ratios in the lower panels,
shown for bins with nonzero entries,
indicate the fractional Poisson uncertainties in the observed counts.
The gray bands show the fractional Poisson uncertainties in the control region counts,
added in quadrature with the systematic uncertainties.
For purposes of illustration,
results from the T6tbLL and T5btbtLL models are shown,
where the first and second numbers in parentheses indicate the squark (or gluino) mass
and the LSP mass, respectively, in \GeV.
}
\label{fig:srbaseline}
\end{figure*}

Figures~\ref{fig:baseline1} and~\ref{fig:baseline2}
show a comparison in the baseline region between the data and pre-fit background prediction
for the \njets, \nbjets, \ptmisshard, \nelec, \nmuon, and \mdedx distributions.
The results for long tracks are shown in Fig.~\ref{fig:baseline1} and
for short tracks in Fig.~\ref{fig:baseline2}.
Figure~\ref{fig:srbaseline} presents the pre-fit (left) and post-fit (right)
background predictions in comparison to the data in the 49 individual~SRs.
The corresponding numerical results for the pre-fit predictions
are given in Table \ref{tab:results-run2}.
For purposes of illustration,
these figures and table include example results for the T6tbLL and T5btbtLL models,
with choices for squark (or gluino) mass,
LSP mass,
and chargino $c\tau$ value as indicated in the legends or table header.
The data are found to be consistent with the background-only hypothesis,
and therefore no evidence for new physics is observed.
A slight systematic over-prediction is seen in the short-track category,
consistent with the presence of spurious-particle contamination
in the showering background sidebands.
The background uncertainty model accounts for this effect via a nuisance parameter
that reduces the total genuine-particle background yield for short track bins to nearly zero.

Upper limits on the production cross sections of the considered simplified models
are computed using a maximum likelihood fit.
The likelihood is a product of Poisson functions,
one for each SR,
accounting for the expected yields and evaluated using the observed event counts.
The mean expected background and signal yields are constrained using gamma functions
that account for the observed event counts in each sideband control region
and for the simulated signal event counts.
Other sources of uncertainty are accounted for using log-normal functions,
with one nuisance parameter per source of uncertainty. 
Limits are determined under the asymptotic approximation~\cite{Cowan:2010js}
of the \CLs criterion described in Refs.~\cite{Junk:1999kv,Read:2002hq}.
All 49 SRs are used in evaluating the limits for each signal model point.
For the models of gluino pair production and top or bottom squark pair production,
limits are derived in the mass plane of the mother particle and the LSP for different choices
of chargino~$c\tau$.
We evaluate 95\% confidence level (\CL) upper limits on the signal cross sections.
The approximate NNLO+NNLL cross section is used to determine corresponding mass exclusion curves.
Expected limits are computed using the background-only hypothesis
in place of the observed numbers of events.

For the strong production models,
the sensitivity of our results depends on the mass difference \dmstrong
between the gluino or squark and the LSP.
The limits weaken when \dmstrong is small because of the resulting
small boost (and thus short decay length) of the chargino in the detector frame.
The limits also weaken for small LSP masses
because of the significant boost of the chargino,
which increasingly fails to decay within the tracker volume as \dmstrong increases.
The strongest constraints on the cross section reside
at varying intermediate values of \dmstrong,
depending on the chargino~$c\tau$.

\begin{table*}[htp]
\topcaption{
Predicted pre-fit background and uncertainties in the 49 search regions (SRs).
Statistical and bin-wise systematic uncertainties are added in quadrature.
The control region (CR) counts
corresponding to each background category are given in the column
to the left of the respective column. 
The numbers in parentheses for the signal points indicate
the squark (or gluino) mass in \GeV,
the LSP mass in \GeV,
and $c\tau$ for the chargino in~cm, respectively.
}
\renewcommand{\arraystretch}{1.25}
\cmsTable{
\begin{scotch}{ccccccccccc}
 \multirow{2}{*}{\shortstack{SR\\number}} &  \multirow{2}{*}{CR} & \multirow{2}{*}{Spurious} &
 \multirow{2}{*}{CR} & \multirow{2}{*}{Showering} &
 \multirow{2}{*}{CR} & \multirow{2}{*}{Muon} & \multirow{2}{*}{Total bkg.} & \multirow{2}{*}{\shortstack{T6tbLL\\(1000, 900, 200)}}
   & \multirow{2}{*}{\shortstack{T5btbtLL\\(1500, 1100, 10)}} & \multirow{2}{*}{Observed}\\
 & & & & & & & & & & \\
\hline
1 & 80  & 6.40 $\pm$ 0.72  & 56 & 17.2 $\pm$ 2.3       & 313 & 0.23 $\pm$ 0.01 &  23.9 $\pm$ 2.4       & 0.10 $\pm$ 0.06 & 0.00 $\pm$ 0.00 & 33\\
2 & 14  & 1.12 $\pm$ 0.30  & 4  & 1.23 $\pm$ 0.61      & 17  & 0.01 $\pm$ 0.01 &  2.36 $\pm$ 0.68      & 0.80 $\pm$ 0.17 & 0.00 $\pm$ 0.00 & 3\\
3 & 170 & 58.9 $\pm$ 4.5   & 33 & 8.1 $\pm$ 1.4        & 0   & 0.00 $\pm$ 0.00 &  67.0 $\pm$ 4.7       & 0.04 $\pm$ 0.03 & 0.00 $\pm$ 0.00 & 51\\
4 & 23  & 8.0  $\pm$ 1.7   & 3  & 0.74 $\pm$ 0.43      & 0   & 0.00 $\pm$ 0.00 &  8.7 $\pm$ 1.7        & 0.17 $\pm$ 0.07 & 0.00 $\pm$ 0.00 & 5\\
5 & 44  & 3.52 $\pm$ 0.53  & 24 & 7.4 $\pm$ 1.5        & 183 & 0.13 $\pm$ 0.01 &  11.0 $\pm$ 1.6       & 0.77 $\pm$ 0.16 & 0.03 $\pm$ 0.03 & 10\\
6 & 7   & 0.56 $\pm$ 0.21  & 1  & 0.31 $\pm$ 0.31      & 9   & 0.01 $\pm$ 0.01 &  0.87 $\pm$ 0.37      & 4.86 $\pm$ 0.38 & 0.00 $\pm$ 0.00 & 1\\
7 & 92  & 31.9 $\pm$ 3.3   & 23 & 5.7 $\pm$ 1.2        & 0   & 0.00 $\pm$ 0.00 &  37.5 $\pm$ 3.5       & 0.19 $\pm$ 0.07 & 0.00 $\pm$ 0.00 & 25\\
8 & 10  & 3.5  $\pm$ 1.1   & 2  & 0.49 $\pm$ 0.35      & 0   & 0.00 $\pm$ 0.00 &  4.0 $\pm$ 1.2        & 1.02 $\pm$ 0.15 & 0.03 $\pm$ 0.02 & 1\\
9 & 4   & 0.32 $\pm$ 0.16  & 8  & 2.46 $\pm$ 0.87      & 37  & 0.03 $\pm$ 0.03 &  2.81 $\pm$ 0.88      & 0.00 $\pm$ 0.00 & 0.00 $\pm$ 0.00 & 3\\
10 & 1  & 0.08 $\pm$ 0.08  & 0  & 0.00$^{+0.57}_{-0.00}$ & 0   & 0.00 $\pm$ 0.00 &  0.08$^{+0.57}_{-0.08}$ & 0.06 $\pm$ 0.04 & 0.00 $\pm$ 0.00 & 0\\
11 & 10 & 3.5  $\pm$ 1.1   & 4  & 0.98 $\pm$ 0.49      & 0   & 0.00 $\pm$ 0.00 &  4.5 $\pm$ 1.2        & 0.00 $\pm$ 0.00 & 0.00 $\pm$ 0.00 & 4\\
12 & 2  & 0.69 $\pm$ 0.49  & 0  & 0.00$^{+0.90}_{-0.00}$ & 0   & 0.00 $\pm$ 0.00 &  0.7$^{+1.0}_{-0.7}$    & 0.00 $\pm$ 0.00 & 0.11 $\pm$ 0.11 & 0\\
13 & 10 & 0.80 $\pm$ 0.25  & 15 & 4.6 $\pm$ 1.2        & 157 & 0.12 $\pm$ 0.01 &  5.5 $\pm$ 1.2        & 0.14 $\pm$ 0.07 & 0.18 $\pm$ 0.10 & 7\\
14 & 1  & 0.08 $\pm$ 0.08  & 0  & 0.00$^{+0.57}_{-0.00}$ & 5   & 0.00 $\pm$ 0.00 &  0.08$^{+0.57}_{-0.08}$ & 1.19 $\pm$ 0.18 & 0.96 $\pm$ 0.27 & 0\\
15 & 31 & 10.7 $\pm$ 1.9   & 12 & 2.95 $\pm$ 0.85      & 0   & 0.00 $\pm$ 0.00 &  13.7 $\pm$ 2.1       & 0.03 $\pm$ 0.02 & 0.61 $\pm$ 0.17 & 9\\
16 & 5  & 1.73 $\pm$ 0.77  & 0  & 0.00$^{+0.90}_{-0.00}$ & 0   & 0.00 $\pm$ 0.00 &  1.7 $\pm$ 1.2        & 0.18 $\pm$ 0.06 & 3.41 $\pm$ 0.42 & 3\\
17 & 1  & 0.08 $\pm$ 0.08  & 4  & 1.23 $\pm$ 0.61      & 186 & 0.14 $\pm$ 0.01 &  1.45 $\pm$ 0.62      & 0.07 $\pm$ 0.05 & 0.00 $\pm$ 0.00 & 0\\
18 & 0  & 0.01 $\pm$ 0.01  & 0  & 0.00$^{+0.57}_{-0.00}$ & 12  & 0.01 $\pm$ 0.01 &  0.02$^{+0.57}_{-0.02}$ & 0.67 $\pm$ 0.18 & 0.00 $\pm$ 0.00 & 0\\
19 & 7  & 2.42 $\pm$ 0.92  & 2  & 0.49 $\pm$ 0.35      & 0   & 0.00 $\pm$ 0.00 &  2.92 $\pm$ 0.98      & 0.00 $\pm$ 0.00 & 0.00 $\pm$ 0.00 & 4\\
20 & 3  & 1.04 $\pm$ 0.60  & 1  & 0.25 $\pm$ 0.25      & 0   & 0.00 $\pm$ 0.00 &  1.28 $\pm$ 0.65      & 0.07 $\pm$ 0.04 & 0.00 $\pm$ 0.00 & 1\\
21 & 2  & 0.16 $\pm$ 0.11  & 4  & 1.23 $\pm$ 0.61      & 198 & 0.15 $\pm$ 0.01 &  1.54 $\pm$ 0.63      & 0.61 $\pm$ 0.14 & 1.19 $\pm$ 0.35 & 2\\
22 & 0  & 0.03 $\pm$ 0.03  & 0  & 0.00$^{+0.57}_{-0.00}$ & 10  & 0.01 $\pm$ 0.01 &  0.04$^{+0.57}_{-0.04}$ & 4.81 $\pm$ 0.40 & 1.91 $\pm$ 0.40 & 0\\
23 & 9  & 3.1  $\pm$ 1.0   & 5  & 1.23 $\pm$ 0.55      & 0   & 0.00 $\pm$ 0.00 &  4.4 $\pm$ 1.2        & 0.10 $\pm$ 0.05 & 0.94 $\pm$ 0.16 & 6\\
24 & 1  & 0.35 $\pm$ 0.35  & 0  & 0.00$^{+0.90}_{-0.00}$ & 0   & 0.00 $\pm$ 0.00 &  0.35$^{+0.97}_{-0.35}$ & 1.22 $\pm$ 0.19 & 4.10 $\pm$ 0.51 & 0\\
25 & 6  & 0.48 $\pm$ 0.20  & 1  & 0.31 $\pm$ 0.31      & 23  & 0.02 $\pm$ 0.02 &  0.80 $\pm$ 0.36      & 0.02 $\pm$ 0.02 & 0.09 $\pm$ 0.09 & 3\\
26 & 4  & 0.32 $\pm$ 0.16  & 0  & 0.00$^{+0.57}_{-0.00}$ & 1   & 0.00 $\pm$ 0.00 &  0.32$^{+0.59}_{-0.32}$ & 0.08 $\pm$ 0.05 & 0.00 $\pm$ 0.00 & 0\\
27 & 42 & 14.6 $\pm$ 2.2   & 4  & 0.98 $\pm$ 0.49      & 0   & 0.00 $\pm$ 0.00 &  15.5 $\pm$ 2.3       & 0.00 $\pm$ 0.00 & 0.00 $\pm$ 0.00 & 13\\
28 & 8  & 2.77 $\pm$ 0.98  & 1  & 0.25 $\pm$ 0.25      & 0   & 0.00 $\pm$ 0.00 &  3.0 $\pm$ 1.0        & 0.02 $\pm$ 0.02 & 0.03 $\pm$ 0.03 & 2\\
29 & 3  & 0.24 $\pm$ 0.14  & 0  & 0.00$^{+0.57}_{-0.00}$ & 19  & 0.01 $\pm$ 0.01 &  0.25$^{+0.58}_{-0.25}$ & 0.00 $\pm$ 0.00 & 0.00 $\pm$ 0.00 & 0\\
30 & 3  & 0.24 $\pm$ 0.14  & 0  & 0.00$^{+0.57}_{-0.00}$ & 0   & 0.00 $\pm$ 0.00 &  0.24$^{+0.58}_{-0.24}$ & 0.00 $\pm$ 0.00 & 0.00 $\pm$ 0.00 & 0\\
31 & 8  & 2.77 $\pm$ 0.98  & 1  & 0.25 $\pm$ 0.25      & 0   & 0.00 $\pm$ 0.00 &  3.0 $\pm$ 1.0        & 0.00 $\pm$ 0.00 & 0.00 $\pm$ 0.00 & 2\\
32 & 3  & 1.04 $\pm$ 0.60  & 0  & 0.00$^{+0.90}_{-0.00}$ & 0   & 0.00 $\pm$ 0.00 &  1.0$^{+1.1}_{-1.0}$    & 0.00 $\pm$ 0.00 & 0.03 $\pm$ 0.02 & 0\\
33 & 3  & 0.24 $\pm$ 0.14  & 4  & 1.23 $\pm$ 0.61      & 155 & 0.11 $\pm$ 0.01 &  1.58 $\pm$ 0.63      & 0.06 $\pm$ 0.04 & 0.48 $\pm$ 0.25 & 2\\
34 & 0  & 0.04 $\pm$ 0.04  & 0  & 0.00$^{+0.57}_{-0.00}$ & 6   & 0.00 $\pm$ 0.00 &  0.05$^{+0.57}_{-0.05}$ & 0.71 $\pm$ 0.16 & 0.59 $\pm$ 0.27 & 0\\
35 & 20 & 6.9 $\pm$ 1.6    & 2  & 0.49 $\pm$ 0.35      & 0   & 0.00 $\pm$ 0.00 &  7.4 $\pm$ 1.6        & 0.00 $\pm$ 0.00 & 0.31 $\pm$ 0.18 & 4\\
36 & 2  & 0.69 $\pm$ 0.49  & 1  & 0.25 $\pm$ 0.25      & 0   & 0.00 $\pm$ 0.00 &  0.94 $\pm$ 0.55      & 0.12 $\pm$ 0.06 & 0.76 $\pm$ 0.22 & 1\\
37 & 9  & 0.72 $\pm$ 0.24  & 0  & 0.00$^{+0.57}_{-0.00}$ & 7   & 0.01 $\pm$ 0.01 &  0.73 $\pm$ 0.61      & 0.03 $\pm$ 0.03 & 0.00 $\pm$ 0.00 & 0\\
38 & 3  & 0.24 $\pm$ 0.14  & 0  & 0.00$^{+0.57}_{-0.00}$ & 1   & 0.00 $\pm$ 0.00 &  0.24$^{+0.58}_{-0.24}$ & 0.04 $\pm$ 0.04 & 0.01 $\pm$ 0.01 & 1\\
39 & 39 & 13.5 $\pm$ 2.2   & 5  & 1.23 $\pm$ 0.55      & 0   & 0.00 $\pm$ 0.00 &  14.7 $\pm$ 2.2       & 0.00 $\pm$ 0.00 & 0.00 $\pm$ 0.00 & 10\\
40 & 2  & 0.69 $\pm$ 0.49  & 1  & 0.25 $\pm$ 0.25      & 0   & 0.00 $\pm$ 0.00 &  0.94 $\pm$ 0.55      & 0.00 $\pm$ 0.00 & 0.00 $\pm$ 0.00 & 2\\
41 & 1  & 0.08 $\pm$ 0.08  & 1  & 0.31 $\pm$ 0.31      & 12  & 0.01 $\pm$ 0.01 &  0.40 $\pm$ 0.32      & 0.00 $\pm$ 0.00 & 0.00 $\pm$ 0.00 & 0\\
42 & 0  & 0.01 $\pm$ 0.01  & 0  & 0.00$^{+0.57}_{-0.00}$ & 0   & 0.00 $\pm$ 0.00 &  0.01$^{+0.57}_{-0.01}$ & 0.00 $\pm$ 0.00 & 0.00 $\pm$ 0.00 & 0\\
43 & 6  & 2.08 $\pm$ 0.85  & 3  & 0.74 $\pm$ 0.43      & 0   & 0.00 $\pm$ 0.00 &  2.82 $\pm$ 0.95      & 0.00 $\pm$ 0.00 & 0.00 $\pm$ 0.00 & 4\\
44 & 0  & 0.37 $\pm$ 0.37  & 0  & 0.00$^{+0.90}_{-0.00}$ & 0   & 0.00 $\pm$ 0.00 &  0.37$^{+0.98}_{-0.37}$ & 0.00 $\pm$ 0.00 & 0.03 $\pm$ 0.02 & 0\\
45 & 5  & 0.40 $\pm$ 0.18  & 3  & 0.92 $\pm$ 0.53      & 47  & 0.03 $\pm$ 0.01 &  1.36 $\pm$ 0.56      & 0.27 $\pm$ 0.10 & 0.14 $\pm$ 0.05 & 0\\
46 & 0  & 0.07 $\pm$ 0.07  & 0  & 0.00$^{+0.57}_{-0.00}$ & 3   & 0.00 $\pm$ 0.00 &  0.07$^{+0.57}_{-0.07}$ & 1.12 $\pm$ 0.22 & 0.72 $\pm$ 0.22 & 0\\
47 & 13 & 4.5 $\pm$ 1.3    & 2  & 0.49 $\pm$ 0.35      & 0   & 0.00 $\pm$ 0.00 &  5.0 $\pm$ 1.3        & 0.00 $\pm$ 0.00 & 0.37 $\pm$ 0.14 & 3\\
48 & 2 & 0.69 $\pm$ 0.49   & 0  & 0.00$^{+0.90}_{-0.00}$ & 0   & 0.00 $\pm$ 0.00 &  0.7$^{+1.0}_{-0.7}$    & 0.27 $\pm$ 0.09 & 0.91 $\pm$ 0.19 & 0\\
49 & 1 & 0.35 $\pm$ 0.35   & 0  & 0.00$^{+0.57}_{-0.00}$ & 1   & 0.00 $\pm$ 0.00 &  0.35$^{+0.66}_{-0.35}$ & 4.57 $\pm$ 0.39 & 0.66 $\pm$ 0.24 & 0\\
\end{scotch}
}
\label{tab:results-run2}
\end{table*}

\begin{figure*}[]
\centering
\includegraphics[width=0.49\linewidth]{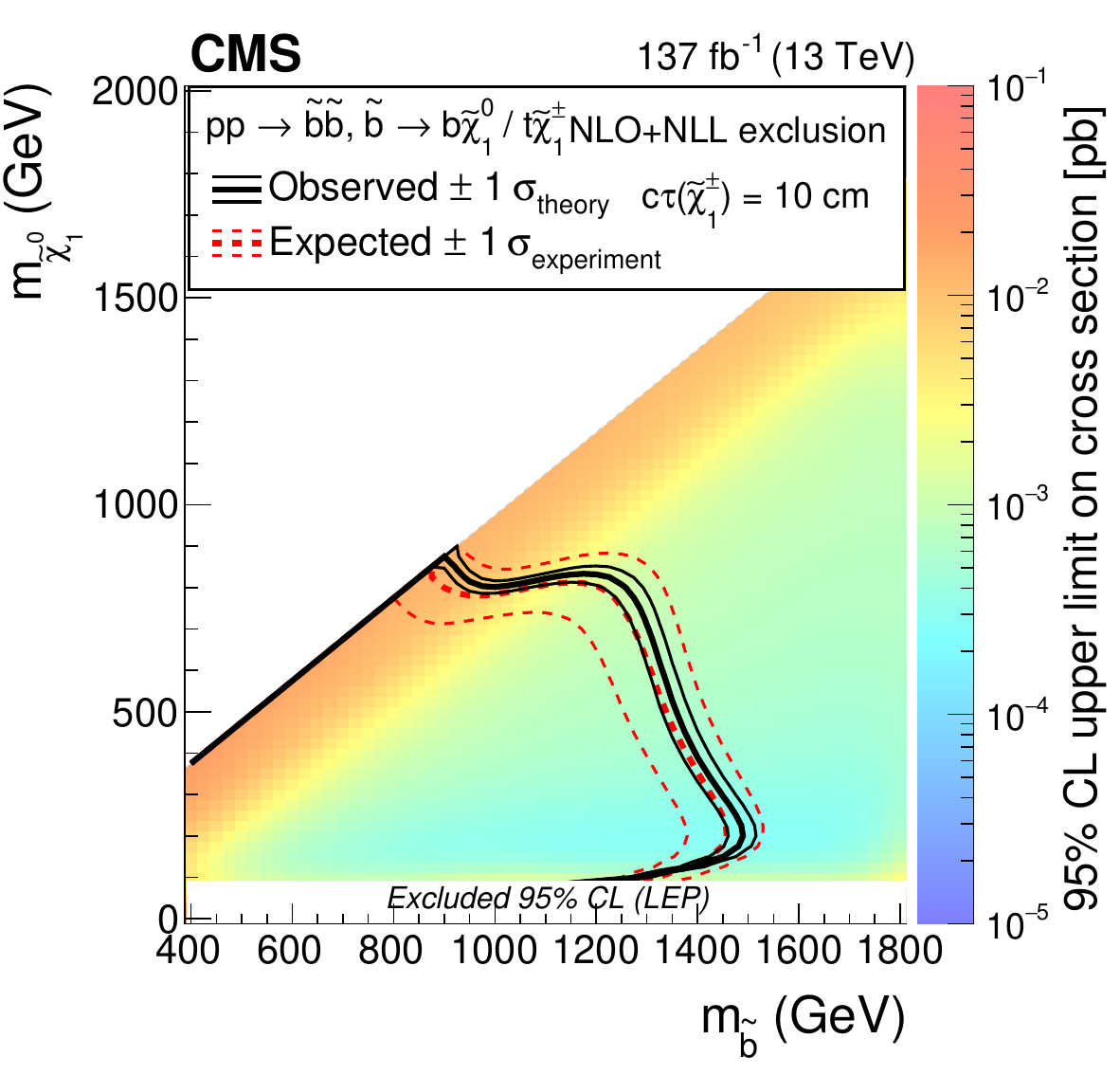}
\includegraphics[width=0.49\linewidth]{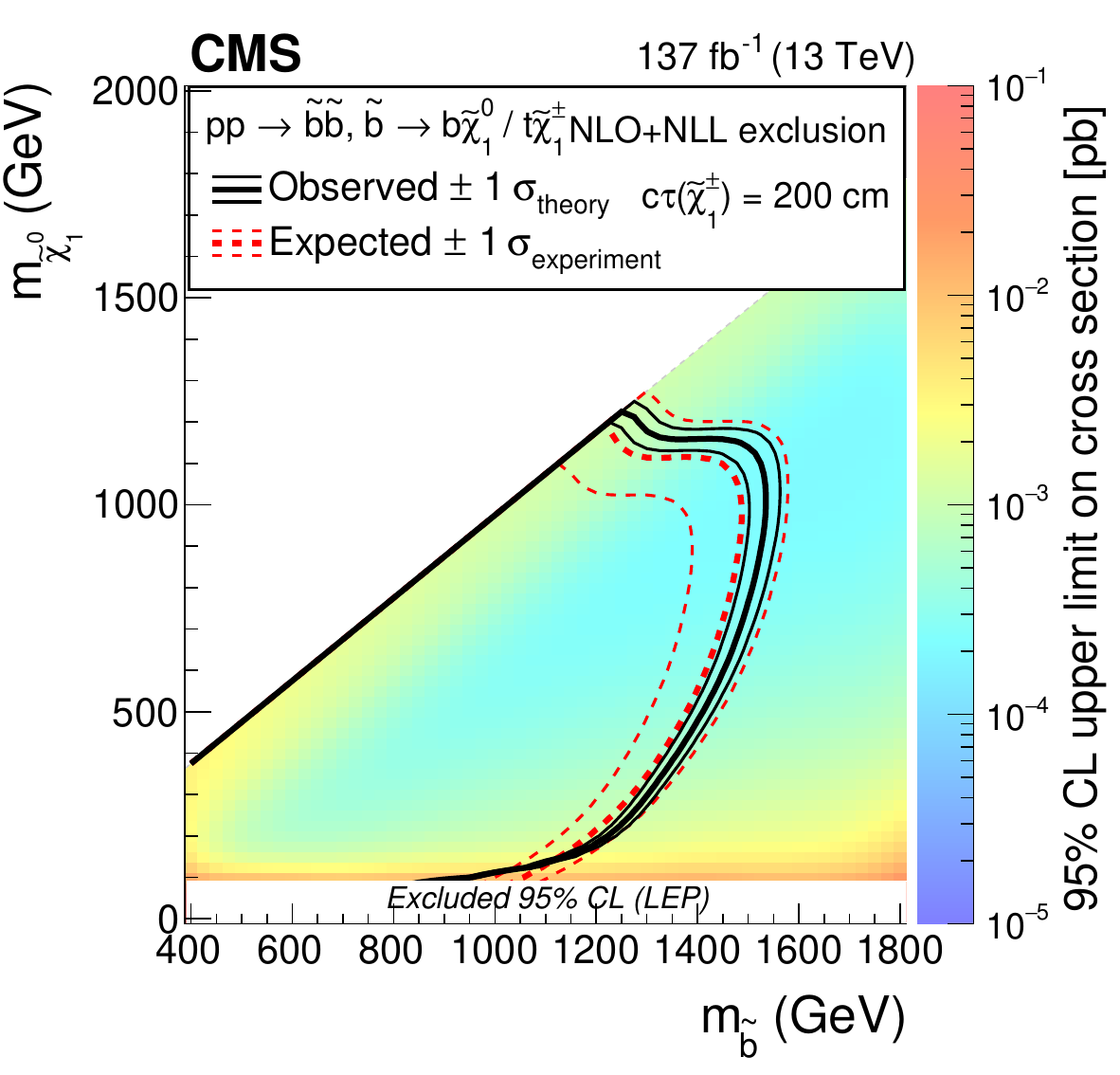} \\
\includegraphics[width=0.49\linewidth]{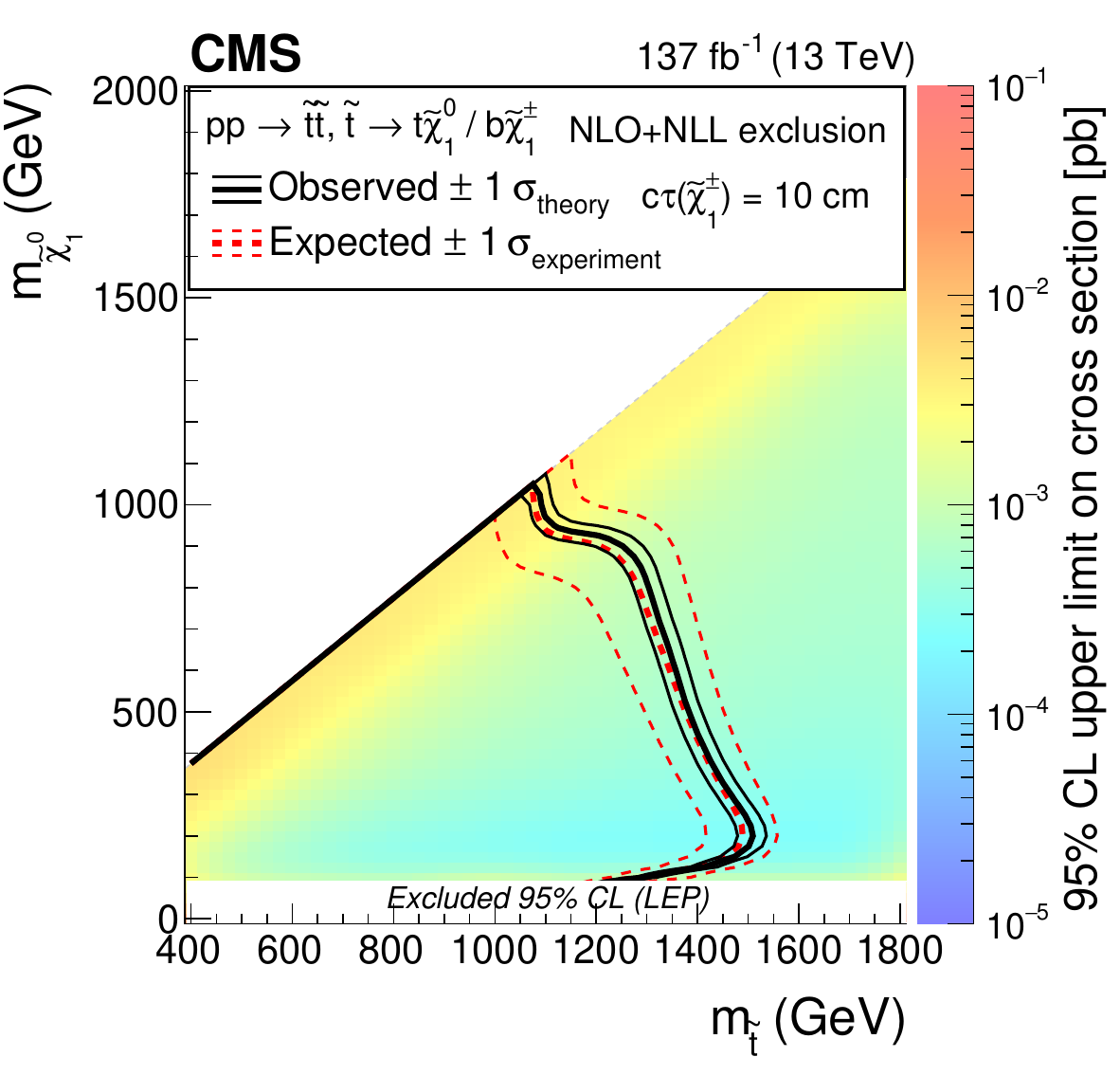}
\includegraphics[width=0.49\linewidth]{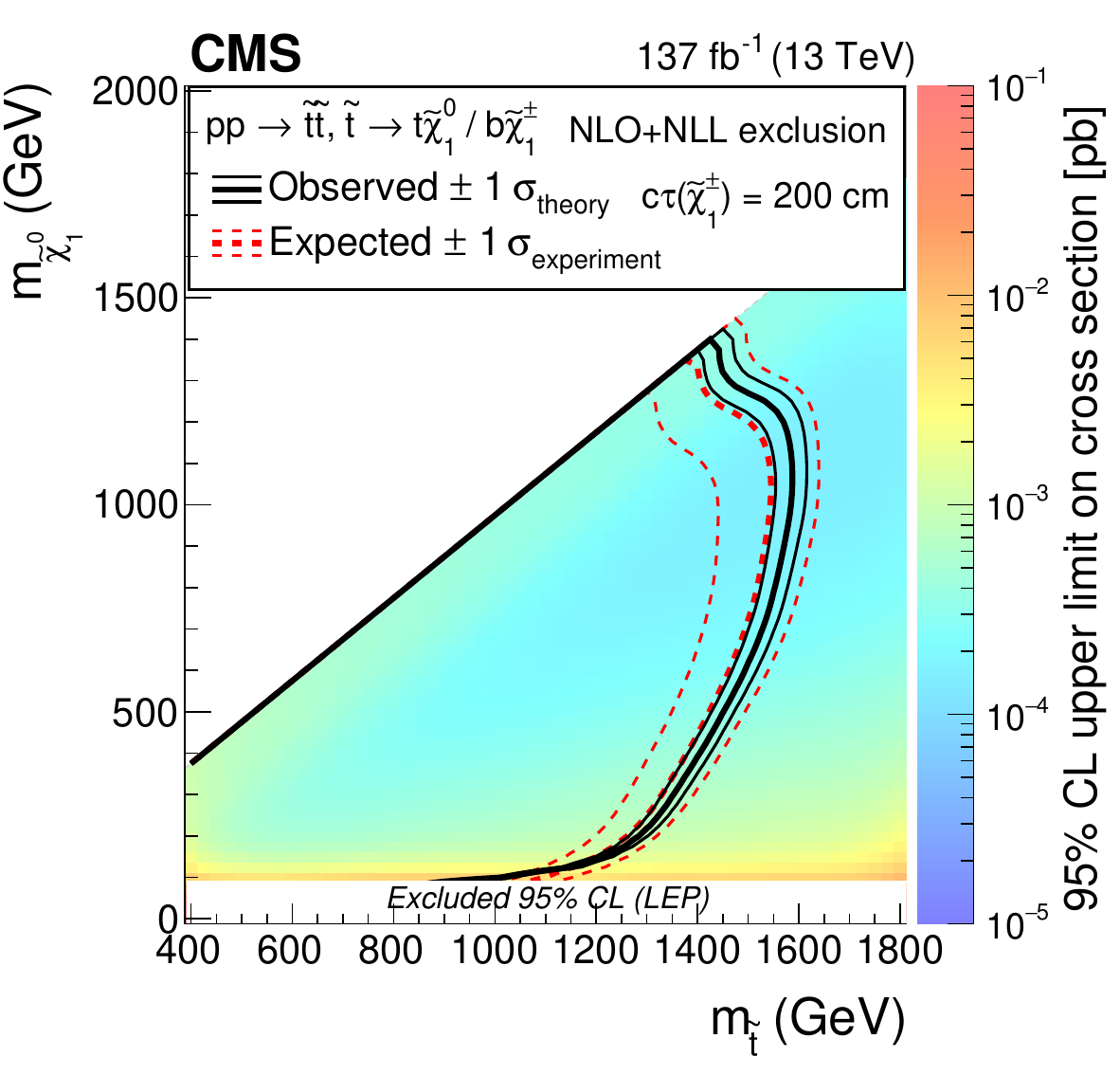} 
\caption{
Observed 95\% \CL upper limits on the signal cross sections (colored area)
versus the bottom or top squark
and neutralino mass for the T6tbLL (upper) and T6btLL (lower) model for a
chargino proper decay length $c\tau$ of 10 (left column) or 200 (right column)\cm.
Also shown are black (red) contours corresponding to the observed (expected) lower limits,
including their uncertainties, on the squark and neutralino masses.
}
\label{fig:T6limits}
\end{figure*}

Upper limits on the cross section for the T6tbLL and T6btLL models
are presented in Fig.~\ref{fig:T6limits}.
For the T6tbLL model,
we exclude bottom squarks below a mass of 900--1540\GeV,
depending on the LSP mass and chargino lifetime.
Charginos and LSPs,
taken to be essentially mass degenerate in our study,
are excluded up to a mass of 850 (1210)\GeV for chargino $c\tau=10$ (200)\cm,
depending on the bottom squark mass.
The analogous limits for the T6btLL model are
1100--1590\GeV for the top squark mass,
and 1050 (1400)\GeV for the chargino and LSP mass.
These results extend the maximum limit on the LSP mass
in the compressed phase space scenario by hundreds of \GeV
compared to the previous study~\cite{CMS:2019ybf},
both for $c\tau=10$ and 200\cm.

\begin{figure*}[]
\centering
\includegraphics[width=0.49\linewidth]{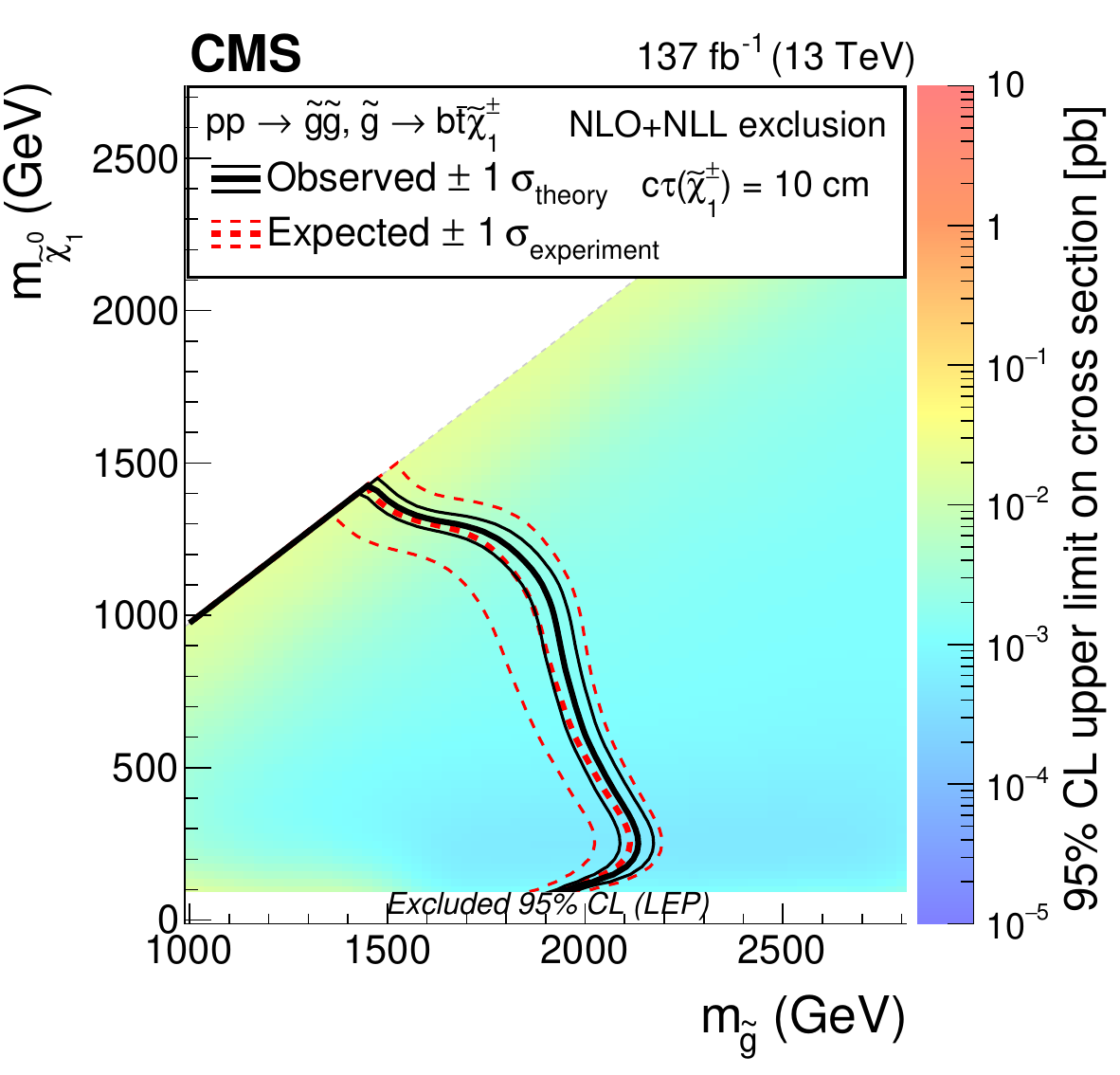}
\includegraphics[width=0.49\linewidth]{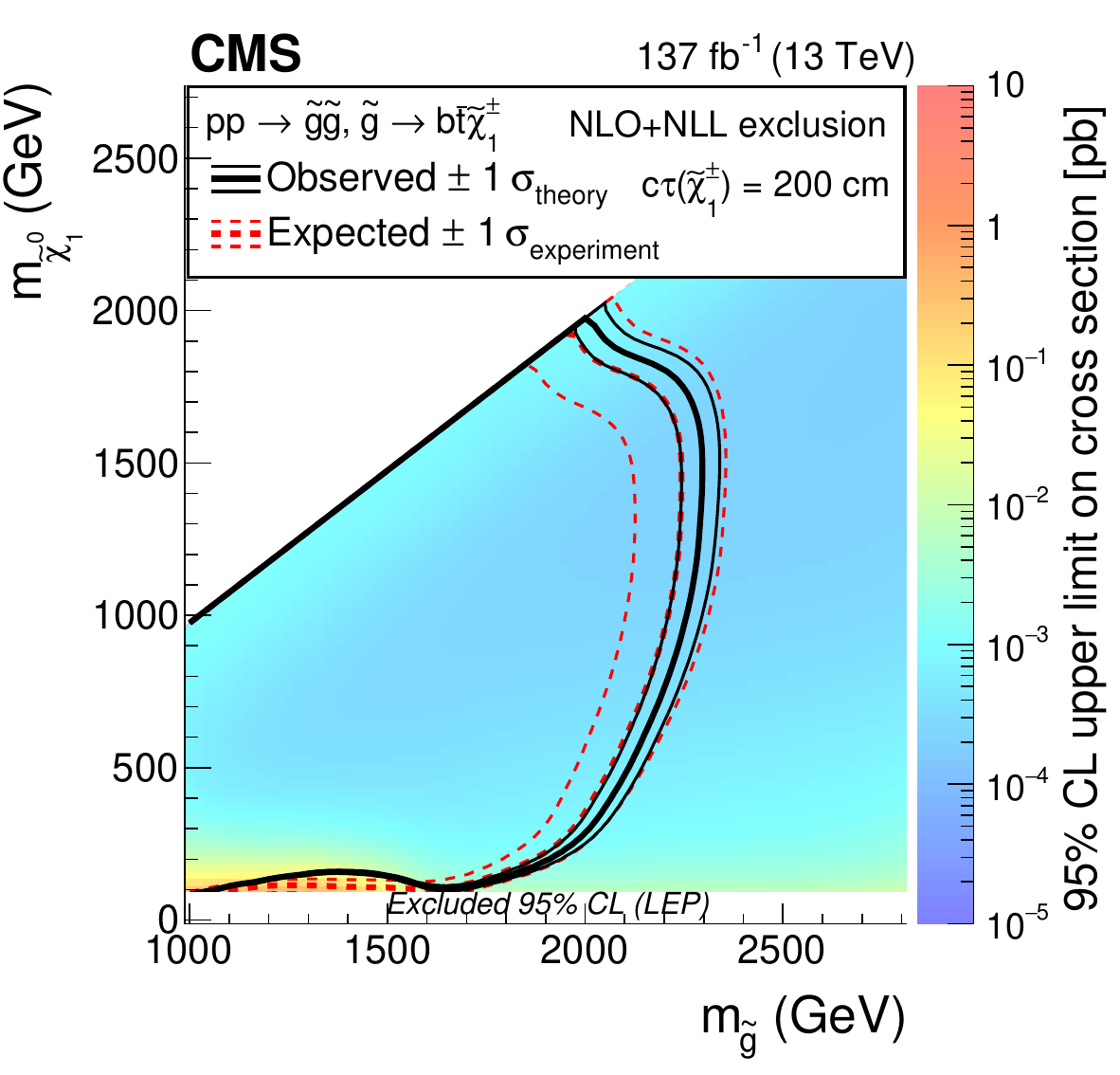}\\
\caption{
Observed 95\% \CL upper limits on the signal cross sections (colored area)
versus the gluino and neutralino mass for the T5btbtLL model for a
chargino proper decay length $c\tau$ of 10 (left column) and 200 (right column)\cm. 
Also shown are black (red) contours corresponding to the observed (expected) lower limits,
including their uncertainties, on the gluino and neutralino masses.
}
\label{fig:T1limits}
\end{figure*}

Upper limits on the cross section for the T5btbtLL model are presented in Fig.~\ref{fig:T1limits}.
Gluinos are excluded below a mass of 1450--2300\GeV,
depending on LSP mass and chargino lifetime.
Charginos and LSPs are excluded up to a maximum of 1950\GeV,
depending on the gluino mass and chargino lifetime.

In Figs.~\ref{fig:T6limits} and~~\ref{fig:T1limits},
the white bands indicate the regions previously
excluded by the CERN LEP (Large Electron-Positron Collider) experiments~\cite{LEP2WG:2002}.

{\tolerance=2000
We examined the change in sensitivity introduced by our inclusion
of the electron+DTk and muon+DTk channels in the analysis.
We find that,
compared to using the fully hadronic channel alone,
addition of these channels leads to an improvement
of around 40\GeV in the top squark mass limit
and of around 100\GeV in both the bottom squark and gluino mass limits.
\par}

\begin{figure*}[]
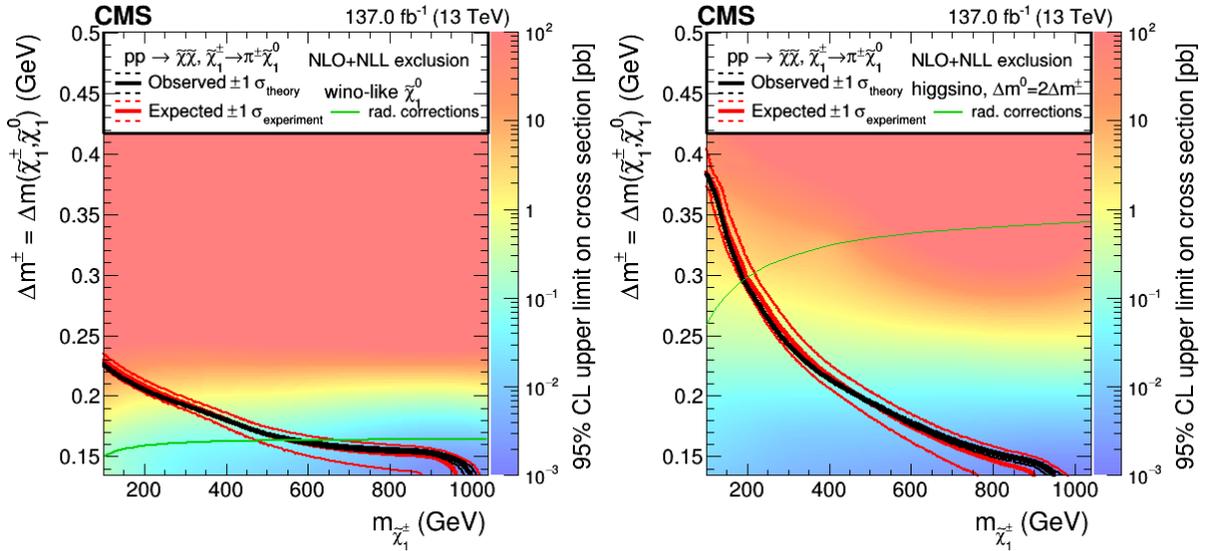

\centering
\includegraphics[width=0.49\linewidth]{Figure_012-a.pdf}
\includegraphics[width=0.49\linewidth]{Figure_012-b.pdf}
\caption{
Observed 95\% \CL upper limits on the signal cross sections (colored area)
versus the chargino-LSP mass difference and the mass of the chargino for the wino (left) and
higgsino (right) DM models. The black contours indicate the boundary where the observed 
upper limit equals the cross section of fully degenerate electroweakino production.
The corresponding expected limits are shown by the red contours.
The green lines represent the set of model points corresponding to the pure wino
and pure higgsino models where only radiative corrections to the mass 
splitting are assumed.
Chargino lifetimes are based on two-loop calculations. 
}
\label{fig:higgsino-limit}
\end{figure*}

Finally, limits on pure wino~\cite{Ibe:2012sx} and higgsino~\cite{Fukuda:2017jmk} DM models 
are presented in Fig.~\ref{fig:higgsino-limit}
for the case of a minimal mass splitting \dmplus  between the two lightest SUSY states,
bounded by two-loop radiative corrections.
The chargino lifetime is determined as a function of \dmplus,
as described in Ref.~\cite{Ibe:2022lkl} for the wino case
and in Ref.~\cite{Nagata:2014wma} for the higgsino case.
The production scenario accounts for the TChiWZ, TChiWW, and TChiW models taken together, 
where the first process occurs only in the higgsino case.
The green line indicates the minimum mass splitting required by the radiative corrections.
Chargino and LSP masses are excluded up to 650\GeV for the pure wino model and up to 
190\GeV for the pure higgsino model.

\section{Summary}
\label{sec:summary}

A search for long-lived charginos based on data collected
in proton-proton collisions at $\sqrt{s}=13\TeV$,
corresponding to an integrated luminosity of 137\fbinv,
is presented.
Event yields are studied in 49 nonoverlapping search regions defined by
the number of electrons, muons, jets, and \PQb-tagged jets,
and by the hard missing transverse momentum,
in final states with at least one identified disappearing track.
Further categorization of the SRs is based on the approximate length of the track
and on its \dedx energy loss in the inner tracking detector.
The analysis targets a wide variety of possible production modes appearing in simplified models
of $R$-parity conserving supersymmetry,
including gluino, top squark, bottom squark, and electroweakino pair production.
A machine-learning-based classifier is employed to optimally select disappearing tracks,
while rejecting tracks originating from failures in the reconstruction
or from combinatorial effects.
Background contributions to the SRs are evaluated
based on the observed yields in data control regions.
The observed yields in the SRs are found to be consistent
with the background-only predictions,
and thus no evidence for supersymmetry is found.

In the context of the examined models,
bottom squarks, top squarks, and gluinos with masses as large as 1540, 1590, and 2300\GeV,
respectively,
are excluded.
For bottom squark pair production,
charginos and the lightest supersymmetric particle,
considered to be essentially mass degenerate in our study,
are excluded up to a mass of 850 (1210)\GeV
for a chargino proper decay length $c\tau$ of 10 (200)\cm.
For top squark pair production,
the corresponding limit on the chargino and LSP mass is 1050 (1400)\GeV.
These results extend the maximum limit on the LSP mass
in the compressed phase space scenario by hundreds of \GeV
compared to the previous study~\cite{CMS:2019ybf},
and extend the reach of sensitivity into mass regions where
a pure wino- or pure higgsino-like LSP can account for the observed dark matter relic density.
Limits are also determined for a pure wino dark matter model~\cite{Ibe:2012sx} and a pure 
higgsino dark matter model~\cite{Fukuda:2017jmk}.
In the context of these two models, charginos and LSPs are excluded up to 650\GeV for the wino
model and up to 190\GeV for the higgsino model.

\begin{acknowledgments}
  We congratulate our colleagues in the CERN accelerator departments for the excellent performance of the LHC and thank the technical and administrative staffs at CERN and at other CMS institutes for their contributions to the success of the CMS effort. In addition, we gratefully acknowledge the computing centers and personnel of the Worldwide LHC Computing Grid and other centers for delivering so effectively the computing infrastructure essential to our analyses. Finally, we acknowledge the enduring support for the construction and operation of the LHC, the CMS detector, and the supporting computing infrastructure provided by the following funding agencies: SC (Armenia); BMBWF and FWF (Austria); FNRS and FWO (Belgium); CNPq, CAPES, FAPERJ, FAPERGS, and FAPESP (Brazil); MES and BNSF (Bulgaria); CERN; CAS, MoST, and NSFC (China); MINCIENCIAS (Colombia); MSES and CSF (Croatia); RIF (Cyprus); SENESCYT (Ecuador); MoER, ERC PUT and ERDF (Estonia); Academy of Finland, MEC, and HIP (Finland); CEA and CNRS/IN2P3 (France); SRNSF (Georgia); BMBF, DFG, and HGF (Germany); GSRI (Greece); NKFIH (Hungary); DAE and DST (India); IPM (Iran); SFI (Ireland); INFN (Italy); MSIP and NRF (Republic of Korea); MES (Latvia); LAS (Lithuania); MOE and UM (Malaysia); BUAP, CINVESTAV, CONACYT, LNS, SEP, and UASLP-FAI (Mexico); MOS (Montenegro); MBIE (New Zealand); PAEC (Pakistan); MES and NSC (Poland); FCT (Portugal); MESTD (Serbia); MCIN/AEI and PCTI (Spain); MOSTR (Sri Lanka); Swiss Funding Agencies (Switzerland); MST (Taipei); MHESI and NSTDA (Thailand); TUBITAK and TENMAK (Turkey); NASU (Ukraine); STFC (United Kingdom); DOE and NSF (USA).
  
  \hyphenation{Rachada-pisek} Individuals have received support from the Marie-Curie program and the European Research Council and Horizon 2020 Grant, contract Nos.\ 675440, 724704, 752730, 758316, 765710, 824093, and COST Action CA16108 (European Union); the Leventis Foundation; the Alfred P.\ Sloan Foundation; the Alexander von Humboldt Foundation; the Science Committee, project no. 22rl-037 (Armenia); the Belgian Federal Science Policy Office; the Fonds pour la Formation \`a la Recherche dans l'Industrie et dans l'Agriculture (FRIA-Belgium); the Agentschap voor Innovatie door Wetenschap en Technologie (IWT-Belgium); the F.R.S.-FNRS and FWO (Belgium) under the ``Excellence of Science -- EOS" -- be.h project n.\ 30820817; the Beijing Municipal Science \& Technology Commission, No. Z191100007219010 and Fundamental Research Funds for the Central Universities (China); the Ministry of Education, Youth and Sports (MEYS) of the Czech Republic; the Shota Rustaveli National Science Foundation, grant FR-22-985 (Georgia); the Deutsche Forschungsgemeinschaft (DFG), under Germany's Excellence Strategy -- EXC 2121 ``Quantum Universe" -- 390833306, and under project number 400140256 - GRK2497; the Hellenic Foundation for Research and Innovation (HFRI), Project Number 2288 (Greece); the Hungarian Academy of Sciences, the New National Excellence Program - \'UNKP, the NKFIH research grants K 124845, K 124850, K 128713, K 128786, K 129058, K 131991, K 133046, K 138136, K 143460, K 143477, 2020-2.2.1-ED-2021-00181, and TKP2021-NKTA-64 (Hungary); the Council of Science and Industrial Research, India; the Latvian Council of Science; the Ministry of Education and Science, project no. 2022/WK/14, and the National Science Center, contracts Opus 2021/41/B/ST2/01369 and 2021/43/B/ST2/01552 (Poland); the Funda\c{c}\~ao para a Ci\^encia e a Tecnologia, grant CEECIND/01334/2018 (Portugal); the National Priorities Research Program by Qatar National Research Fund; MCIN/AEI/10.13039/501100011033, ERDF ``a way of making Europe," and the Programa Estatal de Fomento de la Investigaci{\'o}n Cient{\'i}fica y T{\'e}cnica de Excelencia Mar\'{\i}a de Maeztu, grant MDM-2017-0765 and Programa Severo Ochoa del Principado de Asturias (Spain); the Chulalongkorn Academic into Its 2nd Century Project Advancement Project, and the National Science, Research and Innovation Fund via the Program Management Unit for Human Resources \& Institutional Development, Research and Innovation, grant B05F650021 (Thailand); the Kavli Foundation; the Nvidia Corporation; the SuperMicro Corporation; the Welch Foundation, contract C-1845; and the Weston Havens Foundation (USA).
\end{acknowledgments}

\bibliography{auto_generated} 

\cleardoublepage \appendix\section{The CMS Collaboration \label{app:collab}}\begin{sloppypar}\hyphenpenalty=5000\widowpenalty=500\clubpenalty=5000\input{SUS-21-006-public-authorlist.tex}\end{sloppypar}
\end{document}

%% file: SUS-21-006-public-authorlist.tex
\cmsinstitute{Yerevan Physics Institute, Yerevan, Armenia}
{\tolerance=6000
A.~Hayrapetyan, A.~Tumasyan\cmsAuthorMark{1}\cmsorcid{0009-0000-0684-6742}
\par}
\cmsinstitute{Institut f\"{u}r Hochenergiephysik, Vienna, Austria}
{\tolerance=6000
W.~Adam\cmsorcid{0000-0001-9099-4341}, J.W.~Andrejkovic, T.~Bergauer\cmsorcid{0000-0002-5786-0293}, S.~Chatterjee\cmsorcid{0000-0003-2660-0349}, K.~Damanakis\cmsorcid{0000-0001-5389-2872}, M.~Dragicevic\cmsorcid{0000-0003-1967-6783}, A.~Escalante~Del~Valle\cmsorcid{0000-0002-9702-6359}, P.S.~Hussain\cmsorcid{0000-0002-4825-5278}, M.~Jeitler\cmsAuthorMark{2}\cmsorcid{0000-0002-5141-9560}, N.~Krammer\cmsorcid{0000-0002-0548-0985}, D.~Liko\cmsorcid{0000-0002-3380-473X}, I.~Mikulec\cmsorcid{0000-0003-0385-2746}, J.~Schieck\cmsAuthorMark{2}\cmsorcid{0000-0002-1058-8093}, R.~Sch\"{o}fbeck\cmsorcid{0000-0002-2332-8784}, D.~Schwarz\cmsorcid{0000-0002-3821-7331}, M.~Sonawane\cmsorcid{0000-0003-0510-7010}, S.~Templ\cmsorcid{0000-0003-3137-5692}, W.~Waltenberger\cmsorcid{0000-0002-6215-7228}, C.-E.~Wulz\cmsAuthorMark{2}\cmsorcid{0000-0001-9226-5812}
% ...
\par}
\vskip\cmsinstskip
\dag:~Deceased\\
$^{1}$Also at Yerevan State University, Yerevan, Armenia\\
$^{2}$Also at TU Wien, Vienna, Austria\\
% ...
